\documentclass[draft]{agujournal2019}
\usepackage{url} 
\usepackage[inline]{trackchanges} 
\usepackage{soul}
\usepackage{amsmath}

\usepackage{xcolor}

\DeclareFontFamily{OMS}{oasy}{\skewchar\font48 }
\DeclareFontShape{OMS}{oasy}{m}{n}{%
         <-5.5> oasy5     <5.5-6.5> oasy6
      <6.5-7.5> oasy7     <7.5-8.5> oasy8
      <8.5-9.5> oasy9     <9.5->  oasy10
      }{}
\DeclareFontShape{OMS}{oasy}{b}{n}{%
       <-6> oabsy5
      <6-8> oabsy7
      <8->  oabsy10
      }{}
\DeclareSymbolFont{oasy}{OMS}{oasy}{m}{n}
\SetSymbolFont{oasy}{bold}{OMS}{oasy}{b}{n}

\DeclareMathSymbol{\smallleftarrow}     {\mathrel}{oasy}{"20}
\DeclareMathSymbol{\smallrightarrow}    {\mathrel}{oasy}{"21}
\DeclareMathSymbol{\smallleftrightarrow}{\mathrel}{oasy}{"24}

\usepackage{comment}
\usepackage{float}
\draftfalse

\journalname{JGR: Space Physics}

\begin{document}

%
%

\title{Scaling of electron heating by magnetization during reconnection and applications to dipolarization fronts and super-hot solar flares}




%
%




\authors{M. Hasan Barbhuiya\affil{1}, P. A. Cassak\affil{1}, M. A. Shay\affil{2}, Vadim Roytershteyn\affil{3}, M. Swisdak\affil{4}, Amir Caspi\affil{5}, Andrei Runov\affil{6}, Haoming Liang\affil{7}}

 \affiliation{1}{Department of Physics and Astronomy and the Center for KINETIC Plasma Physics, West Virginia University, WV 26506, USA}
 \affiliation{2}{Department of Physics and Astronomy and the Bartol Research Center, University of Delaware, Newark, DE 19716, USA}
 \affiliation{3}{Space Science Institute, Boulder, CO 80301, USA}
 \affiliation{4}{Institute for Research in Electronics and Applied Physics, University of Maryland, College Park, MD 20742, USA}
 \affiliation{5}{Southwest Research Institute, Boulder, CO 80302, USA}
 \affiliation{6}{Department of Earth and Space Sciences, University of California Los Angeles, CA 90095, USA}
\affiliation{7}{Center for Space Plasma and Aeronomic Research, University of Alabama in Huntsville, Huntsville, AL 35899, USA}






\correspondingauthor{M. Hasan Barbhuiya}{mhb0004@mix.wvu.edu}




\begin{keypoints}
\item We predict major and minor radii of ring distributions during reconnection in terms of upstream parameters and confirm with PIC simulations
\item We find that ring distributions occur at a shoulder (plateau) in the reconnected magnetic field in the simulations
\item The predicted temperatures are comparable to observed values in dipolarization fronts in Earth's magnetotail and in super-hot solar flares 
\end{keypoints}

%
%

%
%


\begin{abstract}
Electron ring velocity space distributions have previously been seen in numerical simulations of magnetic reconnection exhausts and have been suggested to be caused by the magnetization of the electron outflow jet by the compressed reconnected magnetic fields [Shuster et al., {\it Geophys.~Res.~Lett.}, {\bf 41}, 5389 (2014)]. We present a theory of the dependence of the major and minor radii of the ring distributions solely in terms of upstream (lobe) plasma conditions, thereby allowing a prediction of the associated temperature and temperature anisotropy of the rings in terms of upstream parameters.  We test the validity of the prediction using 2.5-dimensional particle-in-cell (PIC) simulations with varying upstream plasma density and temperature, finding excellent agreement between the predicted and simulated values. We confirm the Shuster et al.~suggestion for the cause of the ring distributions, and also find that the ring distributions are located in a region marked by a plateau, or shoulder, in the reconnected magnetic field profile.
The predictions of the temperature are consistent with observed electron temperatures in dipolarization fronts, and may provide an explanation for the generation of plasma with temperatures in the 10s of MK in super-hot solar flares. A possible extension of the model to dayside reconnection is discussed. Since ring distributions are known to excite whistler waves, the present results should be useful for quantifying the generation of whistler waves in reconnection exhausts.
\end{abstract}

 \section*{Plain Language Summary}
Solar flares and geomagnetic substorms are naturally occurring eruptions in space that can impact humans on Earth due to space weather. Both are caused by magnetic reconnection, during which magnetic field lines break and release energy into the surrounding ionized gas (plasma). From past research, we know that electrons near the reconnection site get magnetized in the strong magnetic fields that have already undergone reconnection, leading to a characteristic ring distribution of their velocities where all particles have similar speed in the plane perpendicular to the magnetic field.  We predict the speed of the particles in terms of the ambient properties of the easily measured surrounding plasma, and we confirm the prediction with numerical simulations. We show that the rings are located in a region where there is a leveling off of the magnetic field strength, which is a signature that can be used to identify ring distributions in future satellite measurements. We then use the result to predict temperatures in geomagnetic substorms and solar flares, finding that there is reasonable agreement. This suggests that we can understand the observed temperatures in terms of the ambient plasma properties, which will make it easier to predict these temperatures going forward.

%
%

%


%
%
%
%
\section{Introduction}
\label{sec:intro}

Energy conversion by magnetic reconnection, and its after effects, are of significant importance in numerous magnetospheric and solar processes \cite{Birn07,Gonzalez16}. Two examples are solar flares, which are energetic eruptions in the solar corona caused by reconnection \cite{Priest02}, and geomagnetic storms and substorms, during which energy from the interplanetary magnetic field gets stored and released via reconnection in Earth's magnetotail 
\cite{Angelopoulos08}.  Some of the magnetic energy released during reconnection appears as bulk flow energy of a 
plasma jet. In Earth's magnetotail, the energy in the jet is
ultimately injected into the inner magnetosphere where it can greatly impact magnetospheric dynamics and has important space weather implications \cite{McPherron79,Pulkkinen07}. Analogous dynamics takes place in magnetospheres of other planets \cite{Smith18,Xu21} and in sunward jets that occur during solar flares \cite{Reeves08}.

In Earth's magnetotail, the reconnected magnetic field on the Earthward side of the reconnection site dipolarizes as it releases its stored energy \cite{Fu20}. The Earthward reconnection jet impinges on the pre-existing and relatively dense plasma sheet, which acts as an obstacle to the jet \cite{hesse&birn_1991_JGR}. The jet's kinetic energy compresses the reconnecting magnetic field, producing a dipolarization front (DF) \cite{ohtani_2004_JGR,runov_2009_GRL,sitnov_JGR_2009,runov_2010_Planet_Sci,Runov11,sitnov:2011,Hwang11,Schmid11,runov_2013_JGR,Fu12,Fu13,sitnov:2013}. 
\textcolor{black}{(It has been argued that a more appropriate name for DFs is ``reconnection jet fronts,'' but we retain the name dipolarization fronts to conform to the majority of the literature.)}
Characteristic properties of DFs at the Earthward jet include a steep increase in the magnetic field component $B_z$ normal to the plasma sheet and a steep decrease in plasma density as one goes in the tailward direction.  Here, we use Geocentric Solar Magnetospheric (GSM) coordinates, for which $x$ is Sunward, $y$ is the duskward direction normal to $x$ and Earth's magnetic dipole, and $z$ completes the right-handed coordinate system in the northward direction. DFs have been seen in the mid-tail plasma sheets associated with bursty bulk flows (BBFs) \cite{angelopoulos_1992_JGR}. Energy in the compressed magnetic field in DFs has been observed to convert into particle kinetic energy \cite{angelopoulos_2013_Sci} and particle heating \cite{runov_2015_JGR} while the DFs move Earthward. 

One of the many consequences of DFs, and the focus of this study, is that electrons are significantly heated near the fronts. 
An electron temperature $T_e$ close to 1.8 keV was observed in a DF event by Time History of Events and Macroscale Interactions during Substorms (THEMIS), a factor of $\sim$3 higher than the electron temperature before the spacecraft crossed the DF, with a small perpendicular temperature anisotropy $T_{e,\perp} > T_{e,\|}$, where $\perp$ and $||$ denote the directions perpendicular and parallel to the local magnetic field $\vec{B}$ \cite{runov_2010_Planet_Sci}. Later observations revealed electron temperatures in the DFs in the range of 1--4 keV \cite{runov_2015_JGR}.  Observational studies \cite{Fu11,Pan12,ashour-abdalla_observations_2011,Liu19} attributed such heating to adiabatic processes such as Fermi and betatron acceleration. Moreover, observations of electron velocity distribution functions in DFs reveal various non-isotropic electron pitch-angle distributions (PADs)  \cite{Wu06,Fu12b,Tang21}. So-called pancake PADs have a perpendicular temperature anisotropy
\cite{Wu13}.  They were attributed to betatron acceleration in the compressed magnetic field of the DF
\cite{Xu18}. Also observed are so-called rolling pin PADs, which are a combination of a cigar PAD (with particles moving parallel and antiparallel to the local magnetic field, generated by Fermi acceleration in the bent magnetic field \cite{Wang14}) and a pancake PAD \cite{liu_explaining_2017}. Analytical theory suggests particle distributions with a perpendicular temperature anisotropy are unstable to wave generation, including whistler waves \cite{Gary85}. Whistler waves have been detected near DFs using satellite observations and cause non-adiabatic electron heating through wave-particle interactions \cite{LeContel09,Deng10,Viberg14,Li15,Yoo19}. A later observational study \cite{Grigorenko20} revealed that whistler waves heat electrons to
1--5 keV 
in rolling pin PADs.

Electron dynamics in DFs have also been studied extensively in 
numerical simulations. Motivated by observations, particle-in-cell (PIC) simulations have been used to study two broad classes of DFs: (i) flux rope (FR) type DFs with multiple X-lines, and (ii) flux bundle (FB) type DFs with a single transient X-line \cite{divin:2007,sitnov_JGR_2009, lu_2016_JGR}.
The energization mechanism for electrons in FR-type DFs was found to be repeated reflections between the double peaked $B_z$ structure present 
when there are two X-lines, and is betatron acceleration caused by the compressed $B_z$ in FB-type DFs  \cite{Birn13,lu_2016_JGR}. 
A strong electron temperature anisotropy with $T_{e,\perp} > T_{e,||}$ appears in the magnetic flux pile-up region of FR-type DFs in their PIC simulations and this anisotropy was shown to generate  whistler waves  \cite{fujimoto_2008_whistler}. Electron velocity distribution functions in the electron diffusion region (EDR) and the downstream region were systematically investigated using PIC simulations \cite{Shuster2014,Bessho2014}.  It was shown that the perpendicular temperature anisotropy is associated with electron ring distributions, {\it i.e.,} distributions that are toroidal in velocity space.  They suggested the ring distributions form when electron outflow jets from reconnection get remagnetized by the stronger normal magnetic field $B_z$ in the DF. 
In subsequent studies \cite{shuster_2015, Wang_2016_electron}, it was argued that this magnetization by the reconnected magnetic field heats the electrons downstream of the EDR.
In another PIC simulation study \cite{egedal_2016_PoP}, electron ring distributions were found to grow
in size when moving
downstream from the X-line as a result of betatron heating.
Recent PIC simulations  \cite{huang_formation_2021} suggest that as the DF moves downstream, 
first pancake PADs appear (as a result of betatron acceleration), followed by rolling pin PADs (when particles 
undergo Fermi reflections along with betatron acceleration), and culminating with cigar PAD (when Fermi acceleration becomes the dominating heating mechanism). Thus, electron ring distribution functions are associated with elevated temperatures, wave generation, and subsequent heating via wave-particle interactions in the region of DFs in Earth's magnetotail.

In the solar corona, reconnection during solar flares produces sunward jets (``reconnection outflows'') that have some similarities to DFs \cite{Reeves08}. These jets are associated with both particle acceleration and plasma heating. Solar flares routinely exhibit temperatures of $\sim$10--25~MK ($\sim$0.9--2.2~keV), generally thought to result from collisional energy transfer by particles accelerated to tens or hundreds of keV in or near the reconnection region impacting the dense chromosphere and heating the ambient plasma, whereupon it expands to fill the newly-reconnected flare loop in a process
called chromospheric evaporation \cite{Holman11}. However, a growing body of evidence suggests that the hottest plasmas in the flare thermal distribution are heated directly in the corona  \cite{Fletcher11,Cheung19}. While this likely occurs to some extent in flares of all intensities \cite{Warmuth16}, it appears most pronounced for so-called ``super-hot'' flares, where peak temperatures exceed 30~MK ($\sim$2.6~keV), significantly hotter than the component heated by chromospheric evaporation. Spectroscopic imaging analyses show that the super-hot plasma appears earlier and higher in the flare loop/arcade than the evaporative component \cite{Caspi10, Caspi15}. The densities of the super-hot component are $\sim$10 times smaller than the evaporative component, but $\sim$10 times larger than the background coronal plasma \cite{Caspi10}, suggestive of significant plasma compression. Such super-hot temperatures also appear to be associated exclusively with strong coronal magnetic fields exceeding 100~G \cite{Caspi14} and have a quasi-impulsive time profile, suggesting the mechanism for the heating of the super-hot plasma is directly connected to the magnetic reconnection process itself \cite{Caspi10}. Many super-hot plasma heating mechanisms have been suggested, including Ohmic pre-heating coupled followed by Fermi and betatron acceleration from collapsing magnetic traps \cite{Caspi10b}, gas dynamic shock heating from relaxation of the reconnected magnetic loop \cite{Longcope11, Longcope16}, Fokker-Planck collisions \cite{Allred20}, and others [\cite{Warmuth16} and references therein], but there is not yet a widely-accepted model.


We are not aware of any studies which give a first-principles prediction of the temperatures of the hot electrons downstream of reconnection exhausts as a function of the upstream plasma conditions, \textit{i.e.}, the upstream (lobe) magnetic field, electron temperature and density. Such a prediction requires an understanding of the processes causing the complex electron distribution functions in reconnection exhausts. 
In this study, for reasons justified in what follows, we focus on 
electron ring distributions in the region of the dipolarization front.  Our starting point is the suggestion \cite{Shuster2014,Bessho2014} that electron ring distributions are formed by the remagnetization of electron jets from reconnection. We quantitatively predict the major and minor radii of the ring distributions solely in terms of plasma parameters in the region upstream of the reconnecting region.  In particular, if the ring distributions are formed by the magnetization of electron jets, the major radius is governed by the electron Alfv\'en speed of the electron outflow jet, and the minor radius is governed by the electron thermal speed. To test the predictions, we perform a parametric study using two-dimensional (2D) PIC simulations in which the upstream density and upstream temperature are independently varied. We find ring distributions appear in all ten simulations we perform, and the major and minor radii depend on the upstream plasma parameters in the predicted manner. We further show that the associated electron temperature and temperature anisotropy largely scale according to analytical predictions of the major and minor radii, with the perpendicular temperature in excellent agreement and the parallel temperature being more complicated because there are counterpropagating electron beams along the magnetic field that are not incorporated in the present model. We find the electron ring distributions are associated with the highest electron temperature observed in the simulations, justifying their systematic study here. We confirm that the location at which electron ring distributions appear is associated with the location where the radius of curvature of the magnetic field exceeds the gyroradius based on the bulk flow speed, validating the suggestion by \citeA{Shuster2014} and \citeA{Bessho2014} that the ring distributions form as a result of remagnetization of the electrons.  We also show the ring distributions are suppressed by the presence of a background guide field, as is expected if they are caused by remagnetization. Moreover, we show that electron ring distributions consistently appear where there is a plateau, or shoulder, in the profile of the normal magnetic field $B_z$ downstream of the reconnection exhaust, which may be a useful signature for future observational studies. Finally, we show that the electron temperatures predicted from the theory are comparable to observed temperatures when applied to dipolarization fronts in Earth's magnetotail and super-hot solar flares in the solar corona.

This manuscript is organized as follows. Section~\ref{sec:theo} relates the major and minor radii of the ring distributions to upstream (lobe) plasma parameters and provides the associated analytical expressions of the temperature of ring distributions.  Section~\ref{sec:sims} describes the PIC simulations used in the study. Section~\ref{sec:results} shows the simulation results, revealing ring distributions in all the simulations. Their major and minor radii are extracted and compared to the theory. The location of the ring distributions is related to features in the temperature and magnetic field profiles, and we confirm the rings are caused by remagnetization of the electron outflow jet.
We discuss applications to dipolarization fronts and super-hot solar flares in Section~\ref{sec:discussions}.  We also discuss extending the theory to asymmetric reconnection for dayside magnetopause applications, and discuss implications for direct {\it in situ} observations of ring distributions. The manuscript concludes with Section~\ref{sec:conclusions}, where the key findings and limitations of our study are gathered, and future work is discussed.

\section{Theory}
\label{sec:theo}

We aim to relate the major and minor radii of ring distributions to macroscopic upstream properties of the reconnection process, {\it i.e.,} number density, temperature and magnetic field.  
One form of an ideal ring velocity distribution function $f_{r}(v_\perp,v_{\|})$ is  \cite{wu_1989_a,min&liu_2016a}
  \begin{linenomath*}
  \begin{equation}
    f_{r}\left(v_{\perp}, v_{\|}\right)=\frac{n_{r}}{\pi^{3 / 2} v_{Th}^{3} \Lambda} e^{-\frac{v_{\|}^{2}}{v_{Th}^{2}}} e^{\frac{-\left(v_{\perp}-v_{\perp 0}\right)^{2}}{v_{Th}^{2}}},
  \label{eq:ringVDF}
  \end{equation}
  \end{linenomath*}
where $n_r$ is the number density, $v_{\|}$ and $v_\perp$ are the velocity space coordinates parallel and perpendicular to the central axis of the ring distribution, $v_{\perp0}$ is the major radius of the ring distribution,  
and $v_{Th}$ is the minor radius of the ring distribution, assumed to be Gaussian and isotropic in the parallel and perpendicular directions. 
The normalization factor $\Lambda$, defined by $\Lambda=r \sqrt{\pi} \operatorname{erfc}(-r)+e^{-r^{2}}$, enforces that $n_r = \int d^3v f_{r}$; here $r = v_{\perp0}/v_{Th}$ and erfc($-r$) = $(2/\sqrt{\pi})\int_{-r}^\infty e^{-z^2} dz$ is the complementary error function. 

It was previously suggested \cite{Shuster2014,Bessho2014} that electron ring distributions form when the electron jet from reconnection gets magnetized by the strong normal (reconnected) magnetic field occurring as a result of compression at the dipolarization front. In principle, the same effect can happen for ions, but we only see rings in our simulations for electrons so we focus on them here.
We expect the major radius of the ring distribution
$v_{\perp 0}$ to be the electron outflow speed before the beam gets magnetized, which scales as the electron Alfv\'en speed $c_{Aup,e}$ \cite{Shay01,hoshino01a} based on the reconnecting magnetic field strength $B_{up,e}$ at the upstream edge of the EDR,
  \begin{linenomath*}
  \begin{equation}
    v_{\perp 0}=\frac{B_{up, e}}{\sqrt{4 \pi m_{e} n_{up}}},
  \label{eq:vperp0upOnly}
  \end{equation}
  \end{linenomath*}
where $m_e$ is the electron mass and $n_{up}$ is the density upstream of the EDR which is comparable to the density upstream of the ion diffusion region (IDR) and therefore the upstream (lobe) plasma.

For the minor radius $v_{Th}$,
we propose that
it is governed by the thermal speed $v_{Th}$ of the electron upstream of the reconnection site, {\it i.e.,}
  \begin{linenomath*}
  \begin{equation}
  v_{Th} = \sqrt{\frac{2 k_{B} T_{e,up}}{m_{e}}},
  \label{eq:vThupOnly}
  \end{equation}
  \end{linenomath*}
where $k_B$ is Boltzmann's constant and $T_{e,up}$ is the temperature upstream of the EDR, which is essentially the same as the (lobe) temperature upstream of the IDR at the early times when reconnection that forms a dipolarization front takes place. This effectively assumes that 
the increase in temperature that takes place as electrons flow through the EDR or across separatrices as they go into the exhaust
\cite{Shay14} is small.
Using Eqs.~(\ref{eq:vperp0upOnly}) and (\ref{eq:vThupOnly}), we write the parameter $r$ in terms of upstream parameters as
 \begin{linenomath*}
 \begin{equation}
 r = \frac{B_{up,e}}{\sqrt{8\pi n_{up} k_B T_{e,up}}},
 \label{eq:rupOnly}
 \end{equation}
 \end{linenomath*}
which is related to a form of the upstream electron plasma $\beta_{e,up}$ as $r = \beta_{e,up}^{-1/2}$. Using these expressions, we have the parameters necessary to write Eq.~(\ref{eq:ringVDF}) solely in terms of upstream plasma parameters.

\begin{figure}
    \begin{center}\includegraphics[width=4.5in]{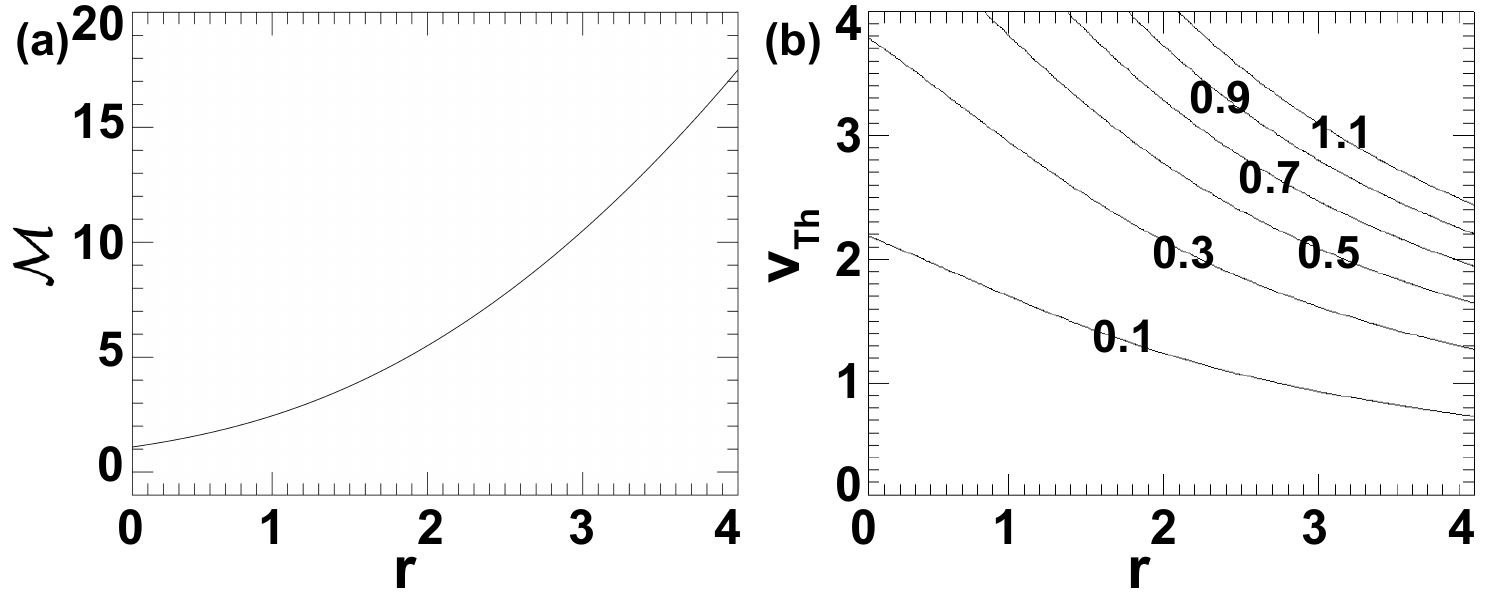}
     \caption{(a) Line plot of $\mathcal{M}$ from Eq.~(\ref{eq:ringVDFTeperpMdef}) as a function of $r = v_{\perp0}/\textcolor{black}{\textcolor{black}{v_{Th}}}$. (b) Contour plot of $T_{\perp}$ from Eq.~(\ref{eq:ring+coreVDFTeperppara}) as a function of $r$ and $v_{Th}$ for a ring population plus a Maxwellian core population assuming $n_M = n_r$ and $v_{Th,M} = v_{Th}$.}
    \label{fig:ContourPlotsTeperp}
    \end{center}
\end{figure}

The perpendicular and parallel temperatures $T_\perp$ and $T_{||}$ associated with the ring distribution in Eq.~(\ref{eq:ringVDF}) are calculated in the standard way using the second velocity moment of $f_r$, {\it i.e.}, $T_{\perp}=[m /(2n_r k_{B})] \int d^{3} v(\vec{v}_{\perp}-\vec{u}_{\perp})^2 f_{r}$  and $T_{||}=[m /(n_r k_{B})] \int d^{3} v(v_{||}-u_{||})^2 f_{r}$,
where 
$\vec{u}_{\perp}$ and $u_{||}$ are the perpendicular and parallel components of the bulk flow velocity calculated from the first velocity moment of the distribution function,
$\vec{u}_\perp = (1/n_r) \int d^3v \vec{v}_\perp f_r$ 
and $u_{\|} = (1/n_r) \int d^3v v_{\|} f_r$.  Since both $\vec{u}_\perp$ and $u_{\|}$ are zero for $f_r$ as given in Eq.~(\ref{eq:ringVDF}), the
resulting $T_{\perp}$ and $T_{\parallel}$ are \cite{wu_1989_a}
\begin{linenomath*}
  \begin{equation}
     T_{\perp}=\mathcal{M}T_{e,up},~T_{\parallel}=T_{e,up}
    \label{eq:ringVDFTeperppara},
  \end{equation}
 \end{linenomath*}
where 
\begin{linenomath*}
  \begin{eqnarray}
    \mathcal{M} & = & \frac{2 e^{-r^{2}}\left(1+r^{2}\right)+\sqrt{\pi} r\left(3+2 r^{2}\right) \operatorname{erfc}(-r)}{2\Lambda} \\ 
    & = & \frac{3}{2} + r^2 - \frac{e^{-r^{2}}}{2\Lambda}. 
  \label{eq:ringVDFTeperpMdef}
  \end{eqnarray}
\end{linenomath*}
A plot of ${\cal M}$ as a function of $r$ is shown in Fig.~\ref{fig:ContourPlotsTeperp}(a).
The effective temperature $T_{{\rm eff}}= (2T_{\perp}+  T_{\|})/3$ is
\begin{linenomath*}
  \begin{equation}
    T_{{\rm eff}}= T_{e,up} \left(\frac{2 \mathcal{M} + 1}{3} \right).
  \label{eq:ringVDFTe}
  \end{equation}
\end{linenomath*}
The temperature anisotropy, defined as $A_{\perp} = T_{\perp}/T_{\parallel}-1$, is 
\begin{linenomath*}
  \begin{equation}
    A_{\perp}= \mathcal{M}-1.
  \label{eq:ringVDFTeAniso}
  \end{equation}
\end{linenomath*}
Thus, Eq.~(\ref{eq:ringVDFTe}) is equivalent to $T_{{\rm eff}}= T_{e,up} (1 + 2 A_\perp/3)$ for this distribution. These expressions give the properties associated with the ring distribution in terms of upstream parameters.

It has been shown \cite{Shuster2014,egedal_2016_PoP} that ring-type distributions in PIC simulations of reconnection are not always ideal like in Eq.~(\ref{eq:ringVDF}); some also have a 
Maxwellian core population. \textcolor{black}{It is possible that this population is related to the initial current sheet population in the simulations, but validating this conjecture is not carried out for the present study. It is not clear if this population is a numerical artifact or also present in Nature. Since it is not the focus of the present study and has been seen in previous independent studies, we simply include it in our analysis to give more accurate comparisons to the simulations.} Thus, we derive the temperatures associated with a distribution $f = f_r + f_M$ that is the sum of the ideal ring distribution $f_r$ from Eq.~(\ref{eq:ringVDF}) and a Maxwellian distribution $f_M =(n_{M}/\pi^{3 / 2} v_{T h, M}^{3}) e^{-v^{2}/v_{Th,M}^{2}}$ with density $n_M$ and temperature $T_M$ associated with the thermal speed
$v_{Th,M}=(2k_B T_M/m)^{1/2}$. The zeroth velocity moment of this distribution gives the total local density as $n=n_r + n_M$.  The temperatures generalizing Eqs.~(\ref{eq:ringVDFTeperppara}) and 
(\ref{eq:ringVDFTe}) are
\begin{linenomath*}
  \begin{equation}
    T_{\perp }=\mathcal{M}\frac{n_r T_{e,up}}{n} + \frac{n_M T_M}{n}, ~ T_{\parallel} =\frac{n_r T_{e,up}}{n}+\frac{n_M T_M}{n} 
  \label{eq:ring+coreVDFTeperppara}
  \end{equation}
\end{linenomath*}
\begin{linenomath*}
  \begin{equation}
     T_{{\rm eff}}=\left(\frac{2 \mathcal{M}+1}{3}\right) \frac{n_r T_{e,up}}{n} + \frac{n_M T_M}{n},
  \label{eq:ring+coreVDFTe}
  \end{equation}
\end{linenomath*}
while the temperature anisotropy in Eq.~(\ref{eq:ringVDFTeAniso}) becomes
\begin{linenomath*}
  \begin{equation}
    A_{\perp}=\frac{(\mathcal{M}-1)n_r T_{e,up}}{n_r T_{e,up} + n_M T_M}.
  \label{eq:ring+coreVDFTeAniso}
  \end{equation}
\end{linenomath*}
A contour plot of $T_{\perp}$ as a function of $r$ and $v_{Th}$ in the limit that $n_M=n_r$ and $v_{Th.M}=v_{Th}$ is shown for reference in Fig.~\ref{fig:ContourPlotsTeperp}(b). These expressions will be useful when we analyze ring distributions in our PIC simulations.

\section{Simulations}
\label{sec:sims}

We use the PIC code {\tt p3d} \cite{zeiler2002} to perform simulations of symmetric antiparallel magnetic reconnection that are 2.5D in position space and 3D in velocity space.  {\tt p3d} employs the trapezoidal leapfrog method \cite{guzdar93a} to advance electromagnetic fields in time 
and the particles are advanced in time using a relativistic Boris stepper \cite{birdsall91a}. The multigrid technique \cite{Trottenberg00} is used to clean the divergence of the electric field every 10 particle time-steps.

In the simulations, lengths are normalized to the ion inertial scale $d_{i0} = c/\omega_{pi0}$ based on a reference density $n_0$ that is the peak density of the initial current sheet population, where $\omega_{pi0} = (4 \pi n_0 e^2 /m_i)^{1/2}$, $e$ is the ion charge, and $c$ is the speed of light. Magnetic fields are normalized to the initial asymptotic upstream reconnecting magnetic field $B_0$. Velocities are normalized to the Alfv\'en speed $c_{A0} = B_0/(4 \pi m_i n_0)^{1/2}$. Times are normalized to the inverse ion cyclotron frequency $\Omega_{ci0}^{-1}= (e B_{0} / m_{i} c)^{-1}$. Temperatures are normalized to $m_i c_{A0}^2/k_B$.  Reduced velocity distribution functions are normalized to $n_0/c^2_{Ao}$.

The simulation coordinate system is defined such that reconnection outflows are along $\pm \hat{x}$ and inflows are along $\pm\hat{z}$, with periodic boundary conditions in both directions. The simulations are initialized with two Harris current sheets and a uniform background plasma population. 
The initial magnetic field profile is 
 \begin{linenomath*}
 \begin{equation}
B_x(z) = \tanh{\left(\frac{z-l_z/4}{w_0}\right)}-\tanh{\left(\frac{z-3l_z/4}{w_0}\right)} -1,
\end{equation}
 \end{linenomath*}
with no initial out-of-plane guide magnetic field unless otherwise stated. Here, $w_0$ is the thickness of the current sheet and $l_z$ is the length of the computational domain in the $\hat{z}$ direction. The temperature and density of the background populations can be varied independently of the current sheet population. The initial electron and ion density profiles are  
 \begin{linenomath*}
 \begin{equation}
 n(z) = \frac{1}{2(T_{e,CS}+T_{i,CS})} \left[ \operatorname{sech}^{2}\left(\frac{z-l_z/4}{w_0}\right) + \operatorname{sech}^{2}\left(\frac{z-3l_z/4}{w_0}\right) \right] + n_{up},
 \end{equation}
 \end{linenomath*}
where $n_{up}$ is the initial density of the background plasma. The current sheet electron temperature $T_{e,CS}$ is uniform with a value of 1/12, and the current sheet ion temperature $T_{i,CS}$ is uniform with a value 5$T_{e,CS}$. 

The speed of light $c$ is 15, and the electron to ion mass ratio is $m_e/m_i = 0.04$. There are $4096 \times 2048$ grid cells in all the simulations, initialized with 100 weighted particles per grid (PPG). A weak initial magnetic perturbation of the form $\delta B_x = -B_{pert} \sin{(2 \pi x/l_x)} \sin{(4\pi z/l_z)}$ and $\delta B_z = B_{pert}l_z/(2 l_x) \cos{(2 \pi x/l_x)} [1-\cos{(4\pi z/l_z)}]$ with $B_{pert} = 0.025$ is used to seed an X- and O-line pair in each of the two current sheets, where $l_x$ is the computational domain size in the $\hat{x}$ direction.

\begin{table}
    \caption{Numerical parameters for two sets of simulations with varying (top) upstream total temperature $T_{TOT,up}$ and (bottom) upstream number density $n_{up}$.  $l_x$ and $l_z$ are system sizes along $\hat{x}$ and $\hat{z}$, respectively, $w_0$ is the initial current sheet thickness, $\Delta x$ is the grid scale along $\hat{x}$ and $\hat{z}$, and $\Delta t$ is the time step.}
 \centering
 \begin{tabular}{ccccccc}
 \hline
  $T_{TOT,up}$ & $l_x \times ~l_z$ & $w_0$ & $\Delta x$ & $\Delta t$  \\
 \hline
    0.2  & 51.20 $\times$ ~25.60 & 0.50 & 0.0125 & 0.00100  \\
    0.4  & 51.20 $\times$ ~25.60 & 0.50 & 0.0125 & 0.00100  \\
    0.6  & 51.20 $\times$ ~25.60 & 0.50 & 0.0125 & 0.00100  \\
    0.8  & 51.20 $\times$ ~25.60 & 0.50 & 0.0125 & 0.00100  \\
    1.0  & 51.20 $\times$ ~25.60 & 0.50 & 0.0125 & 0.00100  \\
 \hline
  $n_{up}$ & $l_x \times ~l_z$ & $w_0$ & $\Delta x$ & $\Delta t$ \\
 \hline
    0.2  & 51.20 $\times$ ~25.60 & 0.50 & 0.0125 & 0.00100  \\
    0.4  & 47.41 $\times$ ~23.71 & 0.46 & 0.0116 & 0.00093  \\
    0.6  & 44.35 $\times$ ~22.17 & 0.43 & 0.0108 & 0.00087  \\
    0.8  & 41.81 $\times$ ~20.91 & 0.41 & 0.0102 & 0.00082  \\
    1.0  & 39.68 $\times$ ~19.84 & 0.39 & 0.0097 & 0.00078  \\
 \hline
 \end{tabular}
 \label{table:setOfSims}
\end{table} 
Two sets of five simulations are performed.  Table~\ref{table:setOfSims} lists relevant simulation parameters, including the system size $l_x \times l_z$,
the initial current sheet half-thickness $w_0$, the grid scale $\Delta x$ in both
directions, and the time step $\Delta t$. In all simulations, the ion to electron temperature ratio $T_{i,up}/T_{e,up}$ of the background plasma is initially 5. One set of simulations has varying $T_{TOT,up} = T_{i,up} + T_{e,up}$, while the initial background density is kept fixed at $n_{up}$ = 0.2. The other set has varying $n_{up}$, with the initial background temperatures kept fixed at $T_{e,up}$ = 1/12 and $T_{i,up}$ = $5 T_{e,up}$. The smallest length scale for each of the simulations is the electron Debye length $\lambda_{De}$ based on the total initial density at the center of the current sheet 1 + $n_{up}$. Thus, $\lambda_{De}$ decreases as $n_{up}$ is increased from 0.2 to 1 by a factor of $(1.2/2)^{1/2}$, \textit{i.e.,} it is $22.5\%$ lower for the $n_{up}=1$ simulation than the $n_{up}=0.2$ simulation. Thus, for the $n_{up} = 1$ simulation, the system size, grid length, initial current sheet thickness, and time step are also reduced by $22.5\%$ (as listed in Table \ref{table:setOfSims}). For other $n_{up}$ values, a similar approach is used to determine their simulation parameters.

\textcolor{black}{Since we use periodic boundary conditions, the minimum system size that allows the ions to fully couple back to the reconnection process is approximately 40 $d_{i0}$ \cite{Pyakurel19}.} Since $l_x$ is smaller than necessary for ions to fully couple back to the reconnected magnetic field, this study focuses on electron dynamics. In some of the simulations, the upper current sheet develops secondary islands which do not coalesce with the primary island by the time the system reaches steady-state. Hence we focus on the lower current sheet.  Finally, we note that the ion and electron inertial lengths $d_i$ and $d_e$ based on the upstream (background) density are related to the length scale used for normalization via $d_i = d_{i0}/\sqrt{n_{up}}$ and $d_{e}=0.2 d_{i}$ for the mass ratio used in the simulations.  Since $n_{up}$ is fixed at 0.2 for the simulations with varying $T_{TOT,up}$, $d_i = 2.24 \ d_{i0}$ and $d_{e}=
0.45 \ d_{i0}$ for those simulations.
For simulations with varying $n_{up}$, the length scales change with $n_{up}$; for example, for $n_{up} = 1$, we have $d_i = d_{i0}$ and $d_{e}=0.2 \ d_{i0}$.
\textcolor{black}{Each simulation is carried out long enough for the reconnection to reach a steady-state, meaning that the reconnection rate becomes approximately constant  in time.}

For plotting reduced electron velocity distribution functions (rEVDFs), which are 2D velocity distributions produced from the full 3D distributions after integrating over one of the three velocity directions, a domain of size $0.5 \ d_{i0} \times 0.5 \ d_{i0}$ centered at the location of interest is used. 
A velocity space bin of size $0.1 \ c_{A0}$ is used in all velocity directions. 

\section{Methods and Results}
\label{sec:results}

\subsection{Presence of ring distributions}
\label{subsec:RingExistence}

A result of this simulation study is that all ten simulations reveal electron ring distributions beyond the downstream edge of the EDR near the region of the dipolarization fronts.  This is ascertained by plotting rEVDFs in the plane perpendicular to the local magnetic field. Since the magnetic field in the region of interest is predominantly in the $\hat{z}$ direction,
we identify $\hat{x} \approx (\hat{u} \times \hat{b}) \times \hat{b} \equiv \perp1$, $\hat{y} \approx \hat{u} \times \hat{b} \equiv \perp2$, and $\hat{z} \approx \hat{b} \equiv \parallel$, where $\hat{b}$ and $\hat{u}$ are the unit vectors along the magnetic field $\vec{B}$ and the bulk flow velocity $\vec{u}$.
Defining the X-line location as $(x_0,z_0)$, the rEVDFs are plotted along the horizontal line $z = z_0$ as a function of $x$ from the X-line to the magnetic island.
At the earliest times in the steady-state reconnection time interval for all simulations, we find 
that rEVDFs near the X-line have striations, and they are rotated by the reconnected magnetic field $B_z$ as one moves in the outflow direction within the EDR. 
Beyond the downstream edge of the EDR, ring-like features begin to arise in the distributions as some electrons complete at least one full gyration around $B_z$, leading to swirls and arcs (not shown), and finally to electron ring distributions for which most electrons complete at least one full gyration. These results are consistent with previous simulation studies \cite{Bessho2014,Shuster2014, shuster_2015,egedal_2016_PoP} 

\begin{figure}
    \begin{center}\includegraphics[width=4in]{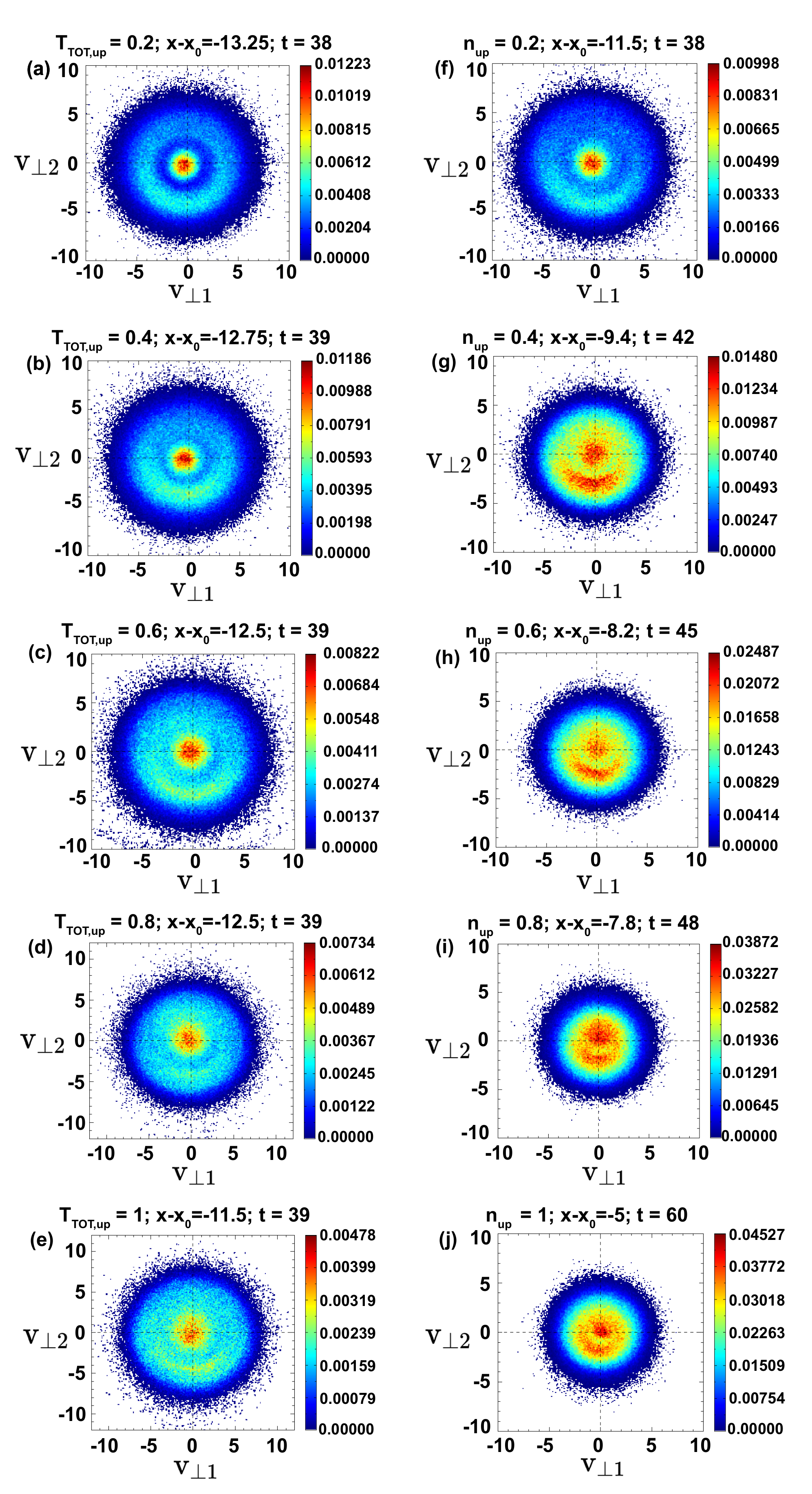}
     \caption{rEVDFs in the $v_{\perp1}-v_{\perp2}$ plane \textcolor{black}{(where velocities are given in units of the normalized Alfv\'en speed $c_{A0}$)} for simulations with (a) - (e) varying $T_{TOT,up}$ and (f) - (j) varying $n_{up}$. The title of each panel gives its background temperature or density as appropriate, the position $x-x_0$ where the rEVDF is measured relative to the X-line, and the time $t$. For all panels, $z-z_0=0$.}
    \label{fig:ringVDFsALLRUNS}
    \end{center}
\end{figure}

The panels of Fig.~\ref{fig:ringVDFsALLRUNS} show rEVDFs as a function of $v_{\perp1}$ and $v_{\perp2}$ for representative ring distributions seen in all ten simulations, with varying $T_{TOT,up}$ on the left from 0.2 to 1 in (a)-(e) and varying $n_{up}$ on the right from 0.2 to 1 in (f)-(j). The title on each panel provides the locations $x-x_0$ and times $t$ at which each rEVDF is plotted. The plotted rEVDFs reveal that there is a noticeable agyrotropy in the ring distributions, but the major and minor radii are well-formed.
\textcolor{black}{It is likely that the cause of the agyrotropy is that not all particles complete one full gyration, as also seen in previous studies \cite{Shuster2014}, but we do not study this feature further in the present study.}
Looking at the rEVDFs in other planes (not shown), we find that along with the ring population and the colder Maxwellian core \textcolor{black}{also seen in previous simulations studies \cite{Shuster2014}}, a population of counterstreaming beams is also present in every simulation \textcolor{black}{in} every rEVDF.  \textcolor{black}{As elevated values of $T_{e,||}$ that would be associated with parallel propagating beams are not seen at the reconnecting magnetic field reversal region in the study by \citeA{Shay14} [see their Fig.~2(d)], we believe it is likely that this population is an artifact due to our simulation size being smaller than in that previous study, leading to accelerated electrons to be transmitted through the boundary to the location we are measuring distributions, but we leave verifying this conjecture for future work.} These rEVDFs reveal that the ring distributions follow clear qualitative trends: with increasing background temperature $T_{TOT,up}$, the rings stay approximately the same size but are thicker in the $v_{\perp1} - v_{\perp2}$ plane [Fig.~\ref{fig:ringVDFsALLRUNS}(a)-(e)], whereas with increasing background density $n_{up}$, the rings shrink in size [Fig.~\ref{fig:ringVDFsALLRUNS}(f)-(j)] while maintaining a similar thickness.

\subsection{Parametric dependence of ring distribution major and minor radii}
\label{subsec:RingPartFitting}

We now quantitatively investigate the parametric dependence of the ring distributions by extracting their major and minor radii from the simulations.
For each distribution in Fig.~\ref{fig:ringVDFsALLRUNS}, we take separate 1D cuts of the rEVDF along $v_{\perp1}=0$ and $v_{\perp2}=0$.  For each 1D cut, we fit three Gaussians to the distribution given by $\sum_{i=1}^3 a_i e^{-[(x-b_i)/c_i]^2}$ using the \textit{Curvefit} tool in \textit{MATLAB R2020a}. The outer two Gaussians are used to fit the ring portion of the distribution and the central Gaussian is used to fit the core. The coefficients $a_i$ are used to calculate $n_r$ and $n_M$, $b_i$ give the bulk flow of each component of the distribution and are related to $v_{\perp0}$, and $c_i$ give the associated thermal speeds $v_{Th}$ and $v_{Th,M}$.

\begin{figure}
    \begin{center}\includegraphics[width=4in]{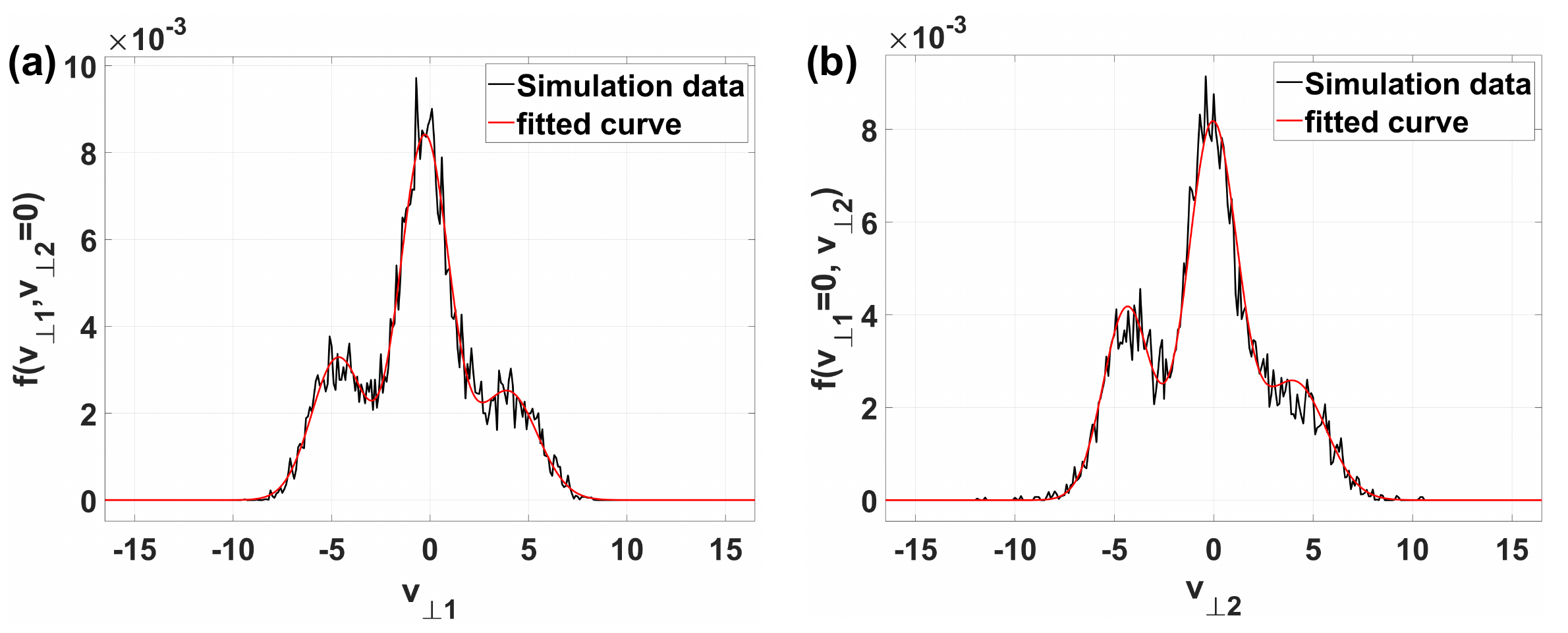}
    \caption{1D cuts of the rEVDF from Fig.~\ref{fig:ringVDFsALLRUNS}(f) for the simulation with $n_{up}$ = 0.2 (black solid curve). The red solid curve is the best fit to three Maxwellians. The cuts are (a) $f(v_{\perp1}$,$v_{\perp2}$=0) and (b) $f(v_{\perp1}$=0,$v_{\perp2})$.}
    \label{fig:ThreeGaussFit}
    \end{center}
\end{figure}

As a case study, 1D cuts and the associated fits are shown in Fig.~\ref{fig:ThreeGaussFit} for the $n_{up} = 0.2$ simulation from Fig.~\ref{fig:ringVDFsALLRUNS}(f). The black curve is the raw distribution function and the red curve is the best fit. Because the rEVDFs are not perfectly symmetric, the best fit coefficients and associated major and minor radii $v_{\perp0}$ and $v_{Th}$ are different in the $v_{\perp1} = 0$ and $v_{\perp2} = 0$ cuts.  We calculate average values for $v_{\perp0}$ and $v_{Th}$ and their standard deviations $\sigma$ derived from propagating the errors in quadrature. The best fit procedure also provides 95\% confidence bounds, which we take as another estimate of the uncertainty of the values. The results of this procedure for all ten simulations are listed in Table \ref{table:OneDfittingData}.  

\begin{table}
    \caption{Data from the fitting method described in Sec.~\ref{subsec:RingPartFitting} for all simulations. The first column gives the value being varied, and $n_r, v_{\perp0}$, and $v_{Th}$ are the ring density, major radius, and minor radius. The $\sigma$ values are standard deviations from the mean from cuts in the $\perp 1$ and $\perp 2$ directions, and 95\% err is the error calculated using 95\% confidence bounds from the fit.}
 \centering
 \begin{tabular}{c c c c c c c c}
 \hline
  $T_{TOT,up}$ & $n_{r}$ & $v_{\perp0}$ & $\sigma_{v_{\perp0}} $ & 95\% err$_{v_{\perp0}}$ & $v_{Th}$ & $\sigma_{v_{Th}}$ & 95\% err$_{v_{Th}}$  \\
 \hline        
    0.2  & 0.30 & 4.29 & 0.19 & 0.15 & 1.47 & 0.05 & 0.22 \\
    0.4  & 0.36 & 4.33 & 0.27 & 0.22 & 1.81 & 0.04 & 0.32 \\
    0.6  & 0.31 & 4.24 & 0.17 & 0.26 & 2.13 & 0.12 & 0.35 \\
    0.8  & 0.33 & 4.23 & 0.19 & 0.49 & 2.41 & 0.05 & 0.57 \\
    1.0  & 0.26 & 4.42 & 0.49 & 0.49 & 2.59 & 0.09 & 0.49 \\
 \hline
  $n_{up}$ & $n_{r}$ & $v_{\perp0}$ & $\sigma_{v_{\perp0}}$ & 95\% err$_{v_{\perp0}}$ & $v_{Th}$ & $\sigma_{v_{Th}}$ & 95\% err$_{v_{Th}}$ \\
 \hline
    0.2  & 0.28 & 4.26 & 0.32 & 0.12 & 1.99 & 0.23 & 0.17 \\
    0.4  & 0.46 & 2.93 & 0.32 & 0.19 & 2.11 & 0.19 & 0.19 \\
    0.6  & 0.91 & 2.52 & 0.29 & 0.19 & 2.08 & 0.15 & 0.17 \\
    0.8  & 1.17 & 1.99 & 0.13 & 0.35 & 1.99 & 0.07 & 0.28 \\
    1.0  & 1.28 & 1.89 & 0.07 & 0.12 & 1.94 & 0.11 & 0.12 \\
 \hline
 \end{tabular}
 \label{table:OneDfittingData}
\end{table}


We now compare the theoretical predictions for the major and minor radii
to the simulation results. For the theoretical predictions, we need to obtain $B_{up,e}, n_{up}$ and $T_{e,up}$ to evaluate $v_{\perp0}$ in Eq.~(\ref{eq:vperp0upOnly}) and $v_{Th}$ in Eq.~(\ref{eq:vThupOnly}).  
We define the upstream edge of the EDR where the electron bulk inflow speed 
starts to differ from the $\hat{z}$ component of the $\vec{E} \times \vec{B}$ velocity. Then, the measured plasma parameters are obtained by averaging quantities over $0.06~d_{i0}$ centered around this location.  We find that the upstream parameters vary in time, changing between the transient time when reconnection onset takes place and when a steady-state is reached. We reason that the dipolarization fronts occur due to jets that arise in the transient initial phase of reconnection. Therefore, we measure the upstream parameters at early times when the reconnection rate starts to increase. For the simulations with varying $T_{TOT,up}$, this time is $t$ = 5 whereas for $n_{up}$ simulations, the time varies from $t$ = 5 for $n_{up} = 0.2$ to $t$ = 10 for $n_{up} = 1$ since increasing $n_{up}$ from 0.2 to 1 decreases the speeds 
by a factor of $5^{1/2}$. 
At the chosen time, we average the desired upstream quantities over five code time units. We find that the data variations are small (within 5\%) during this interval. We also confirm the densities and temperatures do not vary appreciably between the upstream value at the electron layer and the upstream value at the ion layer. The results of this procedure are listed in Table \ref{table:UpstreamData}, along with theoretical predictions of $v_{\perp0}$ using Eq.~(\ref{eq:vperp0upOnly}) and $v_{Th}$ using Eq.~(\ref{eq:vThupOnly}).

\begin{table}
    \caption{Upstream plasma parameters from the simulations using the method described in Sec.~\ref{subsec:RingPartFitting}. The first column gives the value being varied, $B_{up,e}$ is the upstream magnetic field, $n_{up}$ is the upstream density, and $T_{e,up}$ is the upstream temperature at the EDR edge. The last two columns give the theoretical predictions for the major radius $v_{\perp0}$ and minor radius $v_{Th}$ based on the upstream values using Eqs.~(\ref{eq:vperp0upOnly}) and (\ref{eq:vThupOnly}), respectively.}
 \centering
 \begin{tabular}{c c c c c c}
 \hline
  $T_{TOT,up}$ & $B_{up,e}$ & $n_{up}$ & $T_{e,up}$ & Theoretical ${v_{\perp0}}$ & Theoretical $v_{Th}$\\
 \hline        
    0.2  & 0.33 & 0.14 & 0.034 & 4.41 & 1.30 \\
    0.4  & 0.34 & 0.14 & 0.068 & 4.54 & 1.84 \\
    0.6  & 0.33 & 0.14 & 0.10  & 4.41 & 2.24  \\
    0.8  & 0.36 & 0.16 & 0.13  & 4.50 & 2.55  \\
    1.0  & 0.35 & 0.15 & 0.17  & 4.52 & 2.92  \\
 \hline
  $n_{up}$ & $B_{up,e}$ & $n_{up}$ & $T_{e,up}$ & Theoretical ${v_{\perp0}}$ & Theoretical $v_{Th}$\\
 \hline
    0.2  & 0.35 & 0.15 & 0.084 & 4.51 & 2.05 \\
    0.4  & 0.36 & 0.32 & 0.086 & 3.18 & 2.07 \\
    0.6  & 0.38 & 0.51 & 0.087 & 2.66 & 2.08 \\
    0.8  & 0.36 & 0.69 & 0.086 & 2.17 & 2.07 \\
    1.0  & 0.37 & 1.01 & 0.083 & 1.84 & 2.04 \\
 \hline
 \end{tabular}
 \label{table:UpstreamData}
\end{table}

\begin{figure}
    \begin{center}\includegraphics[width=5in]{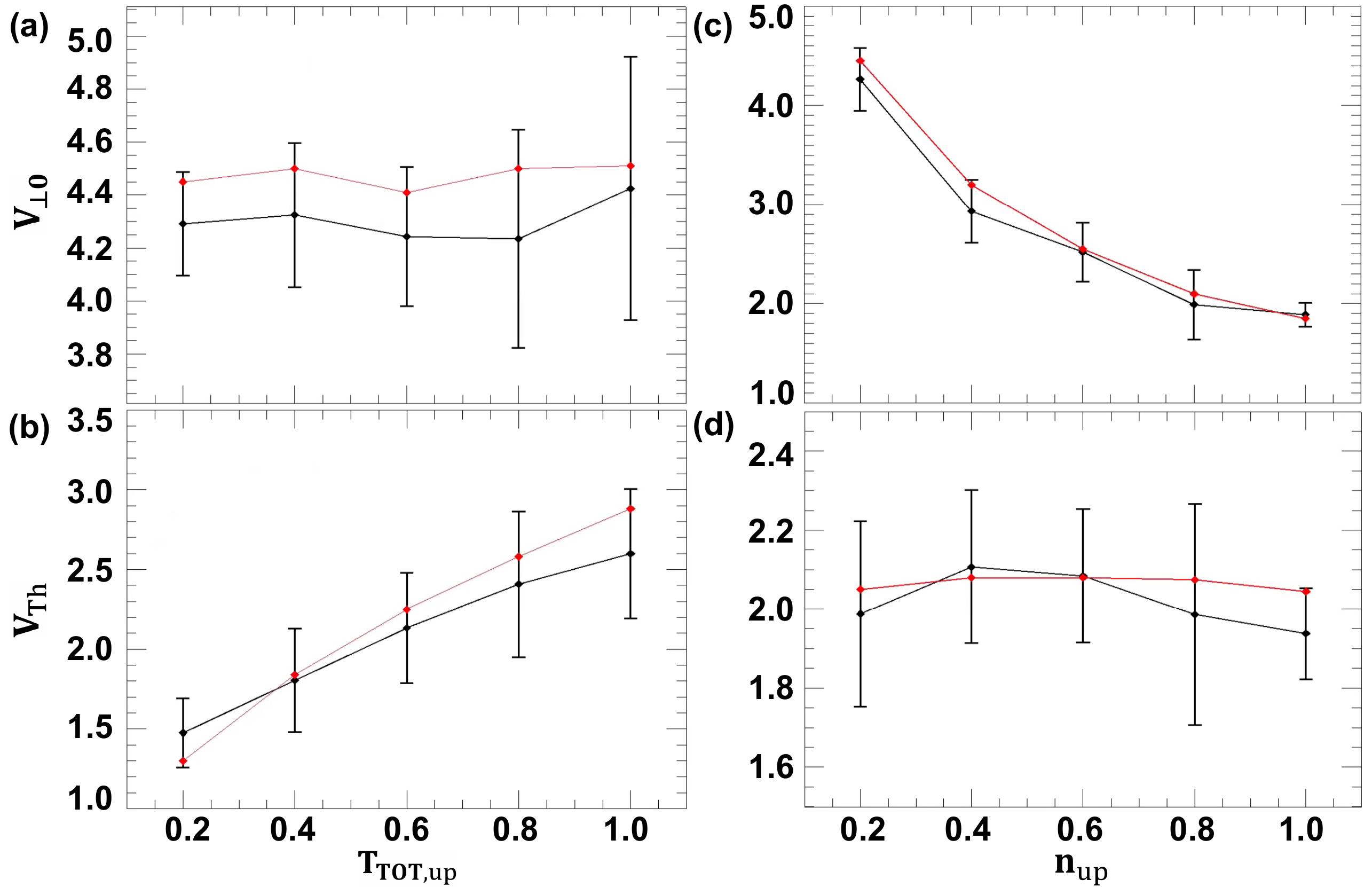}
    \caption{Ring distribution (a) and (c) major radius $v_{\perp,0}$ and (b) and (d) minor radius $v_{Th}$ from simulations with varying (a) and (b) upstream temperature $T_{TOT,up}$ and (c) and (d) upstream density $n_{up}$.  Data in black (with error bars) are from the simulations as given in Table \ref{table:OneDfittingData}, and data in red are from the theoretical predictions in Eqs.~(\ref{eq:vperp0upOnly}) and (\ref{eq:vThupOnly}). \textcolor{black}{Note that the vertical axes of each panel have a different range.}}
    \label{fig:RingParamCompare}
    \end{center}
\end{figure}

The simulation data and theoretical predictions are plotted in Fig.~\ref{fig:RingParamCompare}. The simulation data are displayed as black dots connected by solid black lines. The error bars are the larger of the two errors associated with each measurement given in Table~\ref{table:OneDfittingData}. The theoretical predictions, given in the last two columns of Table \ref{table:UpstreamData}, are displayed as red dots connected by red lines. The simulations with varying upstream temperature are shown in Figs.~\ref{fig:RingParamCompare}(a) and (b), displaying $v_{\perp 0}$ and $v_{Th}$, respectively, as a function of $T_{TOT,up}$. The theoretical results are within the error bars from the simulations, confirming that $v_{\perp 0}$ is not dependent on $T_{e,up}$ while $v_{Th}$ scales as $T_{e,up}^{1/2}$.  Analogous results for the simulations with varying upstream density are shown in Figs.~\ref{fig:RingParamCompare}(c) and (d). The predictions again are within the error bars from the simulations, and confirm the scaling of $v_{\perp 0}$ with $n_{up}^{-1/2}$ and the independence of $v_{Th}$ on $n_{up}$.
In summary, we find excellent agreement between the predicted values of both the major and minor radii of the ring distribution and the measured values from the ten simulations. 

We now compare the electron temperatures associated with the ring distributions with the analytical expressions from Section~\ref{sec:theo} by using 
Eq.~(\ref{eq:ring+coreVDFTeperppara})
and (\ref{eq:ring+coreVDFTe}) to find the predicted $T_{e,\perp},~T_{e,||}$, and $T_{e,{\rm eff}}$.
For the core population parameters, we use the fitting results for the central Gaussian described earlier in this section. We find that the core population thermal speed $v_{Th,M}$ values are not those associated with the upstream electron temperatures, but a study of how the core population parameters scale with upstream plasma parameters is beyond the scope of this work.
In the simulations, ring distributions are seen over a finite region of space, so the presented temperature values are mean values over that range. The error is estimated as the standard deviation of the mean.

\begin{figure}
    \begin{center}\includegraphics[width=3.4in]{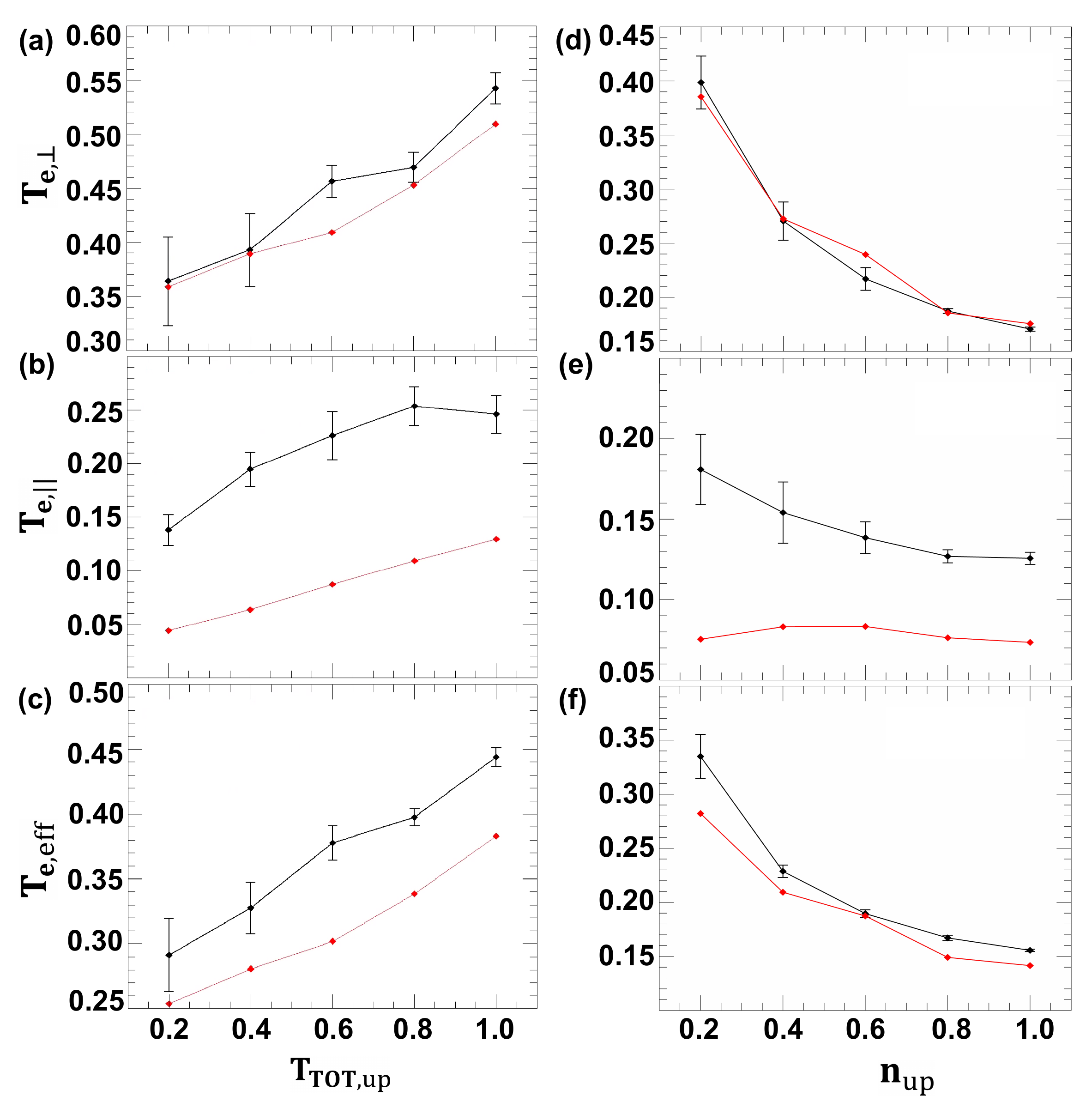}
    \caption{Comparison between predicted electron temperatures $T_{e,\perp}, T_{e,\parallel}$ and $T_{e,{\rm eff}}$ (red lines) and the simulation results (black lines with error bars). (a)-(c) are for the simulations with varying $T_{TOT,up}$, and (d)-(f) are for varying $n_{up}$. \textcolor{black}{Note that the vertical axes of each panel has a different range.}}
    \label{fig:TeCompareTheovsSims}
    \end{center}
\end{figure}

The results are shown in Fig.~\ref{fig:TeCompareTheovsSims}, with simulation results in black and theoretical results in red.  The perpendicular temperatures, in panel (a) for simulations with varying $T_{TOT,up}$ and (d) for simulations with varying $n_{up}$, show excellent agreement between the theory and simulations.  For the parallel electron temperature in panels (b) and (e), we observe a sizable difference between the simulated and predicted values.  This is attributed to our theory not accounting for the parallel propagating counter-streaming beams mentioned in the previous subsection.
However, 
we do find some qualitative agreement.
Since $T_{e,||}$ has a smaller weight than $T_{e,\perp}$ in $T_{e,{\rm eff}}$, we find good qualitative agreement between simulation results and predicted values of $T_{e,{\rm eff}}$ for all ten simulations, shown in panels (c) and (f).  The results for varying $n_{up}$ in panel (f) have very good quantitative agreement, as well. In summary, we find that the temperature in the region where rings are present increases with increasing upstream temperature 
and decreases with increasing upstream density, 
and the model based on ring distributions is quite effective at predicting the scaling and the absolute perpendicular temperatures.

\subsection{Relation of ring distributions to temperature and magnetic field profiles}
\label{subsec:Overview}

\begin{figure}
    \begin{center}\includegraphics[width=6.1in]{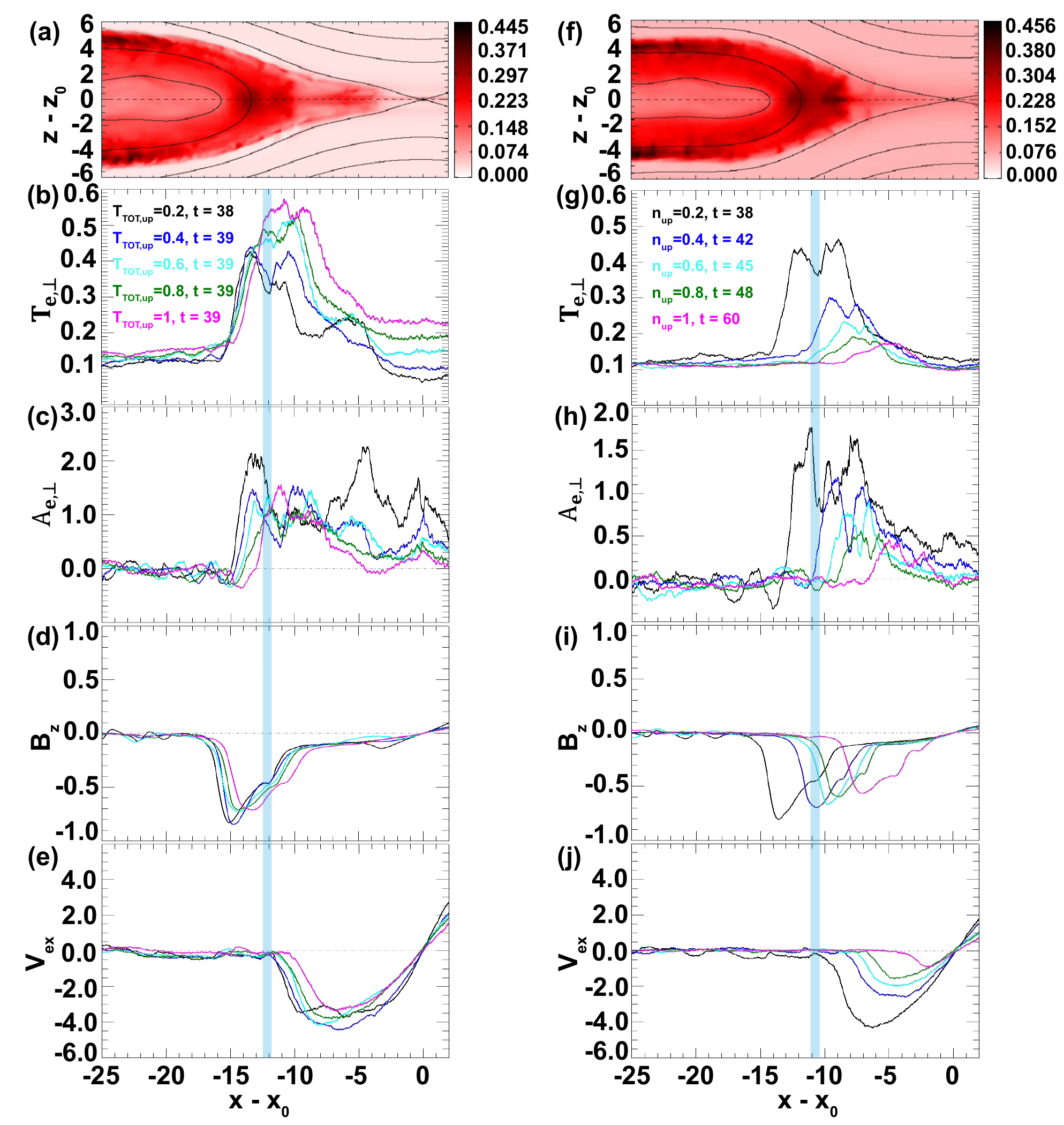}
     \caption{Profiles of plasma parameters downstream of the reconnection site. 
     (a) 2D plot of electron temperature $T_{e,{\rm eff}}$ for the $T_{TOT,up} = 0.2$ simulation at $t$ = 38. Horizontal cuts through the X-line as a function of $x-x_0$ of (b) perpendicular electron temperature $T_{e,\perp}$, (c) electron temperature anisotropy $A_{e,\perp}$, (d) reconnected magnetic field $B_z$, and (e) horizontal velocity $V_{ex}$ for the simulations with varying $T_{TOT,up}$. Panels (g) to (j) repeat (b) to (e), but for the simulations with varying $n_{up}$. Panel (f) shows $T_e$ for the $n_{up} = 0.2$ simulation at $t$ = 38. The vertical blue lines highlight the shoulder in the reconnected magnetic field $B_z$ for the $T_{TOT,up}=0.2$ and $n_{up}=0.2$ simulations.}
    \label{fig:TePeakScaling}
    \end{center}
\end{figure}
 


We now consider the location of the electron ring distributions in relation to the plasma parameter profiles in the region downstream of the EDR. Some plasma parameter profiles in the downstream region are shown in Fig.~\ref{fig:TePeakScaling}. Panels (a) and (f) show 2D plots of $T_{e,{\rm eff}}$ from the $T_{TOT,up}$ = 0.2 simulation and the $n_{up}$ = 0.2 simulation, both at $t=38$. In both cases, the highest electron temperatures observed in the simulation are in the dipolarization front region, between positions $x-x_0$ of -10 and -15. There are also high temperature regions along the separatrix, but these are potentially impacted by the periodic boundary conditions of the simulation and are not treated further here. From previous work \cite{Fu12b,egedal_2016_PoP}, we expect higher temperatures to arise from betatron acceleration of the electrons in the compressed magnetic field. However, the rEVDFs at later times during the steady-state time period (not shown) reveal the ring distributions do not increase in size in our simulations. We believe we do not observe this because our computational domain is smaller than in the previous study, preventing ions from coupling back to the magnetic field in the exhaust region.

The rest of the panels show comparisons of horizontal cuts of various quantities along the line $z = z_0$ for all $T_{TOT,up}$ (left plots) and $n_{up}$ (right plots) simulations. The times $t$ that each profile is taken are given in panels (b) and (g). Panels (b) and (g) show the perpendicular electron temperature $T_{e,\perp}$, revealing similar profiles 
for each upstream temperature with peak values near the dipolarization front, increasing with upstream temperature and decreasing with higher density.  Panels (c) and (h) show the temperature anisotropy $A_{e,\perp}$.
We observe strong electron temperature anisotropies with all the upstream temperature simulations having similar values. We also find a systematic reduction in $A_{e,\perp}$ with increasing upstream densities in the dipolarization front region.

Panels (d) and (i) show the reconnected magnetic field $B_z$. The profiles have the characteristic appearance of a dipolarization front, with a sharply peaked value at the front that decreases towards the X-line.  Importantly, in all simulations, we observe a plateau, or shoulder, in $B_z$ that occurs upstream of the dipolarization front.  Blue vertical lines are used to highlight the shoulder in $B_z$ for the $T_{TOT,up}=0.2$ and $n_{up}=0.2$ simulations. We find that for all the simulations, the $B_z$ shoulder is spatially correlated with the regions of high $T_{e,\perp}$ and $A_{e,\perp}$. 

Finally, panels (e) and (j) show the horizontal electron velocity $V_{ex}$, showing the characteristic increase in speed with distance from the X-line before rolling over and decreasing for all simulations as electron outflows exit the EDR. The horizontal velocity is close to zero in the region of peaked perpendicular temperature and the shoulder in $B_z$.
The spatial profiles in Fig.~\ref{fig:TePeakScaling} are very similar to previous simulations by \citeA{fujimoto_2008_whistler} (see their Figure 2), \textit{i.e.,} the peak in $A_{e,\perp}$ (due to an enhancement in $T_{e,\perp}$) appears in the magnetic pileup region where the electron outflow speed goes to zero. 


We now discuss the locations of the ring distributions relative to these profiles.  We find that the ring distributions shown in Fig.~\ref{fig:ringVDFsALLRUNS} are co-located with the shoulder region of $B_z$ for all simulations. For simulations with increasing upstream temperature, the shoulder regions in $B_z$ are in similar locations and the ring distributions accordingly appear over a similar region in all five simulations (see the location of the ring distributions in the left column of Fig.~\ref{fig:ringVDFsALLRUNS}). However, as upstream density is increased, the shoulder in $B_z$ appears closer to the X-line and so does the location of ring distributions (see the location of ring distributions in the right column of Fig.~\ref{fig:ringVDFsALLRUNS}). For all simulations, we find that the shoulder in $B_z$ has an extent of $\sim 1~d_{i0}$, with a field strength of $\sim 0.5 B_0$.

A possible mechanism for the presence of a shoulder in $B_z$ at the location where there are ring distributions is the diamagnetic effect of the electrons that are magnetized by the strong reconnected magnetic field. The associated current reduces the magnetic field strength in the region where rings are present and increase the field strength outside. This change to the magnetic field appears as a plateau on the $B_z$ profile as it ramps up with distance from the X-line. 

To estimate the amount by which the reconnected magnetic field decreases in the presence of ring distributions, we use conservation of energy. Using Eq.~(\ref{eq:ringVDFTeperpMdef}) and (\ref{eq:ring+coreVDFTeperppara}) to rewrite Eq.~(\ref{eq:ring+coreVDFTe}) for the effective temperature of electrons as an energy equation gives
\begin{linenomath*}
 \begin{equation}
    \frac{3}{2} k_B T_{e,\mathrm{eff}} \simeq \frac{3}{2} k_B T_{e,up} + \frac{1}{2}m_e c_{Aup,e}^2 + \left(1 - \frac{e^{-r^2}}{2 \Lambda} \right) k_B T_{e,up}.
    \label{eq:energyconvRing}
    \end{equation}   
\end{linenomath*}

The left-hand side blue gives the plasma energy at the location where rings are seen because the electron bulk speed vanishes so all energy is thermal. The first two terms on the right-hand side approximately describe the thermal plus kinetic energy of electrons as they leave the EDR. The last term on the right side is associated with the thermal energy arising from the generation of the ring distribution. This extra energy is approximately the energy that is lost by the magnetic field as it decreases due to diamagnetism of the remagnetized electrons.  The term in parentheses goes from 0.5 to 1 as $r = v_{\perp 0}/v_{Th}$ goes from 0 to $\infty$.
In order to conserve total energy, we expect the magnetic field energy to decrease by
\begin{linenomath*}
\begin{equation}
    \Delta \left(\frac{B^2}{8 \pi}\right) \sim  \left(1 - \frac{e^{-r^2}}{2 \Lambda} \right) k_B T_{e,up},
    \label{eq:Bzshoulder}
\end{equation}
\end{linenomath*}

where $\Delta (B^2/8 \pi)$ is the change in magnetic field energy.
Assuming the change in the magnetic field is weak, this decrease is approximately $B \Delta B/(4 \pi)$ where $\Delta B$ is the change in the magnetic field.

In the normalized units of our simulations, $B \simeq 0.5$ at the shoulder, and $r \ge 1$ so $(1-e^{-r^2}/(2 \Lambda))$ is close to 1. For the varying $n_{up}$ simulations where $T_{e,up}$ is kept fixed at 0.0833, this prediction gives a change in magnetic field of $\Delta B \simeq 0.2$. For the varying $T_{TOT,up}$ simulations where $T_{e,up}$ goes from 0.033 to 0.167, this prediction gives a change in magnetic field of $\Delta B \sim 0.1 - 0.3$. From the profiles of $B_z$ in Fig.~\ref{fig:TePeakScaling}(d) and (i), we find that the difference of the profile from a linearly increasing ramp away from the X-line is approximately 0.1 - 0.3, in reasonable agreement with the prediction.
\subsection{Confirmation that ring distributions are caused by remagnetization}
\label{subsec:ElecRemag}

We now confirm the proposed model that electron rings are associated with their remagnetization in the reconnected magnetic field \cite{Shuster2014,Bessho2014}.  We calculate two quantities as a function of $x$: (1) the magnetic field radius of curvature $R_c = |(\hat{b} \cdot \nabla) \hat{b}|^{-1}$, where $\hat{b}$ is the unit vector along the local magnetic field, and (2) the electron gyroradius $\rho_{{\rm bfs}} = V_{ex} / \Omega_{ce}$ based on the horizontal bulk flow speed $V_{ex}$ and the local electron gyrofrequency $\Omega_{ce} = e B / m_e c$. The bulk flow speed is the appropriate speed because the ring distributions are proposed to be formed by outflowing electron beams that get remagnetized. 
The condition for remagnetization is 
$\sqrt{\kappa} = R_c/\rho_{{\rm bfs}} \approx 1$ \cite{Buchner89}. 

\begin{figure}
    \begin{center}\includegraphics[width=2.8in]{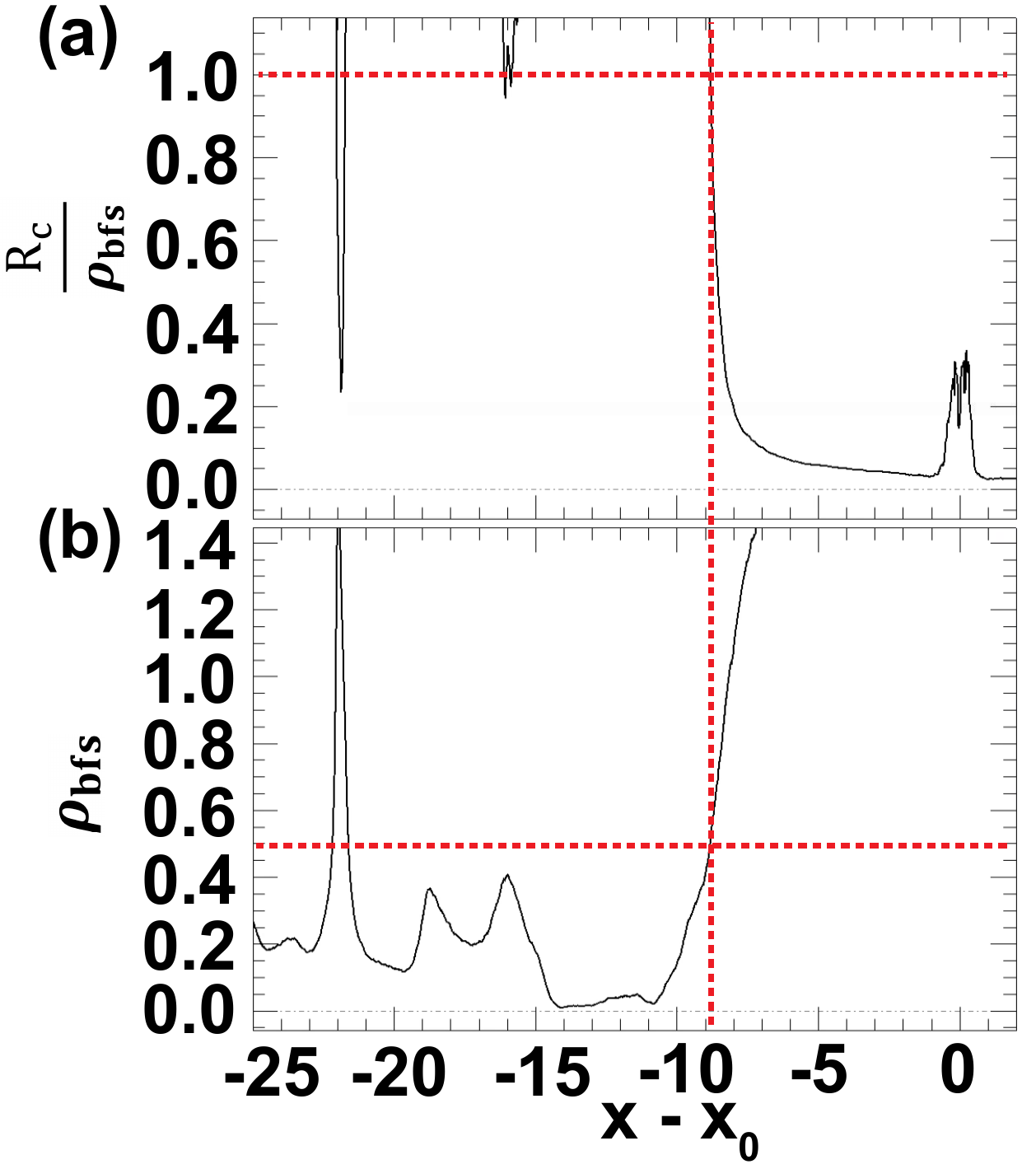}
     \caption{(a) Ratio of the magnetic field radius of curvature $R_c$ to the electron gyroradius $\rho_{{\rm bfs}}$ based on the bulk flow speed as a function of $x-x_0$ in a horizontal cut through the X-line for the $n_{up}=0.2$ simulation at $t=38$. The horizontal red dashed line at $R_c / \rho_{{\rm bfs}} = 1$ is where electrons are expected to remagnetize. The vertical dashed red line marks the $x-x_0$ location where this condition is met. (b) $\rho_{{\rm bfs}}$ vs.~$x-x_0$ for the same simulation, with the horizontal dashed line marking the value of $\rho_{{\rm bfs}}$ where electrons remagnetize.}
    \label{fig:ElecRemag}
    \end{center}
\end{figure}

We plot $R_c/\rho_{{\rm bfs}}$ 
as a function of $x-x_0$ in Fig.~\ref{fig:ElecRemag}(a) 
for the $n_{up}=0.2$ simulation at $t=38$. A horizontal red dashed line marks where $R_c / \rho_{{\rm bfs}} = 1$, which is $x - x_0 \approx -9$ as marked by the vertical red dashed line. Fig.~\ref{fig:ElecRemag}(b) shows $\rho_{{\rm bfs}}$ as a function of $x-x_0$. Its value where $R_c / \rho_{{\rm bfs}} = 1$ is $\approx 0.5 \ d_{i0}$, which for this simulation is $\approx 1.1 \ d_e$.

We now compare this to the location where ring distributions are observed in this simulation. Ring distributions are seen throughout the blue shaded region of Fig.~\ref{fig:TePeakScaling}(g)-(j).  This is located $\simeq 2 d_e$ downstream of the location where $R_c / \rho_{{\rm bfs}} = 1$.
Since the gyroradius of the electron beam is $\sim 1d_e$, the ring distributions are observed one gyro-diameter downstream of the location where the remagnetization condition is first met.  This same behavior is seen in each of the other nine simulations studied here (not shown).  This confirms that the remagnetization of the electron outflow jet is responsible for the generation of the ring distributions.

A further test that the ring distributions are caused by remagnetization of electron exhaust beams is that they should cease to be present with the addition of a sufficiently strong out of plane (guide) magnetic field.  
To test this, we perform simulations with initial guide fields $B_g$ of $0.05$ and $0.25$ for $n_{up}=0.2$, with all other parameters the same as before. A similar analysis as shown in Fig.~\ref{fig:ElecRemag} (not shown) reveals that for the $B_g=0.05$ simulation, $R_c/\rho_{{\rm bfs}}$ is very similar to the no guide field case, \textit{i.e.,} away from the X-line, $R_c/\rho_{{\rm bfs}}$ increases and then crosses 1 signalling remagnetization of the electron outflow jet. The plasma parameter profiles are similar to those seen in Fig.~\ref{fig:TePeakScaling} for the no guide field case (not shown). A scan of rEVDFs as described in previous sections shows ring distributions in the region of a $B_z$ shoulder (not shown). However, for the $B_g=0.25$ simulation, $R_c/\rho_{{\rm bfs}}$ (not shown) is never less than 1 in the downstream region,   implying that electrons are never demagnetized so no remagnetization occurs downstream. We also find no presence of ring EVDFs (not shown) in our scan.  This provides additional evidence that the rings are formed by magnetization of the electron exhaust beams.

\section{Discussion and Applications}
\label{sec:discussions}

The results of this research are potentially useful for a variety of reasons.  By relating the properties of the ring distribution to the upstream (lobe) plasma parameters in Sec.~\ref{sec:theo}, we can make quantitative predictions of the electron temperatures achieved downstream of reconnection exhausts, such as a dipolarization front or a solar flare reconnection outflow. We can also approximately account for the betatron acceleration that is expected to occur following the generation of ring distributions \cite{Fu12b,egedal_2016_PoP}.  We characteristically see the $B_z$ shoulder at a magnetic field strength of about 0.5 as shown in Fig.~\ref{fig:TePeakScaling}, and it further compresses to a strength of 1.  If betatron acceleration were to occur and assuming that the magnetic moment is conserved, we expect the perpendicular temperature to increase by a factor of $\sim 2$ from our predicted values. 

To apply the theory to real systems, we also need to estimate the magnetic field $B_{up,e}$ at the upstream edge of the electron layer from the asymptotic magnetic field strength $B_{up}$.  There is no widely accepted theory for this, so we discuss two possible options.  In model 1, we use   \begin{linenomath*}
  \begin{equation}
  B_{up,e} \approx 2\left(\frac{m_e}{m_i}\right)^{1/2} B_{up}, 
  \end{equation}  
\end{linenomath*}
which captures that the electron outflow velocity at the EDR is often observed to be approximately twice the ion Alfv\'en speed. In model 2, we use \cite{liu_FirstPrinciple_2022}
\begin{linenomath*}
  \begin{equation}
  B_{up,e} \approx \left(\frac{m_e}{m_i}\right)^{1/4} B_{up},
    \end{equation}  
\end{linenomath*}
which follows from conservation of magnetic flux at the electron and ion layers.

We first consider Earth's magnetotail, where there is typically only a weak guide field and typical plasma parameters may be taken as $B_{up} \approx 20$ nT, $n_{up} \approx 0.1$ cm$^{-3}$, and $T_{e,up} \approx 700$ eV, although there is significant uncertainty in all three values.  Using the expressions in Sec.~\ref{sec:theo}, we find the predicted $v_{\perp0}$ to be $(2.8 - 9.2) \times 10^8$ cm/s.  Here and in what follows, the first number in the provided range is using model 1 and the second is using model 2. We also get $v_{Th}=1.6 \times 10^9$ cm/s, so the perpendicular and effective temperatures associated with ring distributions is $T_{\perp} = 890-1270$ eV and $T_{{\rm eff}} = 850-1100$ eV, with an anisotropy of $A_{e,\perp} = 0.2-0.7$.  For comparison, the DF studied in Fig.~4 of \citeA{runov_2010_Planet_Sci} had electron temperatures reaching about 1800 eV with perpendicular temperature $T_{e,\perp}\sim 2000$ eV. 
Doubling our prediction to account for betatron acceleration, we find the predicted values are broadly consistent with the observations. 

We next consider implications for reconnection in solar flares. The presence of a guide field may suppress the mechanism in the present study entirely. However, a range of guide fields is observed including examples with little to no guide field \cite{Qiu17}. Moreover, a leading model for the observed heating from MHD simulation studies also requires a low guide field strength \cite{Longcope10,Longcope16}. We assume typical values of a background coronal temperature of $T_{e,up} = 1$ MK, a density of $n_{up} = 10^9$ cm$^{-3}$, and an ambient magnetic field for a large flare of $B \sim 100$ G, the latter of which is consistent with values inferred from radio and other measurements for large flares \cite{Asai2006,krucker10a,Caspi14}. The associated upstream magnetic field at the electron layer is estimated to be $B_{up,e} = 4.6-15.6$ G using model 1 and 2. Then, the predicted major and minor radii of the ring distributions are $v_{\perp0}=1.4-4.6\times10^9$ cm/s and $v_{Th}=5.5\times10^8$ cm/s. This implies $r = 3-8$, $A_{e,\perp} = 7-70$, and $T_{e,\perp} = 8-70$ MK. Since the coronal plasma $\beta$ is small, $r$ is significantly larger than 1, much higher than its magnetotail counterpart, leading to a much more dramatic increase in temperature due to remagnetizing the electrons.  Taking an asymptotic expansion for the large $r$ limit of Eq.~(\ref{eq:ringVDFTeperpMdef}) gives
\begin{linenomath*}
  \begin{equation}
    \mathcal{M} \approx \frac{3}{2} + r^2.
  \end{equation}
\end{linenomath*}
Using Eqs.~(\ref{eq:vThupOnly}) and (\ref{eq:rupOnly}) for $v_{Th}$ and $r$, Eq.~(\ref{eq:ringVDFTe}) gives an expression for $T_{{\rm eff}}$ for large $r$ as
\begin{linenomath*}
  \begin{equation}
    T_{e,{\rm eff}} \approx T_{e,up} \left(\frac{4}{3}+\frac{B_{up,e}^2}{12 \pi n_{up} k_BT_{e,up}} \right).
  \label{eq:ringVDFTelarge_r}
  \end{equation}
\end{linenomath*}
Evaluating this expression in terms of the typical coronal parameters provided above, we get 
\begin{linenomath*}
  \begin{equation}
    T_{e,{\rm eff}} = 1.33 {\rm \ MK} \left(\frac{T_{e,up}}{1 {\rm \ MK}}\right) + (4.2 {\rm \ MK} - 45 {\rm \ MK}) \left(\frac{B_{up}}{100 {\rm \ G}}\right)^2 \left(\frac{n_{up}}{10^9 {\rm \ cm}^{-3}}\right)^{-1},
  \label{eq:ringVDFTelarge_r_normalized}
  \end{equation}
\end{linenomath*}
where the range in the second term is for model 1 and 2 of $B_{up,e}$.  Therefore, the predicted effective temperature is $T_{e,{\rm eff}} = 5-46$ MK using models 1 and 2 for the typical coronal parameters employed here.  This relation predicts a scaling dependence of the temperature approximately as $B_{up}^2.$
The temperatures predicted here, even when doubled to account for betatron acceleration, are in the same range as the 10s of MK observed during super-hot flares \cite{Caspi10,Caspi14,Warmuth16}. The heating mechanism in our models is the reconnection process, significant heating occurs for magnetic fields starting at about 100~G, and there is an increase in temperature with magnetic field strength. These features are broadly consistent with the relationships derived from a statistical study of X-ray observations of intense flares \cite{Caspi14}. We therefore suggest it may be possible that the super-hot temperatures in such flares are generated by electron beams getting magnetized in reconnected fields, and potentially also subsequently heated further by betatron acceleration as the reconnected magnetic field continues to compress.  This compression likely leads to higher densities than the ambient coronal value, as has been previously suggested \cite{Caspi10b,Longcope11}. 
The proposed mechanism would also help explain the observed association of super-hot temperatures with coronal non-thermal emission and energy content \cite{Caspi15,Warmuth16}. \textcolor{black}{Significant future studies to further explore the viability of the present model for explaining observed temperatures in super-hot solar flares is needed, including a parametric test of Eq.~(\ref{eq:ringVDFTelarge_r_normalized}), determining whether this mechanism is consistent with the high level of compression seen in observations, studying if the small regions where the ring distributions are generated can transmit to the large scales endemic to solar flares, and determining whether guide field strengths in solar flares would magnetize the ring distributions.}

The results of this study could also be applicable to Earth's dayside magnetopause, where ring distributions and whistler mode generation were recently observed both in simulations of asymmetric reconnection with a guide field and in Magnetospheric Multiscale (MMS) Mission observations \cite{Yoo19,Choi2022}. The theory presented in this study is exclusively for symmetric reconnection, but dayside reconnection is typically asymmetric. We expect the mechanism for ring distribution generation to be similar in asymmetric reconnection.  We hypothesize that in asymmetric reconnection, the speed that sets the major radius $v_{\perp0}$ in Eq.~(\ref{eq:vperp0upOnly}) becomes the asymmetric version of the Alfv\'en speed that controls the outflow speed of asymmetric reconnection,
  \begin{linenomath*}
  \begin{equation}
    v_{\perp 0,asym}=\frac{B_{up,asym,e}}{\sqrt{4 \pi m_{e} n_{up,asym}}},
  \label{eq:vperp0upOnlyasym}
  \end{equation}
  \end{linenomath*}
and the thermal speed that sets the minor radius is replaced by
  \begin{linenomath*}
  \begin{equation}
  v_{Th,asym} = \sqrt{\frac{2 k_{B} T_{e,up,asym}}{m_{e}}},
  \label{eq:vThupOnlyasym}
  \end{equation}
  \end{linenomath*}
where $B_{up,asym,e} = B_{up,1,e}B_{up,2,e}/(B_{up,1,e}+B_{up,2,e})$ and $n_{up,asym} = (n_{up,1} B_{up,2,e} + n_{up,2} B_{up,1,e})/(B_{up,1,e}+B_{up,2,e})$ \cite{Cassak07d,Cassak08b} and $T_{e,up,asym} = (T_{e,up,1} n_{up,1} B_{up,2,e} + T_{e,up,2} n_{up,2} B_{up,1,e})/(n_{up,2} B_{up,1,e}+n_{up,1} B_{up,2,e})$ \cite{Shay14}.  It is beyond the scope of the present study to test this hypothesis, but it would be interesting to do so for future work.

We now discuss implications for direct measurements of ring distributions in reconnection events, especially in dipolarization fronts that are accessible to {\it in situ} observations.  The simulations suggest that the physical size of the region where ring distributions are present is relatively small.  In the simulations, the range over which rings are seen is about $1 \ d_i$, corresponding to approximately $720$ km (based on a lobe density of 0.1 cm$^{-3}$) in Earth's magnetotail.  Temporally, we expect that they appear transiently at the dipolarization front.  Simulations of reconnection in large domains do not reveal temperature peaks in the downstream region in the steady-state \cite{Shay14}.  Moreover, since ring distributions are unstable to wave generation \cite{Gary85}, they are expected to rapidly decay, making their direct observation even more challenging. \textcolor{black}{It is also challenging to observe ring distributions when the major radius is smaller than the minor radius, {\it i.e.,} when $r<1$. For typical parameters in Earth’s magnetotail, $r$ is theoretically expected to be approximately 0.2 - 0.6, so \textit{in situ} observations of rings might be challenging but can be potentially possible. Rings are more likely to be identifiable in large $r$ (low electron plasma beta) systems.}

To illustrate the challenges of direct measurement of a ring distribution, we describe an unsuccessful attempt to identify one in Earth's magnetotail using the THEMIS spacecraft \cite{Angelopoulos2009}. On February 27, 2009, four of the five THEMIS spacecraft traversed a DF between 0750 and 0800 UT \cite{runov_2010_Planet_Sci}, and burst mode data were available during this time. Their Figs.~4 and 5 reveal classic signatures of a DF, with a significant decrease in density and an increase in $B_z$ (in GSM coordinates). The P1 (THEMIS B) spacecraft passed through the DF at 07:51:26~UT, shown on the left side of their Fig.~4, with the vertical dashed line denoting the DF.  Immediately upstream of the DF (around 07:51:30~UT), the electron temperature in both directions perpendicular to the magnetic field exceeds the parallel electron temperature, making this location a candidate for having an electron ring distribution. 

To determine whether there is an electron ring distribution at this time, we investigate the EVDFs in the time interval when $T_{e,\perp} > T_{e,||}$. The distributions are averaged over two spacecraft spin periods (6~s), between 07:51:30 and 07:51:36~UT, to get better statistics than a single spin. The low-energy cutoff due to spacecraft charging is $\sim 60$~eV, which is smaller than the predicted major radius for this event, so we expect it to be ostensibly possible to resolve a ring distribution if it is present. Two-dimensional cuts of the EVDF are produced from recombined ElectroStatic Analyzer (ESA) and Solid State Telescope (SST) data in this time range (not shown). Clear signatures of counterstreaming electron beams along the magnetic field are seen in both $\perp-\parallel$ planes. When the raw data is smoothed, a weak signature of what appears to be a ring population is seen. 
However, a closer examination of the uncombined ESA-only burst mode data with no smoothing reveals that the weak ring population signal is not present in the $\perp1-\perp2$ cut 
where it should be, judging from the $\perp-\parallel$ plane cuts.  

There are a few reasons for the misidentification of a ring distribution structure.  In the $\perp1-\perp2$ plane, there is a substantial population of low-energy particles which are of ionospheric origin. When the distribution function is smoothed, this population gives the appearance of a ring. However, the ionospheric population is not what would cause the appearance of a ring distribution by the mechanism studied here and must be excluded. The reason that $T_{e,\perp} > T_{e,\|}$ for this distribution is that the more diffuse magnetotail population is
rather elongated in the $\perp$ directions. 
To determine if this higher-energy magnetotail population is part of a ring distribution, we look at the $\|-\perp$ planes. Because of the strong field-aligned counterpropagating beams, it makes it difficult to tell if removing that population would leave a ring in the high-energy population, but the population in question does not clearly disappear for more field-aligned angles.  Consequently, we are unable to definitively claim there is an electron ring distribution in this particular THEMIS event.  
We suggest that observing a ring distribution in situ likely requires higher temporal resolution than available to THEMIS, but it may be accessible to MMS \cite{Schmid16,Liu18,Zhao19,Grigorenko20,Ma20} which has a much higher temporal resolution.

\section{Conclusions}
\label{sec:conclusions}

The appearance of ring distributions of electrons has been previously identified in particle-in-cell simulations near dipolarization fronts  \cite{Shuster2014,Bessho2014} and for dayside reconnection \cite{Choi2022}. It was suggested that they are caused by remagnetization of the electrons in the reconnected magnetic field \cite{Shuster2014,Bessho2014}. In this study, we carry out a theoretical and numerical analysis that verifies and quantifies this prediction. Our analysis gives the major and minor radii of the ring distribution in terms of upstream conditions that dictate the properties of the reconnection, \textit{i.e.,} the plasma density, electron temperature, and reconnecting magnetic field strength. In particular, the major radius is given by the electron Alfv\'en speed based on the magnetic field and density upstream of the electron current layer, while the minor radius is governed by the electron thermal speed in the upstream region.

We employ 2.5D PIC simulations to test our predictions using five simulations with varying upstream temperature (with the upstream density held fixed) and five simulations with varying upstream density (with the upstream temperature held fixed). We find ring distributions in all 10 simulations.
We extract the major and minor radii of the ring distributions for all ten simulations
by fitting Gaussians to 1D cuts of the reduced distributions. We find that the major radius $v_{\perp0}$ is independent of upstream temperature but decreases for increasing upstream density, while the minor radius $v_{Th}$ increases for increasing upstream temperature and is independent of upstream density. The results are qualitatively and quantitatively consistent with the theoretical predictions, with agreement within one standard deviation of the theoretical predictions for all simulations.

Next, we use the major and minor radii of the ring distributions to compare the electron temperature associated with ring distributions to analytical predictions. We find that the predicted and measured perpendicular electron temperature agrees very well, within 12\%. 
The parallel electron temperature is consistently different by about a factor of 2 between theory and simulation because the simulated plasma also contains counterstreaming beams in the parallel direction that are omitted from the analytical model. Since the perpendicular electron temperature contributes to the total electron temperature more than the parallel, the simulated total temperature is within 20\% of the theoretical predictions. 

By investigating the plasma parameter profiles in the region where the ring distributions are observed, we find the ring distributions, and their associated perpendicular temperature anisotropy, are spatially coincident with a plateau, or shoulder, in the profile of the reconnected magnetic field $B_z$. The shoulder in $B_z$ is present where the ring distributions are because the remagnetized electrons are diamagnetic, thereby slightly lowering $B_z$ within the electron orbit and slightly increasing $B_z$ outside the orbit, thereby setting up a plateau in the $B_z$ profile. A simple calculation using conservation of energy reproduces the approximate perturbed magnetic field due to this effect.

We show that the ring distributions appear approximately two electron gyroradii (one diameter of the gyromotion) downstream from the location where the electrons are remagnetized by the strong reconnected magnetic field, {\it i.e.,} the location where the radius of curvature of the magnetic field exceeds the gyroradius of the electrons based on the bulk flow speed. This result is consistent with the prediction that the ring distributions are associated with reconnection jets that are remagnetized by the reconnected field in a dipolarization front \cite{Shuster2014,Bessho2014}.  We further confirm this by showing that the ring distributions become weaker and then are completely suppressed as an increasingly strong guide field is added.

Finally, we discuss applications of the present results in magnetospheric and solar settings. For dipolarization fronts in Earth's magnetotail, the electron temperatures predicted by the scaling analysis presented here are in the few keV range (when subsequent heating via betatron acceleration is accounted for), which is comparable to the observed electron temperatures.  When applied to solar flares, we predict electron temperatures up to 10s of MK for very energetic flares, and an increase in temperature with the square of the reconnecting magnetic field.  Such temperatures are consistent with those observed in super-hot flares, which are highly likely to come from the coronal reconnection process but for which there is not yet a widely accepted mechanism for their production.  We further motivate a possible extension of the present work to antiparallel asymmetric systems, which may be important for applications to the dayside magnetopause.

The direct {\it in situ} measurement of ring distributions in the magnetotail is expected to be difficult, but potentially possible. Various characteristic pitch-angle distributions have been observed in dipolarization fronts \cite{liu_explaining_2017,liu_rapid_2017} and studied using simulations \cite{huang_formation_2021}. 
It is possible that pancakes and/or the perpendicular features of rolling pins are ring distributions,
and testing this would be interesting future work. We note that a pitch-angle distribution plot of a ring distribution would have a pancake-type structure, but it is not possible using a pitch-angle distribution plot to confirm the lack of low energy particles that is characteristic of a ring distribution.
Rather, a direct investigation of the velocity distribution function is required.
Based on a case study using THEMIS observations, we find that it is difficult to identify ring distributions. Higher temporal resolution, such as that afforded by MMS, would facilitate their identification.

It is known that the significant anisotropy arising in ring distributions makes them unstable to the generation of waves, especially whistlers \cite{Gary85,Umeda2007, fujimoto_2008_whistler, Winske&Daughton2012_whistler}. More broadly, \citeA{Grigorenko20} showed that electrons at 1--5 keV with a perpendicular temperature anisotropy generate whistler waves near DFs. By knowing the major and minor radii of the ring distributions in terms of upstream parameters, the temperature anisotropy can be calculated, which allows for a quantitative estimate of the linear growth rate of these modes.  Such information is an important aspect of understanding particle acceleration and heating as a result of wave-particle interactions \cite{roytershteyn&delzanno2018}.

While whistler waves associated with temperature anisotropies are regularly measured {\it in situ} in Earth's magnetosphere, much less has been studied for the possibility of whistler wave generation associated with solar flares. There has been theoretical work on understanding whistler wave generation in solar coronal loops \cite{vocks_whistler_2006}. In their work, the whistlers are generated from loss cone distributions rather than the mechanism discussed here.  Since the characteristic length scale for the ring distributions is $d_e$, we expect the frequency of whistler waves associated with ring distributions to be comparable to the electron cyclotron frequency $\Omega_{ce} = e B / m_e c$. For the characteristic solar flare plasma parameters used here, we find that the whistler frequencies would be at least on the order of 0.3 GHz. Interestingly, an observational study has seen a long-lived source at 0.327 GHz \cite{aurass_gle_2006}. Whether the mechanism discussed here can account for observed frequencies and whether this can be used as remote evidence in favor of the model presented here would be an interesting topic for future work.

There are many avenues for future work. The present simulations are two-dimensional; we do not expect the fundamental aspects of the results to change in three dimensions, especially given that there is no guide field in the system studied here, but it would be interesting to confirm that 3D effects known to occur in magnetotail-type settings \cite{pritchett:2013,Sitnov14} do not alter the conclusions.  The initial conditions of the present simulations did not include an equilibrium normal magnetic field, which is important for magnetotail reconnection \cite{Lembege82}; we do not anticipate this normal magnetic field would appreciably change the results herein, but it should be verified. The simulation domain size we employ is too small to allow ions to fully couple back to the plasma, so future work should confirm that the results are valid for larger system sizes. For dayside magnetopause applications, the proposed generalization incorporating asymmetries needs to be tested. For solar corona applications, electron-ion collisions may need to be taken into account, and observations should be used to test the functional dependence of the temperature on the magnetic field strength during solar flares predicted here, as well as whether a guide field suppresses such high temperatures. The physical size of the region where electrons are remagnetized is expected from the simulations to be relatively small, so questions about how ring distributions thermalize and whether they control the temperature over a greater volume, as would be necessary to explain the temperatures seen in super-hot flares, would be excellent topics for future work.
Future work to quantify the rate of production of anisotropy-driven wave modes such as whistlers and their interaction with the downstream plasma would be important for applications.

\acknowledgments
M.H.B. acknowledges insightful discussions with Benjamin Woods. P.A.C. gratefully acknowledges support from NSF Grant PHY-1804428, NASA Grant 80NSSC19M0146, and DOE Grant DE-SC0020294.  M.A.S. acknowledges NASA LWS Grant 80NSSC20K1813 and NSF AGS-2024198. V.R. acknowledges DOE grant DE‐SC001931. A.C. was supported by NASA grants NNX17AI71G, 80NSSC19K0287 and 80NSSC22M0111, and by NSF grant 1841039. H.L. acknowledges the partial support of a NASA Parker Solar Probe contract SV4-84017, an NSF EPSCoR RII-Track-1 Cooperative Agreement OIA-1655280, a NASA IMAP subaward under NASA contract 80GSFC19C0027, and NASA awards 80NSSC20K1783 and 80NSSC21K0003. This research uses resources of the National Energy Research Scientific Computing Center (NERSC), a DOE Office of Science User Facility supported by the Office of Science of the US Department of Energy under Contract no. DE-AC02-05CH11231. Simulation data used in this manuscript are available on Zenodo (https://doi.org/10.5281/zenodo.6383101).


%
%

\bibliography{DiFrpaper}

\begin{thebibliography}{}

\bibitem [\protect \citeauthoryear {%
Allred%
, Alaoui%
, Kowalski%
\BCBL {}\ \BBA {} Kerr%
}{%
Allred%
\ \protect \BOthers {.}}{%
{\protect \APACyear {2020}}%
}]{%
Allred20}
\APACinsertmetastar {%
Allred20}%
\begin{APACrefauthors}%
Allred, J\BPBI C.%
, Alaoui, M.%
, Kowalski, A\BPBI F.%
\BCBL {}\ \BBA {} Kerr, G\BPBI S.%
\end{APACrefauthors}%
\unskip\
\newblock
\APACrefYearMonthDay{2020}{}{}.
\newblock
{\BBOQ}\APACrefatitle {Modeling the Transport of Nonthermal Particles in Flares
  Using {F}okker–{P}lanck Kinetic Theory} {Modeling the transport of
  nonthermal particles in flares using {F}okker–{P}lanck kinetic
  theory}.{\BBCQ}
\newblock
\APACjournalVolNumPages{Ap. J.}{902}{}{16}.
\PrintBackRefs{\CurrentBib}

\bibitem [\protect \citeauthoryear {%
Angelopoulos%
}{%
Angelopoulos%
}{%
{\protect \APACyear {2009}}%
}]{%
Angelopoulos2009}
\APACinsertmetastar {%
Angelopoulos2009}%
\begin{APACrefauthors}%
Angelopoulos, V.%
\end{APACrefauthors}%
\unskip\
\newblock
\APACrefYearMonthDay{2009}{}{}.
\newblock
{\BBOQ}\APACrefatitle {The {THEMIS} Mission} {The {THEMIS} mission}.{\BBCQ}
\newblock
\BIn{} J\BPBI L.~Burch\ \BBA {} V.~Angelopoulos\ (\BEDS), \APACrefbtitle {The
  {THEMIS} Mission} {The {THEMIS} mission}\ (\BPGS\ 5--34).
\newblock
\APACaddressPublisher{New York, NY}{Springer New York}.
\newblock
\begin{APACrefDOI} \doi{10.1007/978-0-387-89820-9_2} \end{APACrefDOI}
\PrintBackRefs{\CurrentBib}

\bibitem [\protect \citeauthoryear {%
Angelopoulos%
\ \protect \BOthers {.}}{%
Angelopoulos%
\ \protect \BOthers {.}}{%
{\protect \APACyear {1992}}%
}]{%
angelopoulos_1992_JGR}
\APACinsertmetastar {%
angelopoulos_1992_JGR}%
\begin{APACrefauthors}%
Angelopoulos, V.%
, Baumjohann, W.%
, Kennel, C\BPBI F.%
, Coroniti, F\BPBI V.%
, Kivelson, M\BPBI G.%
, Pellat, R.%
\BDBL {}Paschmann, G.%
\end{APACrefauthors}%
\unskip\
\newblock
\APACrefYearMonthDay{1992}{}{}.
\newblock
{\BBOQ}\APACrefatitle {Bursty bulk flows in the inner central plasma sheet}
  {Bursty bulk flows in the inner central plasma sheet}.{\BBCQ}
\newblock
\APACjournalVolNumPages{Journal of Geophysical Research: Space
  Physics}{97}{A4}{4027--4039}.
\newblock
\begin{APACrefDOI} \doi{https://doi.org/10.1029/91JA02701} \end{APACrefDOI}
\PrintBackRefs{\CurrentBib}

\bibitem [\protect \citeauthoryear {%
Angelopoulos%
\ \protect \BOthers {.}}{%
Angelopoulos%
\ \protect \BOthers {.}}{%
{\protect \APACyear {2008}}%
}]{%
Angelopoulos08}
\APACinsertmetastar {%
Angelopoulos08}%
\begin{APACrefauthors}%
Angelopoulos, V.%
, McFadden, J\BPBI P.%
, Larson, D.%
, Carlson, C\BPBI W.%
, Mende, S\BPBI B.%
, Frey, H.%
\BDBL {}Kepko, L.%
\end{APACrefauthors}%
\unskip\
\newblock
\APACrefYearMonthDay{2008}{}{}.
\newblock
{\BBOQ}\APACrefatitle {Tail Reconnection Triggering Substorm Onset} {Tail
  reconnection triggering substorm onset}.{\BBCQ}
\newblock
\APACjournalVolNumPages{Science}{321}{5891}{931-935}.
\newblock
\begin{APACrefDOI} \doi{10.1126/science.1160495} \end{APACrefDOI}
\PrintBackRefs{\CurrentBib}

\bibitem [\protect \citeauthoryear {%
Angelopoulos%
\ \protect \BOthers {.}}{%
Angelopoulos%
\ \protect \BOthers {.}}{%
{\protect \APACyear {2013}}%
}]{%
angelopoulos_2013_Sci}
\APACinsertmetastar {%
angelopoulos_2013_Sci}%
\begin{APACrefauthors}%
Angelopoulos, V.%
, Runov, A.%
, Zhou, X\BHBI Z.%
, Turner, D\BPBI L.%
, Kiehas, S\BPBI A.%
, Li, S\BHBI S.%
\BCBL {}\ \BBA {} Shinohara, I.%
\end{APACrefauthors}%
\unskip\
\newblock
\APACrefYearMonthDay{2013}{}{}.
\newblock
{\BBOQ}\APACrefatitle {Electromagnetic {Energy} {Conversion} at {Reconnection}
  {Fronts}} {Electromagnetic {Energy} {Conversion} at {Reconnection}
  {Fronts}}.{\BBCQ}
\newblock
\APACjournalVolNumPages{Science}{341}{6153}{1478--1482}.
\newblock
\begin{APACrefDOI} \doi{10.1126/science.1236992} \end{APACrefDOI}
\PrintBackRefs{\CurrentBib}

\bibitem [\protect \citeauthoryear {%
{Asai}%
\ \protect \BOthers {.}}{%
{Asai}%
\ \protect \BOthers {.}}{%
{\protect \APACyear {2006}}%
}]{%
Asai2006}
\APACinsertmetastar {%
Asai2006}%
\begin{APACrefauthors}%
{Asai}, A.%
, {Nakajima}, H.%
, {Shimojo}, M.%
, {White}, S\BPBI M.%
, {Hudson}, H\BPBI S.%
\BCBL {}\ \BBA {} {Lin}, R\BPBI P.%
\end{APACrefauthors}%
\unskip\
\newblock
\APACrefYearMonthDay{2006}{}{}.
\newblock
{\BBOQ}\APACrefatitle {{Preflare Nonthermal Emission Observed in Microwaves and
  Hard X-Rays}} {{Preflare Nonthermal Emission Observed in Microwaves and Hard
  X-Rays}}.{\BBCQ}
\newblock
\APACjournalVolNumPages{Publications of the Astronomical Society of
  Japan}{58}{}{L1-L5}.
\newblock
\begin{APACrefDOI} \doi{10.1093/pasj/58.1.L1} \end{APACrefDOI}
\PrintBackRefs{\CurrentBib}

\bibitem [\protect \citeauthoryear {%
Ashour-Abdalla%
\ \protect \BOthers {.}}{%
Ashour-Abdalla%
\ \protect \BOthers {.}}{%
{\protect \APACyear {2011}}%
}]{%
ashour-abdalla_observations_2011}
\APACinsertmetastar {%
ashour-abdalla_observations_2011}%
\begin{APACrefauthors}%
Ashour-Abdalla, M.%
, El-Alaoui, M.%
, Goldstein, M\BPBI L.%
, Zhou, M.%
, Schriver, D.%
, Richard, R.%
\BDBL {}Hwang, K\BHBI J.%
\end{APACrefauthors}%
\unskip\
\newblock
\APACrefYearMonthDay{2011}{}{}.
\newblock
{\BBOQ}\APACrefatitle {Observations and simulations of non-local acceleration
  of electrons in magnetotail magnetic reconnection events} {Observations and
  simulations of non-local acceleration of electrons in magnetotail magnetic
  reconnection events}.{\BBCQ}
\newblock
\APACjournalVolNumPages{Nature Physics}{7}{4}{360--365}.
\newblock
\begin{APACrefDOI} \doi{10.1038/nphys1903} \end{APACrefDOI}
\PrintBackRefs{\CurrentBib}

\bibitem [\protect \citeauthoryear {%
Aurass%
, Mann%
, Rausche%
\BCBL {}\ \BBA {} Warmuth%
}{%
Aurass%
\ \protect \BOthers {.}}{%
{\protect \APACyear {2006}}%
}]{%
aurass_gle_2006}
\APACinsertmetastar {%
aurass_gle_2006}%
\begin{APACrefauthors}%
Aurass, H.%
, Mann, G.%
, Rausche, G.%
\BCBL {}\ \BBA {} Warmuth, A.%
\end{APACrefauthors}%
\unskip\
\newblock
\APACrefYearMonthDay{2006}{}{}.
\newblock
{\BBOQ}\APACrefatitle {The {GLE} on {Oct}. 28, 2003 – Radio {diagnostics} of
  {relativistic} {electron} and {proton} {injection}} {The {GLE} on {Oct}. 28,
  2003 – radio {diagnostics} of {relativistic} {electron} and {proton}
  {injection}}.{\BBCQ}
\newblock
\APACjournalVolNumPages{Astronomy \& Astrophysics}{457}{2}{681--692}.
\newblock
\begin{APACrefDOI} \doi{10.1051/0004-6361:20065238} \end{APACrefDOI}
\PrintBackRefs{\CurrentBib}

\bibitem [\protect \citeauthoryear {%
Bessho%
, Chen%
, Shuster%
\BCBL {}\ \BBA {} Wang%
}{%
Bessho%
\ \protect \BOthers {.}}{%
{\protect \APACyear {2014}}%
}]{%
Bessho2014}
\APACinsertmetastar {%
Bessho2014}%
\begin{APACrefauthors}%
Bessho, N.%
, Chen, L\BHBI J.%
, Shuster, J\BPBI R.%
\BCBL {}\ \BBA {} Wang, S.%
\end{APACrefauthors}%
\unskip\
\newblock
\APACrefYearMonthDay{2014}{12}{}.
\newblock
{\BBOQ}\APACrefatitle {Electron distribution functions in the electron
  diffusion region of magnetic reconnection: Physics behind the fine
  structures} {Electron distribution functions in the electron diffusion region
  of magnetic reconnection: Physics behind the fine structures}.{\BBCQ}
\newblock
\APACjournalVolNumPages{Geophysical Research Letters}{41}{}{8688-8695}.
\newblock
\begin{APACrefDOI} \doi{10.1002/2014gl062034} \end{APACrefDOI}
\PrintBackRefs{\CurrentBib}

\bibitem [\protect \citeauthoryear {%
Birdsall%
\ \BBA {} Langdon%
}{%
Birdsall%
\ \BBA {} Langdon%
}{%
{\protect \APACyear {1991}}%
}]{%
birdsall91a}
\APACinsertmetastar {%
birdsall91a}%
\begin{APACrefauthors}%
Birdsall, C\BPBI K.%
\BCBT {}\ \BBA {} Langdon, A\BPBI B.%
\end{APACrefauthors}%
\unskip\
\newblock
\APACrefYearMonthDay{1991}{}{}.
\newblock
{\BBOQ}\APACrefatitle {Plasma Physics via Computer Simulation} {Plasma physics
  via computer simulation}.{\BBCQ}
\newblock
\BIn{} (\BCHAP~15).
\newblock
\APACaddressPublisher{Philadelphia}{Institute of Physics Publishing}.
\PrintBackRefs{\CurrentBib}

\bibitem [\protect \citeauthoryear {%
Birn%
, Hesse%
, Nakamura%
\BCBL {}\ \BBA {} Zaharia%
}{%
Birn%
\ \protect \BOthers {.}}{%
{\protect \APACyear {2013}}%
}]{%
Birn13}
\APACinsertmetastar {%
Birn13}%
\begin{APACrefauthors}%
Birn, J.%
, Hesse, M.%
, Nakamura, R.%
\BCBL {}\ \BBA {} Zaharia, S.%
\end{APACrefauthors}%
\unskip\
\newblock
\APACrefYearMonthDay{2013}{}{}.
\newblock
{\BBOQ}\APACrefatitle {Particle acceleration in dipolarization events}
  {Particle acceleration in dipolarization events}.{\BBCQ}
\newblock
\APACjournalVolNumPages{Journal of Geophysical Research: Space
  Physics}{118}{5}{1960-1971}.
\newblock
\begin{APACrefDOI} \doi{https://doi.org/10.1002/jgra.50132} \end{APACrefDOI}
\PrintBackRefs{\CurrentBib}

\bibitem [\protect \citeauthoryear {%
Birn%
\ \BBA {} Priest%
}{%
Birn%
\ \BBA {} Priest%
}{%
{\protect \APACyear {2007}}%
}]{%
Birn07}
\APACinsertmetastar {%
Birn07}%
\begin{APACrefauthors}%
Birn, J.%
\BCBT {}\ \BBA {} Priest, E.%
\end{APACrefauthors}%
\ (\BEDS).
\unskip\
\newblock
\APACrefYear{2007}.
\newblock
\APACrefbtitle {Reconnection of Magnetic Fields} {Reconnection of magnetic
  fields}.
\newblock
\APACaddressPublisher{}{Cambridge University Press}.
\PrintBackRefs{\CurrentBib}

\bibitem [\protect \citeauthoryear {%
B{\"u}chner%
\ \BBA {} Zelenyi%
}{%
B{\"u}chner%
\ \BBA {} Zelenyi%
}{%
{\protect \APACyear {1989}}%
}]{%
Buchner89}
\APACinsertmetastar {%
Buchner89}%
\begin{APACrefauthors}%
B{\"u}chner, J.%
\BCBT {}\ \BBA {} Zelenyi, L\BPBI M.%
\end{APACrefauthors}%
\unskip\
\newblock
\APACrefYearMonthDay{1989}{}{}.
\newblock
{\BBOQ}\APACrefatitle {Regular and chaotic charged particle motion in
  magnetotaillike field reversals: 1. Basic theory of trapped motion} {Regular
  and chaotic charged particle motion in magnetotaillike field reversals: 1.
  basic theory of trapped motion}.{\BBCQ}
\newblock
\APACjournalVolNumPages{Journal of Geophysical Research: Space
  Physics}{94}{A9}{11821-11842}.
\newblock
\begin{APACrefDOI} \doi{https://doi.org/10.1029/JA094iA09p11821}
  \end{APACrefDOI}
\PrintBackRefs{\CurrentBib}

\bibitem [\protect \citeauthoryear {%
Caspi%
}{%
Caspi%
}{%
{\protect \APACyear {2010}}%
}]{%
Caspi10b}
\APACinsertmetastar {%
Caspi10b}%
\begin{APACrefauthors}%
Caspi, A.%
\end{APACrefauthors}%
\unskip\
\newblock
\APACrefYear{2010}.
\unskip\
\newblock
\APACrefbtitle {Super-hot ({T} {$>$} 30 {MK}) Thermal Plasma in Solar Flares}
  {Super-hot ({T} {$>$} 30 {MK}) thermal plasma in solar flares}\
  \APACtypeAddressSchool {\BUPhD}{}{}.
\unskip\
\newblock
\APACaddressSchool {}{University of California, Berkeley}.
\PrintBackRefs{\CurrentBib}

\bibitem [\protect \citeauthoryear {%
Caspi%
, Krucker%
\BCBL {}\ \BBA {} Lin%
}{%
Caspi%
\ \protect \BOthers {.}}{%
{\protect \APACyear {2014}}%
}]{%
Caspi14}
\APACinsertmetastar {%
Caspi14}%
\begin{APACrefauthors}%
Caspi, A.%
, Krucker, S.%
\BCBL {}\ \BBA {} Lin, R\BPBI P.%
\end{APACrefauthors}%
\unskip\
\newblock
\APACrefYearMonthDay{2014}{}{}.
\newblock
{\BBOQ}\APACrefatitle {Statistical Properties of Super-Hot Solar Flares}
  {Statistical properties of super-hot solar flares}.{\BBCQ}
\newblock
\APACjournalVolNumPages{Ap. J.}{781}{}{43}.
\PrintBackRefs{\CurrentBib}

\bibitem [\protect \citeauthoryear {%
Caspi%
\ \BBA {} Lin%
}{%
Caspi%
\ \BBA {} Lin%
}{%
{\protect \APACyear {2010}}%
}]{%
Caspi10}
\APACinsertmetastar {%
Caspi10}%
\begin{APACrefauthors}%
Caspi, A.%
\BCBT {}\ \BBA {} Lin, R\BPBI P.%
\end{APACrefauthors}%
\unskip\
\newblock
\APACrefYearMonthDay{2010}{}{}.
\newblock
{\BBOQ}\APACrefatitle {{RHESSI} Line and Continuum Observations of Super-Hot
  Flare Plasma} {{RHESSI} line and continuum observations of super-hot flare
  plasma}.{\BBCQ}
\newblock
\APACjournalVolNumPages{Ap. J. Lett.}{725}{}{L161}.
\PrintBackRefs{\CurrentBib}

\bibitem [\protect \citeauthoryear {%
Caspi%
, Shih%
, McTiernan%
\BCBL {}\ \BBA {} Krucker%
}{%
Caspi%
\ \protect \BOthers {.}}{%
{\protect \APACyear {2015}}%
}]{%
Caspi15}
\APACinsertmetastar {%
Caspi15}%
\begin{APACrefauthors}%
Caspi, A.%
, Shih, A\BPBI Y.%
, McTiernan, J\BPBI M.%
\BCBL {}\ \BBA {} Krucker, S.%
\end{APACrefauthors}%
\unskip\
\newblock
\APACrefYearMonthDay{2015}{}{}.
\newblock
{\BBOQ}\APACrefatitle {Hard X-ray Imaging of Individual Spectral Components in
  Solar Flares} {Hard x-ray imaging of individual spectral components in solar
  flares}.{\BBCQ}
\newblock
\APACjournalVolNumPages{Ap. J. Lett.}{811}{}{L1}.
\PrintBackRefs{\CurrentBib}

\bibitem [\protect \citeauthoryear {%
Cassak%
\ \BBA {} Shay%
}{%
Cassak%
\ \BBA {} Shay%
}{%
{\protect \APACyear {2007}}%
}]{%
Cassak07d}
\APACinsertmetastar {%
Cassak07d}%
\begin{APACrefauthors}%
Cassak, P\BPBI A.%
\BCBT {}\ \BBA {} Shay, M\BPBI A.%
\end{APACrefauthors}%
\unskip\
\newblock
\APACrefYearMonthDay{2007}{}{}.
\newblock
{\BBOQ}\APACrefatitle {Scaling of Asymmetric Magnetic Reconnection: General
  Theory and Collisional Simulations} {Scaling of asymmetric magnetic
  reconnection: General theory and collisional simulations}.{\BBCQ}
\newblock
\APACjournalVolNumPages{Phys.~Plasmas}{14}{}{102114}.
\PrintBackRefs{\CurrentBib}

\bibitem [\protect \citeauthoryear {%
Cassak%
\ \BBA {} Shay%
}{%
Cassak%
\ \BBA {} Shay%
}{%
{\protect \APACyear {2008}}%
}]{%
Cassak08b}
\APACinsertmetastar {%
Cassak08b}%
\begin{APACrefauthors}%
Cassak, P\BPBI A.%
\BCBT {}\ \BBA {} Shay, M\BPBI A.%
\end{APACrefauthors}%
\unskip\
\newblock
\APACrefYearMonthDay{2008}{}{}.
\newblock
{\BBOQ}\APACrefatitle {Scaling of Asymmetric {H}all Reconnection} {Scaling of
  asymmetric {H}all reconnection}.{\BBCQ}
\newblock
\APACjournalVolNumPages{Geophys.~Res.~Lett.}{35}{}{L19102}.
\PrintBackRefs{\CurrentBib}

\bibitem [\protect \citeauthoryear {%
{Cheung}%
\ \protect \BOthers {.}}{%
{Cheung}%
\ \protect \BOthers {.}}{%
{\protect \APACyear {2019}}%
}]{%
Cheung19}
\APACinsertmetastar {%
Cheung19}%
\begin{APACrefauthors}%
{Cheung}, M\BPBI C\BPBI M.%
, {Rempel}, M.%
, {Chintzoglou}, G.%
, {Chen}, F.%
, {Testa}, P.%
, {Mart{\'\i}nez-Sykora}, J.%
\BDBL {}{McIntosh}, S\BPBI W.%
\end{APACrefauthors}%
\unskip\
\newblock
\APACrefYearMonthDay{2019}{{\APACmonth{11}}}{}.
\newblock
{\BBOQ}\APACrefatitle {{A comprehensive three-dimensional radiative
  magnetohydrodynamic simulation of a solar flare}} {{A comprehensive
  three-dimensional radiative magnetohydrodynamic simulation of a solar
  flare}}.{\BBCQ}
\newblock
\APACjournalVolNumPages{Nature Astronomy}{3}{}{160-166}.
\newblock
\begin{APACrefDOI} \doi{10.1038/s41550-018-0629-3} \end{APACrefDOI}
\PrintBackRefs{\CurrentBib}

\bibitem [\protect \citeauthoryear {%
Choi%
, Bessho%
, Wang%
, Chen%
\BCBL {}\ \BBA {} Hesse%
}{%
Choi%
\ \protect \BOthers {.}}{%
{\protect \APACyear {2022}}%
}]{%
Choi2022}
\APACinsertmetastar {%
Choi2022}%
\begin{APACrefauthors}%
Choi, S.%
, Bessho, N.%
, Wang, S.%
, Chen, L\BHBI J.%
\BCBL {}\ \BBA {} Hesse, M.%
\end{APACrefauthors}%
\unskip\
\newblock
\APACrefYearMonthDay{2022}{}{}.
\newblock
{\BBOQ}\APACrefatitle {Whistler waves generated by nongyrotropic and gyrotropic
  electron beams during asymmetric guide field reconnection} {Whistler waves
  generated by nongyrotropic and gyrotropic electron beams during asymmetric
  guide field reconnection}.{\BBCQ}
\newblock
\APACjournalVolNumPages{Physics of Plasmas}{29}{1}{012903}.
\newblock
\begin{APACrefDOI} \doi{10.1063/5.0059884} \end{APACrefDOI}
\PrintBackRefs{\CurrentBib}

\bibitem [\protect \citeauthoryear {%
Deng%
\ \protect \BOthers {.}}{%
Deng%
\ \protect \BOthers {.}}{%
{\protect \APACyear {2010}}%
}]{%
Deng10}
\APACinsertmetastar {%
Deng10}%
\begin{APACrefauthors}%
Deng, X.%
, Ashour-Abdalla, M.%
, Zhou, M.%
, Walker, R.%
, El-Alaoui, M.%
, Angelopoulos, V.%
\BDBL {}Schriver, D.%
\end{APACrefauthors}%
\unskip\
\newblock
\APACrefYearMonthDay{2010}{}{}.
\newblock
{\BBOQ}\APACrefatitle {Wave and particle characteristics of earthward electron
  injections associated with dipolarization fronts} {Wave and particle
  characteristics of earthward electron injections associated with
  dipolarization fronts}.{\BBCQ}
\newblock
\APACjournalVolNumPages{Journal of Geophysical Research: Space
  Physics}{115}{A9}{{A09225}}.
\newblock
\begin{APACrefDOI} \doi{https://doi.org/10.1029/2009JA015107} \end{APACrefDOI}
\PrintBackRefs{\CurrentBib}

\bibitem [\protect \citeauthoryear {%
Divin%
, Sitnov%
, Swisdak%
\BCBL {}\ \BBA {} Drake%
}{%
Divin%
\ \protect \BOthers {.}}{%
{\protect \APACyear {2007}}%
}]{%
divin:2007}
\APACinsertmetastar {%
divin:2007}%
\begin{APACrefauthors}%
Divin, A\BPBI V.%
, Sitnov, M\BPBI I.%
, Swisdak, M.%
\BCBL {}\ \BBA {} Drake, J\BPBI F.%
\end{APACrefauthors}%
\unskip\
\newblock
\APACrefYearMonthDay{2007}{}{}.
\newblock
{\BBOQ}\APACrefatitle {{Reconnection onset in the magnetotail: Particle
  simulations with open boundary conditions}} {{Reconnection onset in the
  magnetotail: Particle simulations with open boundary conditions}}.{\BBCQ}
\newblock
\APACjournalVolNumPages{Geophys. Res. Lett.}{34}{9}{09109}.
\newblock
\begin{APACrefDOI} \doi{10.1029/2007GL029292} \end{APACrefDOI}
\PrintBackRefs{\CurrentBib}

\bibitem [\protect \citeauthoryear {%
Egedal%
, Wetherton%
, Daughton%
\BCBL {}\ \BBA {} Le%
}{%
Egedal%
\ \protect \BOthers {.}}{%
{\protect \APACyear {2016}}%
}]{%
egedal_2016_PoP}
\APACinsertmetastar {%
egedal_2016_PoP}%
\begin{APACrefauthors}%
Egedal, J.%
, Wetherton, B.%
, Daughton, W.%
\BCBL {}\ \BBA {} Le, A.%
\end{APACrefauthors}%
\unskip\
\newblock
\APACrefYearMonthDay{2016}{}{}.
\newblock
{\BBOQ}\APACrefatitle {Processes setting the structure of the electron
  distribution function within the exhausts of anti-parallel reconnection}
  {Processes setting the structure of the electron distribution function within
  the exhausts of anti-parallel reconnection}.{\BBCQ}
\newblock
\APACjournalVolNumPages{Physics of Plasmas}{23}{12}{122904}.
\newblock
\begin{APACrefDOI} \doi{10.1063/1.4972135} \end{APACrefDOI}
\PrintBackRefs{\CurrentBib}

\bibitem [\protect \citeauthoryear {%
{Fletcher}%
\ \protect \BOthers {.}}{%
{Fletcher}%
\ \protect \BOthers {.}}{%
{\protect \APACyear {2011}}%
}]{%
Fletcher11}
\APACinsertmetastar {%
Fletcher11}%
\begin{APACrefauthors}%
{Fletcher}, L.%
, {Dennis}, B\BPBI R.%
, {Hudson}, H\BPBI S.%
, {Krucker}, S.%
, {Phillips}, K.%
, {Veronig}, A.%
\BDBL {}{Temmer}, M.%
\end{APACrefauthors}%
\unskip\
\newblock
\APACrefYearMonthDay{2011}{}{}.
\newblock
{\BBOQ}\APACrefatitle {{An Observational Overview of Solar Flares}} {{An
  Observational Overview of Solar Flares}}.{\BBCQ}
\newblock
\APACjournalVolNumPages{Space Science Reviews}{159}{1-4}{19-106}.
\newblock
\begin{APACrefDOI} \doi{10.1007/s11214-010-9701-8} \end{APACrefDOI}
\PrintBackRefs{\CurrentBib}

\bibitem [\protect \citeauthoryear {%
H.~Fu%
\ \protect \BOthers {.}}{%
H.~Fu%
\ \protect \BOthers {.}}{%
{\protect \APACyear {2020}}%
}]{%
Fu20}
\APACinsertmetastar {%
Fu20}%
\begin{APACrefauthors}%
Fu, H.%
, Grigorenko, E\BPBI E.%
, Gabrielse, C.%
, Liu, C.%
, Lu, S.%
, Hwang, K\BPBI J.%
\BDBL {}Chen, F.%
\end{APACrefauthors}%
\unskip\
\newblock
\APACrefYearMonthDay{2020}{}{}.
\newblock
{\BBOQ}\APACrefatitle {Magnetotail dipolarization fronts and particle
  acceleration: A review} {Magnetotail dipolarization fronts and particle
  acceleration: A review}.{\BBCQ}
\newblock
\APACjournalVolNumPages{Sci. China Earth Sci.}{63}{}{235-256}.
\newblock
\begin{APACrefDOI} \doi{10.1007/s11430-019-9551-y} \end{APACrefDOI}
\PrintBackRefs{\CurrentBib}

\bibitem [\protect \citeauthoryear {%
H\BPBI S.~Fu%
\ \protect \BOthers {.}}{%
H\BPBI S.~Fu%
\ \protect \BOthers {.}}{%
{\protect \APACyear {2013}}%
}]{%
Fu13}
\APACinsertmetastar {%
Fu13}%
\begin{APACrefauthors}%
Fu, H\BPBI S.%
, Cao, J\BPBI B.%
, Khotyaintsev, Y\BPBI V.%
, Sitnov, M\BPBI I.%
, Runov, A.%
, Fu, S\BPBI Y.%
\BDBL {}Huang, S\BPBI Y.%
\end{APACrefauthors}%
\unskip\
\newblock
\APACrefYearMonthDay{2013}{}{}.
\newblock
{\BBOQ}\APACrefatitle {Dipolarization fronts as a consequence of transient
  reconnection: In situ evidence} {Dipolarization fronts as a consequence of
  transient reconnection: In situ evidence}.{\BBCQ}
\newblock
\APACjournalVolNumPages{Geophysical Research Letters}{40}{23}{6023-6027}.
\newblock
\begin{APACrefDOI} \doi{https://doi.org/10.1002/2013GL058620} \end{APACrefDOI}
\PrintBackRefs{\CurrentBib}

\bibitem [\protect \citeauthoryear {%
H\BPBI S.~Fu%
, Khotyaintsev%
, André%
\BCBL {}\ \BBA {} Vaivads%
}{%
H\BPBI S.~Fu%
\ \protect \BOthers {.}}{%
{\protect \APACyear {2011}}%
}]{%
Fu11}
\APACinsertmetastar {%
Fu11}%
\begin{APACrefauthors}%
Fu, H\BPBI S.%
, Khotyaintsev, Y\BPBI V.%
, André, M.%
\BCBL {}\ \BBA {} Vaivads, A.%
\end{APACrefauthors}%
\unskip\
\newblock
\APACrefYearMonthDay{2011}{}{}.
\newblock
{\BBOQ}\APACrefatitle {Fermi and betatron acceleration of suprathermal
  electrons behind dipolarization fronts} {Fermi and betatron acceleration of
  suprathermal electrons behind dipolarization fronts}.{\BBCQ}
\newblock
\APACjournalVolNumPages{Geophysical Research Letters}{38}{16}{{L16104}}.
\newblock
\begin{APACrefDOI} \doi{https://doi.org/10.1029/2011GL048528} \end{APACrefDOI}
\PrintBackRefs{\CurrentBib}

\bibitem [\protect \citeauthoryear {%
H\BPBI S.~Fu%
, Khotyaintsev%
, Vaivads%
, André%
\BCBL {}\ \BBA {} Huang%
}{%
H\BPBI S.~Fu%
, Khotyaintsev%
, Vaivads%
, André%
\BCBL {}\ \BBA {} Huang%
}{%
{\protect \APACyear {2012}}%
}]{%
Fu12}
\APACinsertmetastar {%
Fu12}%
\begin{APACrefauthors}%
Fu, H\BPBI S.%
, Khotyaintsev, Y\BPBI V.%
, Vaivads, A.%
, André, M.%
\BCBL {}\ \BBA {} Huang, S\BPBI Y.%
\end{APACrefauthors}%
\unskip\
\newblock
\APACrefYearMonthDay{2012}{}{}.
\newblock
{\BBOQ}\APACrefatitle {Occurrence rate of earthward-propagating dipolarization
  fronts} {Occurrence rate of earthward-propagating dipolarization
  fronts}.{\BBCQ}
\newblock
\APACjournalVolNumPages{Geophysical Research Letters}{39}{10}{{L10101}}.
\newblock
\begin{APACrefDOI} \doi{https://doi.org/10.1029/2012GL051784} \end{APACrefDOI}
\PrintBackRefs{\CurrentBib}

\bibitem [\protect \citeauthoryear {%
H\BPBI S.~Fu%
, Khotyaintsev%
, Vaivads%
, André%
, Sergeev%
\BCBL {}\ \protect \BOthers {.}}{%
H\BPBI S.~Fu%
, Khotyaintsev%
, Vaivads%
, André%
, Sergeev%
\BCBL {}\ \protect \BOthers {.}}{%
{\protect \APACyear {2012}}%
}]{%
Fu12b}
\APACinsertmetastar {%
Fu12b}%
\begin{APACrefauthors}%
Fu, H\BPBI S.%
, Khotyaintsev, Y\BPBI V.%
, Vaivads, A.%
, André, M.%
, Sergeev, V\BPBI A.%
, Huang, S\BPBI Y.%
\BDBL {}Daly, P\BPBI W.%
\end{APACrefauthors}%
\unskip\
\newblock
\APACrefYearMonthDay{2012}{}{}.
\newblock
{\BBOQ}\APACrefatitle {Pitch angle distribution of suprathermal electrons
  behind dipolarization fronts: A statistical overview} {Pitch angle
  distribution of suprathermal electrons behind dipolarization fronts: A
  statistical overview}.{\BBCQ}
\newblock
\APACjournalVolNumPages{Journal of Geophysical Research: Space
  Physics}{117}{A12}{{A12221}}.
\newblock
\begin{APACrefDOI} \doi{https://doi.org/10.1029/2012JA018141} \end{APACrefDOI}
\PrintBackRefs{\CurrentBib}

\bibitem [\protect \citeauthoryear {%
Fujimoto%
\ \BBA {} Sydora%
}{%
Fujimoto%
\ \BBA {} Sydora%
}{%
{\protect \APACyear {2008}}%
}]{%
fujimoto_2008_whistler}
\APACinsertmetastar {%
fujimoto_2008_whistler}%
\begin{APACrefauthors}%
Fujimoto, K.%
\BCBT {}\ \BBA {} Sydora, R\BPBI D.%
\end{APACrefauthors}%
\unskip\
\newblock
\APACrefYearMonthDay{2008}{}{}.
\newblock
{\BBOQ}\APACrefatitle {Whistler waves associated with magnetic reconnection}
  {Whistler waves associated with magnetic reconnection}.{\BBCQ}
\newblock
\APACjournalVolNumPages{Geophysical Research Letters}{35}{19}{}.
\newblock
\begin{APACrefDOI} \doi{10.1029/2008GL035201} \end{APACrefDOI}
\PrintBackRefs{\CurrentBib}

\bibitem [\protect \citeauthoryear {%
Gary%
\ \BBA {} Madland%
}{%
Gary%
\ \BBA {} Madland%
}{%
{\protect \APACyear {1985}}%
}]{%
Gary85}
\APACinsertmetastar {%
Gary85}%
\begin{APACrefauthors}%
Gary, S\BPBI P.%
\BCBT {}\ \BBA {} Madland, C\BPBI D.%
\end{APACrefauthors}%
\unskip\
\newblock
\APACrefYearMonthDay{1985}{}{}.
\newblock
{\BBOQ}\APACrefatitle {Electromagnetic electron temperature anisotropy
  instabilities} {Electromagnetic electron temperature anisotropy
  instabilities}.{\BBCQ}
\newblock
\APACjournalVolNumPages{Journal of Geophysical Research: Space
  Physics}{90}{A8}{7607-7610}.
\newblock
\begin{APACrefDOI} \doi{https://doi.org/10.1029/JA090iA08p07607}
  \end{APACrefDOI}
\PrintBackRefs{\CurrentBib}

\bibitem [\protect \citeauthoryear {%
Gonzalez%
\ \BBA {} Parker%
}{%
Gonzalez%
\ \BBA {} Parker%
}{%
{\protect \APACyear {2016}}%
}]{%
Gonzalez16}
\APACinsertmetastar {%
Gonzalez16}%
\begin{APACrefauthors}%
Gonzalez, W.%
\BCBT {}\ \BBA {} Parker, E.%
\end{APACrefauthors}%
\unskip\
\newblock
\APACrefYear{2016}.
\newblock
\APACrefbtitle {Magnetic Reconnection} {Magnetic reconnection}.
\newblock
\APACaddressPublisher{}{Springer}.
\PrintBackRefs{\CurrentBib}

\bibitem [\protect \citeauthoryear {%
Grigorenko%
\ \protect \BOthers {.}}{%
Grigorenko%
\ \protect \BOthers {.}}{%
{\protect \APACyear {2020}}%
}]{%
Grigorenko20}
\APACinsertmetastar {%
Grigorenko20}%
\begin{APACrefauthors}%
Grigorenko, E\BPBI E.%
, Malykhin, A\BPBI Y.%
, Shklyar, D\BPBI R.%
, Fadanelli, S.%
, Lavraud, B.%
, Panov, E\BPBI V.%
\BDBL {}Le~Contel, O.%
\end{APACrefauthors}%
\unskip\
\newblock
\APACrefYearMonthDay{2020}{}{}.
\newblock
{\BBOQ}\APACrefatitle {Investigation of Electron Distribution Functions
  Associated With Whistler Waves at Dipolarization Fronts in the Earth's
  Magnetotail: {MMS} Observations} {Investigation of electron distribution
  functions associated with whistler waves at dipolarization fronts in the
  earth's magnetotail: {MMS} observations}.{\BBCQ}
\newblock
\APACjournalVolNumPages{Journal of Geophysical Research: Space
  Physics}{125}{9}{e2020JA028268}.
\newblock
\begin{APACrefDOI} \doi{https://doi.org/10.1029/2020JA028268} \end{APACrefDOI}
\PrintBackRefs{\CurrentBib}

\bibitem [\protect \citeauthoryear {%
Guzdar%
, Drake%
, McCarthy%
, Hassam%
\BCBL {}\ \BBA {} Liu%
}{%
Guzdar%
\ \protect \BOthers {.}}{%
{\protect \APACyear {1993}}%
}]{%
guzdar93a}
\APACinsertmetastar {%
guzdar93a}%
\begin{APACrefauthors}%
Guzdar, P\BPBI N.%
, Drake, J\BPBI F.%
, McCarthy, D.%
, Hassam, A\BPBI B.%
\BCBL {}\ \BBA {} Liu, C\BPBI S.%
\end{APACrefauthors}%
\unskip\
\newblock
\APACrefYearMonthDay{1993}{}{}.
\newblock
{\BBOQ}\APACrefatitle {Three-dimensional fluid simulations of the nonlinear
  drift-resistive ballooning modes in tokamak edge plasmas} {Three-dimensional
  fluid simulations of the nonlinear drift-resistive ballooning modes in
  tokamak edge plasmas}.{\BBCQ}
\newblock
\APACjournalVolNumPages{Phys. Fluids B}{5}{10}{3712--3727}.
\PrintBackRefs{\CurrentBib}

\bibitem [\protect \citeauthoryear {%
Hesse%
\ \BBA {} Birn%
}{%
Hesse%
\ \BBA {} Birn%
}{%
{\protect \APACyear {1991}}%
}]{%
hesse&birn_1991_JGR}
\APACinsertmetastar {%
hesse&birn_1991_JGR}%
\begin{APACrefauthors}%
Hesse, M.%
\BCBT {}\ \BBA {} Birn, J.%
\end{APACrefauthors}%
\unskip\
\newblock
\APACrefYearMonthDay{1991}{}{}.
\newblock
{\BBOQ}\APACrefatitle {On dipolarization and its relation to the substorm
  current wedge} {On dipolarization and its relation to the substorm current
  wedge}.{\BBCQ}
\newblock
\APACjournalVolNumPages{Journal of Geophysical Research: Space
  Physics}{96}{A11}{19417--19426}.
\newblock
\begin{APACrefDOI} \doi{https://doi.org/10.1029/91JA01953} \end{APACrefDOI}
\PrintBackRefs{\CurrentBib}

\bibitem [\protect \citeauthoryear {%
{Holman}%
\ \protect \BOthers {.}}{%
{Holman}%
\ \protect \BOthers {.}}{%
{\protect \APACyear {2011}}%
}]{%
Holman11}
\APACinsertmetastar {%
Holman11}%
\begin{APACrefauthors}%
{Holman}, G\BPBI D.%
, {Aschwanden}, M\BPBI J.%
, {Aurass}, H.%
, {Battaglia}, M.%
, {Grigis}, P\BPBI C.%
, {Kontar}, E\BPBI P.%
\BDBL {}{Zharkova}, V\BPBI V.%
\end{APACrefauthors}%
\unskip\
\newblock
\APACrefYearMonthDay{2011}{}{}.
\newblock
{\BBOQ}\APACrefatitle {{Implications of X-ray Observations for Electron
  Acceleration and Propagation in Solar Flares}} {{Implications of X-ray
  Observations for Electron Acceleration and Propagation in Solar
  Flares}}.{\BBCQ}
\newblock
\APACjournalVolNumPages{Space Science Reviews}{159}{1-4}{107-166}.
\newblock
\begin{APACrefDOI} \doi{10.1007/s11214-010-9680-9} \end{APACrefDOI}
\PrintBackRefs{\CurrentBib}

\bibitem [\protect \citeauthoryear {%
Hoshino%
, Mukai%
, Terasawa%
\BCBL {}\ \BBA {} Shinohara%
}{%
Hoshino%
\ \protect \BOthers {.}}{%
{\protect \APACyear {2001}}%
}]{%
hoshino01a}
\APACinsertmetastar {%
hoshino01a}%
\begin{APACrefauthors}%
Hoshino, M.%
, Mukai, T.%
, Terasawa, T.%
\BCBL {}\ \BBA {} Shinohara, I.%
\end{APACrefauthors}%
\unskip\
\newblock
\APACrefYearMonthDay{2001}{}{}.
\newblock
{\BBOQ}\APACrefatitle {Suprathermal electron acceleration in magnetic
  reconnection} {Suprathermal electron acceleration in magnetic
  reconnection}.{\BBCQ}
\newblock
\APACjournalVolNumPages{J. Geophys. Res.}{106}{A11}{25,979--25,997}.
\PrintBackRefs{\CurrentBib}

\bibitem [\protect \citeauthoryear {%
Huang%
, Lu%
, Lu%
, Wang%
\BCBL {}\ \BBA {} Wang%
}{%
Huang%
\ \protect \BOthers {.}}{%
{\protect \APACyear {2021}}%
}]{%
huang_formation_2021}
\APACinsertmetastar {%
huang_formation_2021}%
\begin{APACrefauthors}%
Huang, K.%
, Lu, Q.%
, Lu, S.%
, Wang, R.%
\BCBL {}\ \BBA {} Wang, S.%
\end{APACrefauthors}%
\unskip\
\newblock
\APACrefYearMonthDay{2021}{}{}.
\newblock
{\BBOQ}\APACrefatitle {Formation of {Pancake}, {Rolling} {Pin}, and {Cigar}
  {Distributions} of {Energetic} {Electrons} at the {Dipolarization} {Fronts}
  ({DFs}) {Driven} by {Magnetic} {Reconnection}: {A} {Two}-{Dimensional}
  {Particle}-{In}-{Cell} {Simulation}} {Formation of {Pancake}, {Rolling}
  {Pin}, and {Cigar} {Distributions} of {Energetic} {Electrons} at the
  {Dipolarization} {Fronts} ({DFs}) {Driven} by {Magnetic} {Reconnection}: {A}
  {Two}-{Dimensional} {Particle}-{In}-{Cell} {Simulation}}.{\BBCQ}
\newblock
\APACjournalVolNumPages{Journal of Geophysical Research: Space
  Physics}{126}{10}{e2021JA029939}.
\newblock
\begin{APACrefDOI} \doi{10.1029/2021JA029939} \end{APACrefDOI}
\PrintBackRefs{\CurrentBib}

\bibitem [\protect \citeauthoryear {%
Hwang%
, Goldstein%
, Lee%
\BCBL {}\ \BBA {} Pickett%
}{%
Hwang%
\ \protect \BOthers {.}}{%
{\protect \APACyear {2011}}%
}]{%
Hwang11}
\APACinsertmetastar {%
Hwang11}%
\begin{APACrefauthors}%
Hwang, K\BHBI J.%
, Goldstein, M\BPBI L.%
, Lee, E.%
\BCBL {}\ \BBA {} Pickett, J\BPBI S.%
\end{APACrefauthors}%
\unskip\
\newblock
\APACrefYearMonthDay{2011}{}{}.
\newblock
{\BBOQ}\APACrefatitle {Cluster observations of multiple dipolarization fronts}
  {Cluster observations of multiple dipolarization fronts}.{\BBCQ}
\newblock
\APACjournalVolNumPages{Journal of Geophysical Research: Space
  Physics}{116}{A5}{}.
\newblock
\begin{APACrefDOI} \doi{https://doi.org/10.1029/2010JA015742} \end{APACrefDOI}
\PrintBackRefs{\CurrentBib}

\bibitem [\protect \citeauthoryear {%
Krucker%
\ \protect \BOthers {.}}{%
Krucker%
\ \protect \BOthers {.}}{%
{\protect \APACyear {2010}}%
}]{%
krucker10a}
\APACinsertmetastar {%
krucker10a}%
\begin{APACrefauthors}%
Krucker, S.%
, Hudson, H\BPBI S.%
, Glesener, L.%
, White, S\BPBI M.%
, Masuda, S.%
, Wuelser, J\BHBI P.%
\BCBL {}\ \BBA {} Lin, R\BPBI P.%
\end{APACrefauthors}%
\unskip\
\newblock
\APACrefYearMonthDay{2010}{}{}.
\newblock
{\BBOQ}\APACrefatitle {Measurements of the coronal acceleration region of a
  solar flare} {Measurements of the coronal acceleration region of a solar
  flare}.{\BBCQ}
\newblock
\APACjournalVolNumPages{Ap. J.}{714}{}{1108--1119}.
\newblock
\begin{APACrefDOI} \doi{10.1088/0004-637X/714/2/1108} \end{APACrefDOI}
\PrintBackRefs{\CurrentBib}

\bibitem [\protect \citeauthoryear {%
Le~Contel%
\ \protect \BOthers {.}}{%
Le~Contel%
\ \protect \BOthers {.}}{%
{\protect \APACyear {2009}}%
}]{%
LeContel09}
\APACinsertmetastar {%
LeContel09}%
\begin{APACrefauthors}%
Le~Contel, O.%
, Roux, A.%
, Jacquey, C.%
, Robert, P.%
, Berthomier, M.%
, Chust, T.%
\BDBL {}Singer, H.%
\end{APACrefauthors}%
\unskip\
\newblock
\APACrefYearMonthDay{2009}{}{}.
\newblock
{\BBOQ}\APACrefatitle {Quasi-parallel whistler mode waves observed by THEMIS
  during near-earth dipolarizations} {Quasi-parallel whistler mode waves
  observed by themis during near-earth dipolarizations}.{\BBCQ}
\newblock
\APACjournalVolNumPages{Annales Geophysicae}{27}{6}{2259--2275}.
\newblock
\begin{APACrefDOI} \doi{10.5194/angeo-27-2259-2009} \end{APACrefDOI}
\PrintBackRefs{\CurrentBib}

\bibitem [\protect \citeauthoryear {%
Lembege%
\ \BBA {} Pellat%
}{%
Lembege%
\ \BBA {} Pellat%
}{%
{\protect \APACyear {1982}}%
}]{%
Lembege82}
\APACinsertmetastar {%
Lembege82}%
\begin{APACrefauthors}%
Lembege, B.%
\BCBT {}\ \BBA {} Pellat, R.%
\end{APACrefauthors}%
\unskip\
\newblock
\APACrefYearMonthDay{1982}{}{}.
\newblock
{\BBOQ}\APACrefatitle {Stability of a thick two-dimensional quasineutral sheet}
  {Stability of a thick two-dimensional quasineutral sheet}.{\BBCQ}
\newblock
\APACjournalVolNumPages{Phys. Fluids}{25}{}{1995}.
\PrintBackRefs{\CurrentBib}

\bibitem [\protect \citeauthoryear {%
Li%
\ \protect \BOthers {.}}{%
Li%
\ \protect \BOthers {.}}{%
{\protect \APACyear {2015}}%
}]{%
Li15}
\APACinsertmetastar {%
Li15}%
\begin{APACrefauthors}%
Li, H.%
, Zhou, M.%
, Deng, X.%
, Yuan, Z.%
, Guo, L.%
, Yu, X.%
\BDBL {}Huang, S.%
\end{APACrefauthors}%
\unskip\
\newblock
\APACrefYearMonthDay{2015}{}{}.
\newblock
{\BBOQ}\APACrefatitle {A statistical study on the whistler waves behind
  dipolarization fronts} {A statistical study on the whistler waves behind
  dipolarization fronts}.{\BBCQ}
\newblock
\APACjournalVolNumPages{Journal of Geophysical Research: Space
  Physics}{120}{2}{1086-1095}.
\newblock
\begin{APACrefDOI} \doi{https://doi.org/10.1002/2014JA020474} \end{APACrefDOI}
\PrintBackRefs{\CurrentBib}

\bibitem [\protect \citeauthoryear {%
C\BPBI M.~Liu%
\ \BBA {} Fu%
}{%
C\BPBI M.~Liu%
\ \BBA {} Fu%
}{%
{\protect \APACyear {2019}}%
}]{%
Liu19}
\APACinsertmetastar {%
Liu19}%
\begin{APACrefauthors}%
Liu, C\BPBI M.%
\BCBT {}\ \BBA {} Fu, F\BPBI S.%
\end{APACrefauthors}%
\unskip\
\newblock
\APACrefYearMonthDay{2019}{}{}.
\newblock
{\BBOQ}\APACrefatitle {Anchor Point of Electron Acceleration around
  Dipolarization Fronts in Space Plasmas} {Anchor point of electron
  acceleration around dipolarization fronts in space plasmas}.{\BBCQ}
\newblock
\APACjournalVolNumPages{Ap.~J.~Lett.}{873}{}{{L2}}.
\PrintBackRefs{\CurrentBib}

\bibitem [\protect \citeauthoryear {%
C\BPBI M.~Liu%
, Fu%
, Cao%
\BCBL {}\ \protect \BOthers {.}}{%
C\BPBI M.~Liu%
, Fu%
, Cao%
\BCBL {}\ \protect \BOthers {.}}{%
{\protect \APACyear {2017}}%
}]{%
liu_rapid_2017}
\APACinsertmetastar {%
liu_rapid_2017}%
\begin{APACrefauthors}%
Liu, C\BPBI M.%
, Fu, H\BPBI S.%
, Cao, J\BPBI B.%
, Xu, Y.%
, Yu, Y\BPBI Q.%
, Kronberg, E\BPBI A.%
\BCBL {}\ \BBA {} Daly, P\BPBI W.%
\end{APACrefauthors}%
\unskip\
\newblock
\APACrefYearMonthDay{2017}{}{}.
\newblock
{\BBOQ}\APACrefatitle {Rapid {Pitch} {Angle} {Evolution} of {Suprathermal}
  {Electrons} {Behind} {Dipolarization} {Fronts}} {Rapid {Pitch} {Angle}
  {Evolution} of {Suprathermal} {Electrons} {Behind} {Dipolarization}
  {Fronts}}.{\BBCQ}
\newblock
\APACjournalVolNumPages{Geophysical Research Letters}{44}{20}{10,116--10,124}.
\newblock
\begin{APACrefDOI} \doi{10.1002/2017GL075007} \end{APACrefDOI}
\PrintBackRefs{\CurrentBib}

\bibitem [\protect \citeauthoryear {%
C\BPBI M.~Liu%
, Fu%
, Xu%
, Cao%
\BCBL {}\ \BBA {} Liu%
}{%
C\BPBI M.~Liu%
, Fu%
, Xu%
\BCBL {}\ \protect \BOthers {.}}{%
{\protect \APACyear {2017}}%
}]{%
liu_explaining_2017}
\APACinsertmetastar {%
liu_explaining_2017}%
\begin{APACrefauthors}%
Liu, C\BPBI M.%
, Fu, H\BPBI S.%
, Xu, Y.%
, Cao, J\BPBI B.%
\BCBL {}\ \BBA {} Liu, W\BPBI L.%
\end{APACrefauthors}%
\unskip\
\newblock
\APACrefYearMonthDay{2017}{}{}.
\newblock
{\BBOQ}\APACrefatitle {Explaining the rolling-pin distribution of suprathermal
  electrons behind dipolarization fronts} {Explaining the rolling-pin
  distribution of suprathermal electrons behind dipolarization fronts}.{\BBCQ}
\newblock
\APACjournalVolNumPages{Geophysical Research Letters}{44}{13}{6492--6499}.
\newblock
\begin{APACrefDOI} \doi{10.1002/2017GL074029} \end{APACrefDOI}
\PrintBackRefs{\CurrentBib}

\bibitem [\protect \citeauthoryear {%
C\BPBI M.~Liu%
\ \protect \BOthers {.}}{%
C\BPBI M.~Liu%
\ \protect \BOthers {.}}{%
{\protect \APACyear {2018}}%
}]{%
Liu18}
\APACinsertmetastar {%
Liu18}%
\begin{APACrefauthors}%
Liu, C\BPBI M.%
, Fu, H\BPBI S.%
, Xu, Y.%
, Khotyaintsev, Y\BPBI V.%
, Burch, J\BPBI L.%
, Ergun, R\BPBI E.%
\BDBL {}Torbert, R\BPBI B.%
\end{APACrefauthors}%
\unskip\
\newblock
\APACrefYearMonthDay{2018}{}{}.
\newblock
{\BBOQ}\APACrefatitle {Electron-Scale Measurements of Dipolarization Front}
  {Electron-scale measurements of dipolarization front}.{\BBCQ}
\newblock
\APACjournalVolNumPages{Geophysical Research Letters}{45}{10}{4628-4638}.
\newblock
\begin{APACrefDOI} \doi{https://doi.org/10.1029/2018GL077928} \end{APACrefDOI}
\PrintBackRefs{\CurrentBib}

\bibitem [\protect \citeauthoryear {%
Y\BHBI H.~Liu%
\ \protect \BOthers {.}}{%
Y\BHBI H.~Liu%
\ \protect \BOthers {.}}{%
{\protect \APACyear {2022}}%
}]{%
liu_FirstPrinciple_2022}
\APACinsertmetastar {%
liu_FirstPrinciple_2022}%
\begin{APACrefauthors}%
Liu, Y\BHBI H.%
, Cassak, P.%
, Li, X.%
, Hesse, M.%
, Lin, S\BHBI C.%
\BCBL {}\ \BBA {} Genestreti, K.%
\end{APACrefauthors}%
\unskip\
\newblock
\APACrefYearMonthDay{2022}{{\APACmonth{04}}}{}.
\newblock
{\BBOQ}\APACrefatitle {First-principles theory of the rate of magnetic
  reconnection in magnetospheric and solar plasmas} {First-principles theory of
  the rate of magnetic reconnection in magnetospheric and solar
  plasmas}.{\BBCQ}
\newblock
\APACjournalVolNumPages{Communications Physics}{5}{1}{1--9}.
\newblock
\begin{APACrefDOI} \doi{10.1038/s42005-022-00854-x} \end{APACrefDOI}
\PrintBackRefs{\CurrentBib}

\bibitem [\protect \citeauthoryear {%
Longcope%
\ \BBA {} Guidoni%
}{%
Longcope%
\ \BBA {} Guidoni%
}{%
{\protect \APACyear {2011}}%
}]{%
Longcope11}
\APACinsertmetastar {%
Longcope11}%
\begin{APACrefauthors}%
Longcope, D\BPBI W.%
\BCBT {}\ \BBA {} Guidoni, S.%
\end{APACrefauthors}%
\unskip\
\newblock
\APACrefYearMonthDay{2011}{}{}.
\newblock
{\BBOQ}\APACrefatitle {A model for the origin of high density in looptop
  {X}-ray source} {A model for the origin of high density in looptop {X}-ray
  source}.{\BBCQ}
\newblock
\APACjournalVolNumPages{Ap. J.}{740}{}{73}.
\PrintBackRefs{\CurrentBib}

\bibitem [\protect \citeauthoryear {%
Longcope%
, Jardins%
, Carranza-Fulmer%
\BCBL {}\ \BBA {} Qiu%
}{%
Longcope%
\ \protect \BOthers {.}}{%
{\protect \APACyear {2010}}%
}]{%
Longcope10}
\APACinsertmetastar {%
Longcope10}%
\begin{APACrefauthors}%
Longcope, D\BPBI W.%
, Jardins, A\BPBI C\BPBI D.%
, Carranza-Fulmer, T.%
\BCBL {}\ \BBA {} Qiu, J.%
\end{APACrefauthors}%
\unskip\
\newblock
\APACrefYearMonthDay{2010}{}{}.
\newblock
{\BBOQ}\APACrefatitle {A Quantitative Model of Energy Release and Heating by
  Time-dependent, Localized Reconnection in a Flare with Thermal Loop-top
  {X}-ray Source} {A quantitative model of energy release and heating by
  time-dependent, localized reconnection in a flare with thermal loop-top
  {X}-ray source}.{\BBCQ}
\newblock
\APACjournalVolNumPages{Solar Phys.}{267}{}{107}.
\PrintBackRefs{\CurrentBib}

\bibitem [\protect \citeauthoryear {%
Longcope%
, Qiu%
\BCBL {}\ \BBA {} Brewer%
}{%
Longcope%
\ \protect \BOthers {.}}{%
{\protect \APACyear {2016}}%
}]{%
Longcope16}
\APACinsertmetastar {%
Longcope16}%
\begin{APACrefauthors}%
Longcope, D\BPBI W.%
, Qiu, J.%
\BCBL {}\ \BBA {} Brewer, J.%
\end{APACrefauthors}%
\unskip\
\newblock
\APACrefYearMonthDay{2016}{}{}.
\newblock
{\BBOQ}\APACrefatitle {A reconnection-driven model of the hard {X}-ray loop-top
  source from flare 2004 {F}ebruary 26} {A reconnection-driven model of the
  hard {X}-ray loop-top source from flare 2004 {F}ebruary 26}.{\BBCQ}
\newblock
\APACjournalVolNumPages{Ap. J.}{833}{}{211}.
\PrintBackRefs{\CurrentBib}

\bibitem [\protect \citeauthoryear {%
Lu%
, Angelopoulos%
\BCBL {}\ \BBA {} Fu%
}{%
Lu%
\ \protect \BOthers {.}}{%
{\protect \APACyear {2016}}%
}]{%
lu_2016_JGR}
\APACinsertmetastar {%
lu_2016_JGR}%
\begin{APACrefauthors}%
Lu, S.%
, Angelopoulos, V.%
\BCBL {}\ \BBA {} Fu, H.%
\end{APACrefauthors}%
\unskip\
\newblock
\APACrefYearMonthDay{2016}{}{}.
\newblock
{\BBOQ}\APACrefatitle {Suprathermal particle energization in dipolarization
  fronts: {Particle}-in-cell simulations} {Suprathermal particle energization
  in dipolarization fronts: {Particle}-in-cell simulations}.{\BBCQ}
\newblock
\APACjournalVolNumPages{Journal of Geophysical Research: Space
  Physics}{121}{10}{}.
\newblock
\begin{APACrefDOI} \doi{https://doi.org/10.1002/2016JA022815} \end{APACrefDOI}
\PrintBackRefs{\CurrentBib}

\bibitem [\protect \citeauthoryear {%
Ma%
, Zhou%
, Zhong%
\BCBL {}\ \BBA {} Deng%
}{%
Ma%
\ \protect \BOthers {.}}{%
{\protect \APACyear {2020}}%
}]{%
Ma20}
\APACinsertmetastar {%
Ma20}%
\begin{APACrefauthors}%
Ma, W.%
, Zhou, M.%
, Zhong, Z.%
\BCBL {}\ \BBA {} Deng, X.%
\end{APACrefauthors}%
\unskip\
\newblock
\APACrefYearMonthDay{2020}{}{}.
\newblock
{\BBOQ}\APACrefatitle {Electron Acceleration Rate at Dipolarization Fronts}
  {Electron acceleration rate at dipolarization fronts}.{\BBCQ}
\newblock
\APACjournalVolNumPages{Ap.~J.}{903}{}{84}.
\PrintBackRefs{\CurrentBib}

\bibitem [\protect \citeauthoryear {%
McPherron%
}{%
McPherron%
}{%
{\protect \APACyear {1979}}%
}]{%
McPherron79}
\APACinsertmetastar {%
McPherron79}%
\begin{APACrefauthors}%
McPherron, R\BPBI L.%
\end{APACrefauthors}%
\unskip\
\newblock
\APACrefYearMonthDay{1979}{}{}.
\newblock
{\BBOQ}\APACrefatitle {Magnetospheric substorms} {Magnetospheric
  substorms}.{\BBCQ}
\newblock
\APACjournalVolNumPages{Reviews of Geophysics}{17}{4}{657-681}.
\newblock
\begin{APACrefDOI} \doi{https://doi.org/10.1029/RG017i004p00657}
  \end{APACrefDOI}
\PrintBackRefs{\CurrentBib}

\bibitem [\protect \citeauthoryear {%
Min%
\ \BBA {} Liu%
}{%
Min%
\ \BBA {} Liu%
}{%
{\protect \APACyear {2016}}%
}]{%
min&liu_2016a}
\APACinsertmetastar {%
min&liu_2016a}%
\begin{APACrefauthors}%
Min, K.%
\BCBT {}\ \BBA {} Liu, K.%
\end{APACrefauthors}%
\unskip\
\newblock
\APACrefYearMonthDay{2016}{}{}.
\newblock
{\BBOQ}\APACrefatitle {Proton velocity ring‐driven instabilities in the inner
  magnetosphere: Linear theory and particle‐in‐cell simulations} {Proton
  velocity ring‐driven instabilities in the inner magnetosphere: Linear
  theory and particle‐in‐cell simulations}.{\BBCQ}
\newblock
\APACjournalVolNumPages{Journal of Geophysical Research: Space
  Physics}{121}{}{475-491}.
\newblock
\begin{APACrefDOI} \doi{10.1002/2015ja022042} \end{APACrefDOI}
\PrintBackRefs{\CurrentBib}

\bibitem [\protect \citeauthoryear {%
Ohtani%
, Shay%
\BCBL {}\ \BBA {} Mukai%
}{%
Ohtani%
\ \protect \BOthers {.}}{%
{\protect \APACyear {2004}}%
}]{%
ohtani_2004_JGR}
\APACinsertmetastar {%
ohtani_2004_JGR}%
\begin{APACrefauthors}%
Ohtani, S\BHBI i.%
, Shay, M\BPBI A.%
\BCBL {}\ \BBA {} Mukai, T.%
\end{APACrefauthors}%
\unskip\
\newblock
\APACrefYearMonthDay{2004}{}{}.
\newblock
{\BBOQ}\APACrefatitle {Temporal structure of the fast convective flow in the
  plasma sheet: {Comparison} between observations and two-fluid simulations}
  {Temporal structure of the fast convective flow in the plasma sheet:
  {Comparison} between observations and two-fluid simulations}.{\BBCQ}
\newblock
\APACjournalVolNumPages{Journal of Geophysical Research: Space
  Physics}{109}{A3}{}.
\newblock
\begin{APACrefDOI} \doi{https://doi.org/10.1029/2003JA010002} \end{APACrefDOI}
\PrintBackRefs{\CurrentBib}

\bibitem [\protect \citeauthoryear {%
Pan%
, Ashour-Abdalla%
, El-Alaoui%
, Walker%
\BCBL {}\ \BBA {} Goldstein%
}{%
Pan%
\ \protect \BOthers {.}}{%
{\protect \APACyear {2012}}%
}]{%
Pan12}
\APACinsertmetastar {%
Pan12}%
\begin{APACrefauthors}%
Pan, Q.%
, Ashour-Abdalla, M.%
, El-Alaoui, M.%
, Walker, R\BPBI J.%
\BCBL {}\ \BBA {} Goldstein, M\BPBI L.%
\end{APACrefauthors}%
\unskip\
\newblock
\APACrefYearMonthDay{2012}{}{}.
\newblock
{\BBOQ}\APACrefatitle {Adiabatic acceleration of suprathermal electrons
  associated with dipolarization fronts} {Adiabatic acceleration of
  suprathermal electrons associated with dipolarization fronts}.{\BBCQ}
\newblock
\APACjournalVolNumPages{Journal of Geophysical Research: Space
  Physics}{117}{A12}{{A12224}}.
\newblock
\begin{APACrefDOI} \doi{https://doi.org/10.1029/2012JA018156} \end{APACrefDOI}
\PrintBackRefs{\CurrentBib}

\bibitem [\protect \citeauthoryear {%
Priest%
\ \BBA {} Forbes%
}{%
Priest%
\ \BBA {} Forbes%
}{%
{\protect \APACyear {2002}}%
}]{%
Priest02}
\APACinsertmetastar {%
Priest02}%
\begin{APACrefauthors}%
Priest, E\BPBI R.%
\BCBT {}\ \BBA {} Forbes, T\BPBI R.%
\end{APACrefauthors}%
\unskip\
\newblock
\APACrefYearMonthDay{2002}{}{}.
\newblock
{\BBOQ}\APACrefatitle {The magnetic nature of solar flares} {The magnetic
  nature of solar flares}.{\BBCQ}
\newblock
\APACjournalVolNumPages{Astron.~Astrophs.~Rev.}{10}{}{313-377}.
\PrintBackRefs{\CurrentBib}

\bibitem [\protect \citeauthoryear {%
Pritchett%
}{%
Pritchett%
}{%
{\protect \APACyear {2013}}%
}]{%
pritchett:2013}
\APACinsertmetastar {%
pritchett:2013}%
\begin{APACrefauthors}%
Pritchett, P\BPBI L.%
\end{APACrefauthors}%
\unskip\
\newblock
\APACrefYearMonthDay{2013}{}{}.
\newblock
{\BBOQ}\APACrefatitle {{The onset of magnetic reconnection in three
  dimensions}} {{The onset of magnetic reconnection in three
  dimensions}}.{\BBCQ}
\newblock
\APACjournalVolNumPages{Phys. Plasmas}{20}{}{080703}.
\PrintBackRefs{\CurrentBib}

\bibitem [\protect \citeauthoryear {%
Pulkkinen%
}{%
Pulkkinen%
}{%
{\protect \APACyear {2007}}%
}]{%
Pulkkinen07}
\APACinsertmetastar {%
Pulkkinen07}%
\begin{APACrefauthors}%
Pulkkinen, T.%
\end{APACrefauthors}%
\unskip\
\newblock
\APACrefYearMonthDay{2007}{}{}.
\newblock
{\BBOQ}\APACrefatitle {Space Weather: Terrestrial Perspective} {Space weather:
  Terrestrial perspective}.{\BBCQ}
\newblock
\APACjournalVolNumPages{Living Reviews in Solar Physics}{4}{}{1}.
\newblock
\begin{APACrefDOI} \doi{10.12942/lrsp-2007-1} \end{APACrefDOI}
\PrintBackRefs{\CurrentBib}

\bibitem [\protect \citeauthoryear {%
Qiu%
, Longcope%
, Cassak%
\BCBL {}\ \BBA {} Priest%
}{%
Qiu%
\ \protect \BOthers {.}}{%
{\protect \APACyear {2017}}%
}]{%
Qiu17}
\APACinsertmetastar {%
Qiu17}%
\begin{APACrefauthors}%
Qiu, J.%
, Longcope, D\BPBI W.%
, Cassak, P\BPBI A.%
\BCBL {}\ \BBA {} Priest, E\BPBI R.%
\end{APACrefauthors}%
\unskip\
\newblock
\APACrefYearMonthDay{2017}{}{}.
\newblock
{\BBOQ}\APACrefatitle {Elongation of Flare Ribbons} {Elongation of flare
  ribbons}.{\BBCQ}
\newblock
\APACjournalVolNumPages{Ap.~J.}{838}{}{17}.
\PrintBackRefs{\CurrentBib}

\bibitem [\protect \citeauthoryear {%
Reeves%
\ \protect \BOthers {.}}{%
Reeves%
\ \protect \BOthers {.}}{%
{\protect \APACyear {2008}}%
}]{%
Reeves08}
\APACinsertmetastar {%
Reeves08}%
\begin{APACrefauthors}%
Reeves, K\BPBI K.%
, Guild, T\BPBI B.%
, Hughes, W\BPBI J.%
, Korreck, K\BPBI E.%
, Lin, J.%
, Raymond, J.%
\BDBL {}Wiltberger, M.%
\end{APACrefauthors}%
\unskip\
\newblock
\APACrefYearMonthDay{2008}{}{}.
\newblock
{\BBOQ}\APACrefatitle {Posteruptive phenomena in coronal mass ejections and
  substorms: Indicators of a universal process?} {Posteruptive phenomena in
  coronal mass ejections and substorms: Indicators of a universal
  process?}{\BBCQ}
\newblock
\APACjournalVolNumPages{J.~Geophys.~Res.}{113}{}{A00B02}.
\PrintBackRefs{\CurrentBib}

\bibitem [\protect \citeauthoryear {%
Roytershteyn%
\ \BBA {} Delzanno%
}{%
Roytershteyn%
\ \BBA {} Delzanno%
}{%
{\protect \APACyear {2018}}%
}]{%
roytershteyn&delzanno2018}
\APACinsertmetastar {%
roytershteyn&delzanno2018}%
\begin{APACrefauthors}%
Roytershteyn, V.%
\BCBT {}\ \BBA {} Delzanno, G\BPBI L.%
\end{APACrefauthors}%
\unskip\
\newblock
\APACrefYearMonthDay{2018}{}{}.
\newblock
{\BBOQ}\APACrefatitle {Spectral {Approach} to {Plasma} {Kinetic} {Simulations}
  {Based} on {Hermite} {Decomposition} in the {Velocity} {Space}} {Spectral
  {Approach} to {Plasma} {Kinetic} {Simulations} {Based} on {Hermite}
  {Decomposition} in the {Velocity} {Space}}.{\BBCQ}
\newblock
\APACjournalVolNumPages{Frontiers in Astronomy and Space Sciences}{5}{}{}.
\PrintBackRefs{\CurrentBib}

\bibitem [\protect \citeauthoryear {%
Runov%
\ \protect \BOthers {.}}{%
Runov%
\ \protect \BOthers {.}}{%
{\protect \APACyear {2015}}%
}]{%
runov_2015_JGR}
\APACinsertmetastar {%
runov_2015_JGR}%
\begin{APACrefauthors}%
Runov, A.%
, Angelopoulos, V.%
, Gabrielse, C.%
, Liu, J.%
, Turner, D\BPBI L.%
\BCBL {}\ \BBA {} Zhou, X\BHBI Z.%
\end{APACrefauthors}%
\unskip\
\newblock
\APACrefYearMonthDay{2015}{}{}.
\newblock
{\BBOQ}\APACrefatitle {Average thermodynamic and spectral properties of plasma
  in and around dipolarizing flux bundles} {Average thermodynamic and spectral
  properties of plasma in and around dipolarizing flux bundles}.{\BBCQ}
\newblock
\APACjournalVolNumPages{Journal of Geophysical Research: Space
  Physics}{120}{6}{4369--4383}.
\newblock
\begin{APACrefDOI} \doi{https://doi.org/10.1002/2015JA021166} \end{APACrefDOI}
\PrintBackRefs{\CurrentBib}

\bibitem [\protect \citeauthoryear {%
Runov%
\ \protect \BOthers {.}}{%
Runov%
\ \protect \BOthers {.}}{%
{\protect \APACyear {2013}}%
}]{%
runov_2013_JGR}
\APACinsertmetastar {%
runov_2013_JGR}%
\begin{APACrefauthors}%
Runov, A.%
, Angelopoulos, V.%
, Gabrielse, C.%
, Zhou, X\BHBI Z.%
, Turner, D.%
\BCBL {}\ \BBA {} Plaschke, F.%
\end{APACrefauthors}%
\unskip\
\newblock
\APACrefYearMonthDay{2013}{}{}.
\newblock
{\BBOQ}\APACrefatitle {Electron fluxes and pitch-angle distributions at
  dipolarization fronts: {THEMIS} multipoint observations} {Electron fluxes and
  pitch-angle distributions at dipolarization fronts: {THEMIS} multipoint
  observations}.{\BBCQ}
\newblock
\APACjournalVolNumPages{Journal of Geophysical Research: Space
  Physics}{118}{2}{744--755}.
\newblock
\begin{APACrefDOI} \doi{https://doi.org/10.1002/jgra.50121} \end{APACrefDOI}
\PrintBackRefs{\CurrentBib}

\bibitem [\protect \citeauthoryear {%
Runov%
\ \protect \BOthers {.}}{%
Runov%
\ \protect \BOthers {.}}{%
{\protect \APACyear {2010}}%
}]{%
runov_2010_Planet_Sci}
\APACinsertmetastar {%
runov_2010_Planet_Sci}%
\begin{APACrefauthors}%
Runov, A.%
, Angelopoulos, V.%
, Sitnov, M.%
, Sergeev, V\BPBI A.%
, Nakamura, R.%
, Nishimura, Y.%
\BDBL {}Singer, H\BPBI J.%
\end{APACrefauthors}%
\unskip\
\newblock
\APACrefYearMonthDay{2010}{}{}.
\newblock
{\BBOQ}\APACrefatitle {Dipolarization fronts in the magnetotail plasma sheet}
  {Dipolarization fronts in the magnetotail plasma sheet}.{\BBCQ}
\newblock
\APACjournalVolNumPages{Planetary and Space Science}{59}{7}{517--525}.
\newblock
\begin{APACrefDOI} \doi{10.1016/j.pss.2010.06.006} \end{APACrefDOI}
\PrintBackRefs{\CurrentBib}

\bibitem [\protect \citeauthoryear {%
Runov%
\ \protect \BOthers {.}}{%
Runov%
\ \protect \BOthers {.}}{%
{\protect \APACyear {2009}}%
}]{%
runov_2009_GRL}
\APACinsertmetastar {%
runov_2009_GRL}%
\begin{APACrefauthors}%
Runov, A.%
, Angelopoulos, V.%
, Sitnov, M\BPBI I.%
, Sergeev, V\BPBI A.%
, Bonnell, J.%
, McFadden, J\BPBI P.%
\BDBL {}Auster, U.%
\end{APACrefauthors}%
\unskip\
\newblock
\APACrefYearMonthDay{2009}{}{}.
\newblock
{\BBOQ}\APACrefatitle {{THEMIS} observations of an earthward-propagating
  dipolarization front} {{THEMIS} observations of an earthward-propagating
  dipolarization front}.{\BBCQ}
\newblock
\APACjournalVolNumPages{Geophysical Research Letters}{36}{14}{}.
\newblock
\begin{APACrefDOI} \doi{https://doi.org/10.1029/2009GL038980} \end{APACrefDOI}
\PrintBackRefs{\CurrentBib}

\bibitem [\protect \citeauthoryear {%
Runov%
\ \protect \BOthers {.}}{%
Runov%
\ \protect \BOthers {.}}{%
{\protect \APACyear {2011}}%
}]{%
Runov11}
\APACinsertmetastar {%
Runov11}%
\begin{APACrefauthors}%
Runov, A.%
, Angelopoulos, V.%
, Zhou, X\BHBI Z.%
, Zhang, X\BHBI J.%
, Li, S.%
, Plaschke, F.%
\BCBL {}\ \BBA {} Bonnell, J.%
\end{APACrefauthors}%
\unskip\
\newblock
\APACrefYearMonthDay{2011}{}{}.
\newblock
{\BBOQ}\APACrefatitle {A {THEMIS} multicase study of dipolarization fronts in
  the magnetotail plasma sheet} {A {THEMIS} multicase study of dipolarization
  fronts in the magnetotail plasma sheet}.{\BBCQ}
\newblock
\APACjournalVolNumPages{Journal of Geophysical Research: Space
  Physics}{116}{A5}{}.
\newblock
\begin{APACrefDOI} \doi{https://doi.org/10.1029/2010JA016316} \end{APACrefDOI}
\PrintBackRefs{\CurrentBib}

\bibitem [\protect \citeauthoryear {%
Schmid%
\ \protect \BOthers {.}}{%
Schmid%
\ \protect \BOthers {.}}{%
{\protect \APACyear {2016}}%
}]{%
Schmid16}
\APACinsertmetastar {%
Schmid16}%
\begin{APACrefauthors}%
Schmid, D.%
, Nakamura, R.%
, Volwerk, M.%
, Plaschke, F.%
, Narita, Y.%
, Baumjohann, W.%
\BDBL {}Kepko, E\BPBI L.%
\end{APACrefauthors}%
\unskip\
\newblock
\APACrefYearMonthDay{2016}{}{}.
\newblock
{\BBOQ}\APACrefatitle {A comparative study of dipolarization fronts at {MMS}
  and {C}luster} {A comparative study of dipolarization fronts at {MMS} and
  {C}luster}.{\BBCQ}
\newblock
\APACjournalVolNumPages{Geophysical Research Letters}{43}{12}{6012-6019}.
\newblock
\begin{APACrefDOI} \doi{https://doi.org/10.1002/2016GL069520} \end{APACrefDOI}
\PrintBackRefs{\CurrentBib}

\bibitem [\protect \citeauthoryear {%
Schmid%
, Volwerk%
, Nakamura%
, Baumjohann%
\BCBL {}\ \BBA {} Heyn%
}{%
Schmid%
\ \protect \BOthers {.}}{%
{\protect \APACyear {2011}}%
}]{%
Schmid11}
\APACinsertmetastar {%
Schmid11}%
\begin{APACrefauthors}%
Schmid, D.%
, Volwerk, M.%
, Nakamura, R.%
, Baumjohann, W.%
\BCBL {}\ \BBA {} Heyn, M.%
\end{APACrefauthors}%
\unskip\
\newblock
\APACrefYearMonthDay{2011}{}{}.
\newblock
{\BBOQ}\APACrefatitle {A statistical and event study of magnetotail
  dipolarization fronts} {A statistical and event study of magnetotail
  dipolarization fronts}.{\BBCQ}
\newblock
\APACjournalVolNumPages{Annales Geophysicae}{29}{9}{1537--1547}.
\newblock
\begin{APACrefDOI} \doi{10.5194/angeo-29-1537-2011} \end{APACrefDOI}
\PrintBackRefs{\CurrentBib}

\bibitem [\protect \citeauthoryear {%
Sharma~Pyakurel%
\ \protect \BOthers {.}}{%
Sharma~Pyakurel%
\ \protect \BOthers {.}}{%
{\protect \APACyear {2019}}%
}]{%
Pyakurel19}
\APACinsertmetastar {%
Pyakurel19}%
\begin{APACrefauthors}%
Sharma~Pyakurel, P.%
, Shay, M\BPBI A.%
, Phan, T\BPBI D.%
, Matthaeus, W\BPBI H.%
, Drake, J\BPBI F.%
, TenBarge, J\BPBI M.%
\BDBL {}Chasapis, A.%
\end{APACrefauthors}%
\unskip\
\newblock
\APACrefYearMonthDay{2019}{}{}.
\newblock
{\BBOQ}\APACrefatitle {Transition from ion-coupled to electron-only
  reconnection: Basic physics and implications for plasma turbulence}
  {Transition from ion-coupled to electron-only reconnection: Basic physics and
  implications for plasma turbulence}.{\BBCQ}
\newblock
\APACjournalVolNumPages{Physics of Plasmas}{26}{8}{082307}.
\newblock
\begin{APACrefURL} \url{https://doi.org/10.1063/1.5090403} \end{APACrefURL}
\newblock
\begin{APACrefDOI} \doi{10.1063/1.5090403} \end{APACrefDOI}
\PrintBackRefs{\CurrentBib}

\bibitem [\protect \citeauthoryear {%
Shay%
, Drake%
, Rogers%
\BCBL {}\ \BBA {} Denton%
}{%
Shay%
\ \protect \BOthers {.}}{%
{\protect \APACyear {2001}}%
}]{%
Shay01}
\APACinsertmetastar {%
Shay01}%
\begin{APACrefauthors}%
Shay, M\BPBI A.%
, Drake, J\BPBI F.%
, Rogers, B\BPBI N.%
\BCBL {}\ \BBA {} Denton, R\BPBI E.%
\end{APACrefauthors}%
\unskip\
\newblock
\APACrefYearMonthDay{2001}{}{}.
\newblock
{\BBOQ}\APACrefatitle {Alfv\'enic collisionless reconnection and the {H}all
  term} {Alfv\'enic collisionless reconnection and the {H}all term}.{\BBCQ}
\newblock
\APACjournalVolNumPages{J. Geophys. Res.}{106}{}{3751}.
\PrintBackRefs{\CurrentBib}

\bibitem [\protect \citeauthoryear {%
Shay%
\ \protect \BOthers {.}}{%
Shay%
\ \protect \BOthers {.}}{%
{\protect \APACyear {2014}}%
}]{%
Shay14}
\APACinsertmetastar {%
Shay14}%
\begin{APACrefauthors}%
Shay, M\BPBI A.%
, Haggerty, C\BPBI C.%
, Phan, T\BPBI D.%
, Drake, J\BPBI F.%
, Cassak, P\BPBI A.%
, Wu, P.%
\BDBL {}Malakit, K.%
\end{APACrefauthors}%
\unskip\
\newblock
\APACrefYearMonthDay{2014}{}{}.
\newblock
{\BBOQ}\APACrefatitle {Electron heating during magnetic reconnection: A
  simulation scaling study} {Electron heating during magnetic reconnection: A
  simulation scaling study}.{\BBCQ}
\newblock
\APACjournalVolNumPages{Phys.~Plasmas}{21}{}{122902}.
\PrintBackRefs{\CurrentBib}

\bibitem [\protect \citeauthoryear {%
Shuster%
\ \protect \BOthers {.}}{%
Shuster%
\ \protect \BOthers {.}}{%
{\protect \APACyear {2014}}%
}]{%
Shuster2014}
\APACinsertmetastar {%
Shuster2014}%
\begin{APACrefauthors}%
Shuster, J\BPBI R.%
, Chen, L\BHBI J.%
, Daughton, W\BPBI S.%
, Lee, L\BPBI C.%
, Lee, K\BPBI H.%
, Bessho, N.%
\BDBL {}Argall, M\BPBI R.%
\end{APACrefauthors}%
\unskip\
\newblock
\APACrefYearMonthDay{2014}{08}{}.
\newblock
{\BBOQ}\APACrefatitle {Highly structured electron anisotropy in collisionless
  reconnection exhausts} {Highly structured electron anisotropy in
  collisionless reconnection exhausts}.{\BBCQ}
\newblock
\APACjournalVolNumPages{Geophysical Research Letters}{41}{}{5389-5395}.
\newblock
\begin{APACrefDOI} \doi{10.1002/2014gl060608} \end{APACrefDOI}
\PrintBackRefs{\CurrentBib}

\bibitem [\protect \citeauthoryear {%
Shuster%
\ \protect \BOthers {.}}{%
Shuster%
\ \protect \BOthers {.}}{%
{\protect \APACyear {2015}}%
}]{%
shuster_2015}
\APACinsertmetastar {%
shuster_2015}%
\begin{APACrefauthors}%
Shuster, J\BPBI R.%
, Chen, L\BHBI J.%
, Hesse, M.%
, Argall, M\BPBI R.%
, Daughton, W.%
, Torbert, R\BPBI B.%
\BCBL {}\ \BBA {} Bessho, N.%
\end{APACrefauthors}%
\unskip\
\newblock
\APACrefYearMonthDay{2015}{04}{}.
\newblock
{\BBOQ}\APACrefatitle {Spatiotemporal evolution of electron characteristics in
  the electron diffusion region of magnetic reconnection: Implications for
  acceleration and heating} {Spatiotemporal evolution of electron
  characteristics in the electron diffusion region of magnetic reconnection:
  Implications for acceleration and heating}.{\BBCQ}
\newblock
\APACjournalVolNumPages{Geophysical Research Letters}{42}{}{2586-2593}.
\newblock
\begin{APACrefDOI} \doi{10.1002/2015gl063601} \end{APACrefDOI}
\PrintBackRefs{\CurrentBib}

\bibitem [\protect \citeauthoryear {%
Sitnov%
, Buzulukova%
, Swisdak%
, Merkin%
\BCBL {}\ \BBA {} Moore%
}{%
Sitnov%
\ \protect \BOthers {.}}{%
{\protect \APACyear {2013}}%
}]{%
sitnov:2013}
\APACinsertmetastar {%
sitnov:2013}%
\begin{APACrefauthors}%
Sitnov, M\BPBI I.%
, Buzulukova, N.%
, Swisdak, M.%
, Merkin, V\BPBI G.%
\BCBL {}\ \BBA {} Moore, T\BPBI E.%
\end{APACrefauthors}%
\unskip\
\newblock
\APACrefYearMonthDay{2013}{}{}.
\newblock
{\BBOQ}\APACrefatitle {{Spontaneous formation of dipolarization fronts and
  reconnection onset in the magnetotail}} {{Spontaneous formation of
  dipolarization fronts and reconnection onset in the magnetotail}}.{\BBCQ}
\newblock
\APACjournalVolNumPages{Geophys. Res. Lett.}{40}{1}{22--27}.
\newblock
\begin{APACrefDOI} \doi{10.1029/2012GL054701} \end{APACrefDOI}
\PrintBackRefs{\CurrentBib}

\bibitem [\protect \citeauthoryear {%
Sitnov%
\ \protect \BOthers {.}}{%
Sitnov%
\ \protect \BOthers {.}}{%
{\protect \APACyear {2014}}%
}]{%
Sitnov14}
\APACinsertmetastar {%
Sitnov14}%
\begin{APACrefauthors}%
Sitnov, M\BPBI I.%
, Merkin, V\BPBI G.%
, Swisdak, M.%
, Motoba, T.%
, Buzulukova, N.%
, Moore, T\BPBI E.%
\BDBL {}Ohtani, S.%
\end{APACrefauthors}%
\unskip\
\newblock
\APACrefYearMonthDay{2014}{}{}.
\newblock
{\BBOQ}\APACrefatitle {Magnetic reconnection, buoyancy, and flapping motions in
  magnetotail explosions} {Magnetic reconnection, buoyancy, and flapping
  motions in magnetotail explosions}.{\BBCQ}
\newblock
\APACjournalVolNumPages{Journal of Geophysical Research: Space
  Physics}{119}{9}{7151-7168}.
\newblock
\begin{APACrefDOI} \doi{https://doi.org/10.1002/2014JA020205} \end{APACrefDOI}
\PrintBackRefs{\CurrentBib}

\bibitem [\protect \citeauthoryear {%
Sitnov%
\ \BBA {} Swisdak%
}{%
Sitnov%
\ \BBA {} Swisdak%
}{%
{\protect \APACyear {2011}}%
}]{%
sitnov:2011}
\APACinsertmetastar {%
sitnov:2011}%
\begin{APACrefauthors}%
Sitnov, M\BPBI I.%
\BCBT {}\ \BBA {} Swisdak, M.%
\end{APACrefauthors}%
\unskip\
\newblock
\APACrefYearMonthDay{2011}{}{}.
\newblock
{\BBOQ}\APACrefatitle {{Onset of collisionless magnetic reconnection in
  two-dimensional current sheets and formation of dipolarization fronts}}
  {{Onset of collisionless magnetic reconnection in two-dimensional current
  sheets and formation of dipolarization fronts}}.{\BBCQ}
\newblock
\APACjournalVolNumPages{J. Geophys. Res.}{116}{A}{12216}.
\newblock
\begin{APACrefDOI} \doi{10.1029/2011JA016920} \end{APACrefDOI}
\PrintBackRefs{\CurrentBib}

\bibitem [\protect \citeauthoryear {%
Sitnov%
, Swisdak%
\BCBL {}\ \BBA {} Divin%
}{%
Sitnov%
\ \protect \BOthers {.}}{%
{\protect \APACyear {2009}}%
}]{%
sitnov_JGR_2009}
\APACinsertmetastar {%
sitnov_JGR_2009}%
\begin{APACrefauthors}%
Sitnov, M\BPBI I.%
, Swisdak, M.%
\BCBL {}\ \BBA {} Divin, A\BPBI V.%
\end{APACrefauthors}%
\unskip\
\newblock
\APACrefYearMonthDay{2009}{}{}.
\newblock
{\BBOQ}\APACrefatitle {Dipolarization fronts as a signature of transient
  reconnection in the magnetotail} {Dipolarization fronts as a signature of
  transient reconnection in the magnetotail}.{\BBCQ}
\newblock
\APACjournalVolNumPages{Journal of Geophysical Research: Space
  Physics}{114}{A4}{}.
\newblock
\begin{APACrefDOI} \doi{https://doi.org/10.1029/2008JA013980} \end{APACrefDOI}
\PrintBackRefs{\CurrentBib}

\bibitem [\protect \citeauthoryear {%
Smith%
\ \protect \BOthers {.}}{%
Smith%
\ \protect \BOthers {.}}{%
{\protect \APACyear {2018}}%
}]{%
Smith18}
\APACinsertmetastar {%
Smith18}%
\begin{APACrefauthors}%
Smith, A\BPBI W.%
, Jackman, C\BPBI M.%
, Thomsen, M\BPBI F.%
, Sergis, N.%
, Mitchell, D\BPBI G.%
\BCBL {}\ \BBA {} Roussos, E.%
\end{APACrefauthors}%
\unskip\
\newblock
\APACrefYearMonthDay{2018}{}{}.
\newblock
{\BBOQ}\APACrefatitle {Dipolarization Fronts With Associated Energized
  Electrons in Saturn's Magnetotail} {Dipolarization fronts with associated
  energized electrons in saturn's magnetotail}.{\BBCQ}
\newblock
\APACjournalVolNumPages{Journal of Geophysical Research: Space
  Physics}{123}{4}{2714-2735}.
\newblock
\begin{APACrefDOI} \doi{https://doi.org/10.1002/2017JA024904} \end{APACrefDOI}
\PrintBackRefs{\CurrentBib}

\bibitem [\protect \citeauthoryear {%
Tang%
, Wang%
\BCBL {}\ \BBA {} Zhou%
}{%
Tang%
\ \protect \BOthers {.}}{%
{\protect \APACyear {2021}}%
}]{%
Tang21}
\APACinsertmetastar {%
Tang21}%
\begin{APACrefauthors}%
Tang, C\BPBI L.%
, Wang, X.%
\BCBL {}\ \BBA {} Zhou, M.%
\end{APACrefauthors}%
\unskip\
\newblock
\APACrefYearMonthDay{2021}{}{}.
\newblock
{\BBOQ}\APACrefatitle {Electron Pitch Angle Distributions Around Dipolarization
  Fronts at the Off Magnetic Equator} {Electron pitch angle distributions
  around dipolarization fronts at the off magnetic equator}.{\BBCQ}
\newblock
\APACjournalVolNumPages{Journal of Geophysical Research: Space
  Physics}{126}{2}{e2020JA028787}.
\newblock
\begin{APACrefDOI} \doi{https://doi.org/10.1029/2020JA028787} \end{APACrefDOI}
\PrintBackRefs{\CurrentBib}

\bibitem [\protect \citeauthoryear {%
Trottenberg%
, Oosterlee%
\BCBL {}\ \BBA {} Schuller%
}{%
Trottenberg%
\ \protect \BOthers {.}}{%
{\protect \APACyear {2000}}%
}]{%
Trottenberg00}
\APACinsertmetastar {%
Trottenberg00}%
\begin{APACrefauthors}%
Trottenberg, U.%
, Oosterlee, C\BPBI W.%
\BCBL {}\ \BBA {} Schuller, A.%
\end{APACrefauthors}%
\unskip\
\newblock
\APACrefYear{2000}.
\newblock
\APACrefbtitle {Multigrid} {Multigrid}.
\newblock
\APACaddressPublisher{}{Academic Press, San Diego}.
\PrintBackRefs{\CurrentBib}

\bibitem [\protect \citeauthoryear {%
Umeda%
, Ashour-Abdalla%
, Schriver%
, Richard%
\BCBL {}\ \BBA {} Coroniti%
}{%
Umeda%
\ \protect \BOthers {.}}{%
{\protect \APACyear {2007}}%
}]{%
Umeda2007}
\APACinsertmetastar {%
Umeda2007}%
\begin{APACrefauthors}%
Umeda, T.%
, Ashour-Abdalla, M.%
, Schriver, D.%
, Richard, R\BPBI L.%
\BCBL {}\ \BBA {} Coroniti, F\BPBI V.%
\end{APACrefauthors}%
\unskip\
\newblock
\APACrefYearMonthDay{2007}{}{}.
\newblock
{\BBOQ}\APACrefatitle {Particle-in-cell simulation of {Maxwellian} ring
  velocity distribution} {Particle-in-cell simulation of {Maxwellian} ring
  velocity distribution}.{\BBCQ}
\newblock
\APACjournalVolNumPages{Journal of Geophysical Research: Space
  Physics}{112}{A4}{}.
\newblock
\begin{APACrefDOI} \doi{10.1029/2006JA012124} \end{APACrefDOI}
\PrintBackRefs{\CurrentBib}

\bibitem [\protect \citeauthoryear {%
Viberg%
\ \protect \BOthers {.}}{%
Viberg%
\ \protect \BOthers {.}}{%
{\protect \APACyear {2014}}%
}]{%
Viberg14}
\APACinsertmetastar {%
Viberg14}%
\begin{APACrefauthors}%
Viberg, H.%
, Khotyaintsev, Y\BPBI V.%
, Vaivads, A.%
, André, M.%
, Fu, H\BPBI S.%
\BCBL {}\ \BBA {} Cornilleau-Wehrlin, N.%
\end{APACrefauthors}%
\unskip\
\newblock
\APACrefYearMonthDay{2014}{}{}.
\newblock
{\BBOQ}\APACrefatitle {Whistler mode waves at magnetotail dipolarization
  fronts} {Whistler mode waves at magnetotail dipolarization fronts}.{\BBCQ}
\newblock
\APACjournalVolNumPages{Journal of Geophysical Research: Space
  Physics}{119}{4}{2605-2611}.
\newblock
\begin{APACrefDOI} \doi{https://doi.org/10.1002/2014JA019892} \end{APACrefDOI}
\PrintBackRefs{\CurrentBib}

\bibitem [\protect \citeauthoryear {%
Vocks%
\ \BBA {} Mann%
}{%
Vocks%
\ \BBA {} Mann%
}{%
{\protect \APACyear {2006}}%
}]{%
vocks_whistler_2006}
\APACinsertmetastar {%
vocks_whistler_2006}%
\begin{APACrefauthors}%
Vocks, C.%
\BCBT {}\ \BBA {} Mann, G.%
\end{APACrefauthors}%
\unskip\
\newblock
\APACrefYearMonthDay{2006}{}{}.
\newblock
{\BBOQ}\APACrefatitle {Whistler wave excitation by relativistic electrons in
  coronal loops during solar flares} {Whistler wave excitation by relativistic
  electrons in coronal loops during solar flares}.{\BBCQ}
\newblock
\APACjournalVolNumPages{Astronomy \& Astrophysics}{452}{1}{331--337}.
\newblock
\begin{APACrefDOI} \doi{10.1051/0004-6361:20054042} \end{APACrefDOI}
\PrintBackRefs{\CurrentBib}

\bibitem [\protect \citeauthoryear {%
K.~Wang%
\ \protect \BOthers {.}}{%
K.~Wang%
\ \protect \BOthers {.}}{%
{\protect \APACyear {2014}}%
}]{%
Wang14}
\APACinsertmetastar {%
Wang14}%
\begin{APACrefauthors}%
Wang, K.%
, Lin, C\BHBI H.%
, Wang, L\BHBI Y.%
, Hada, T.%
, Nishimura, Y.%
, Turner, D\BPBI L.%
\BCBL {}\ \BBA {} Angelopoulos, V.%
\end{APACrefauthors}%
\unskip\
\newblock
\APACrefYearMonthDay{2014}{}{}.
\newblock
{\BBOQ}\APACrefatitle {Pitch angle distributions of electrons at dipolarization
  sites during geomagnetic activity: {THEMIS} observations} {Pitch angle
  distributions of electrons at dipolarization sites during geomagnetic
  activity: {THEMIS} observations}.{\BBCQ}
\newblock
\APACjournalVolNumPages{Journal of Geophysical Research: Space
  Physics}{119}{12}{9747-9760}.
\newblock
\begin{APACrefDOI} \doi{https://doi.org/10.1002/2014JA020176} \end{APACrefDOI}
\PrintBackRefs{\CurrentBib}

\bibitem [\protect \citeauthoryear {%
S.~Wang%
\ \protect \BOthers {.}}{%
S.~Wang%
\ \protect \BOthers {.}}{%
{\protect \APACyear {2016}}%
}]{%
Wang_2016_electron}
\APACinsertmetastar {%
Wang_2016_electron}%
\begin{APACrefauthors}%
Wang, S.%
, Chen, L.%
, Bessho, N.%
, Kistler, L\BPBI M.%
, Shuster, J\BPBI R.%
\BCBL {}\ \BBA {} Guo, R.%
\end{APACrefauthors}%
\unskip\
\newblock
\APACrefYearMonthDay{2016}{03}{}.
\newblock
{\BBOQ}\APACrefatitle {Electron heating in the exhaust of magnetic reconnection
  with negligible guide field} {Electron heating in the exhaust of magnetic
  reconnection with negligible guide field}.{\BBCQ}
\newblock
\APACjournalVolNumPages{Journal of Geophysical Research: Space
  Physics}{121}{}{2104-2130}.
\newblock
\begin{APACrefDOI} \doi{10.1002/2015ja021892} \end{APACrefDOI}
\PrintBackRefs{\CurrentBib}

\bibitem [\protect \citeauthoryear {%
{Warmuth}%
\ \BBA {} {Mann}%
}{%
{Warmuth}%
\ \BBA {} {Mann}%
}{%
{\protect \APACyear {2016}}%
}]{%
Warmuth16}
\APACinsertmetastar {%
Warmuth16}%
\begin{APACrefauthors}%
{Warmuth}, A.%
\BCBT {}\ \BBA {} {Mann}, G.%
\end{APACrefauthors}%
\unskip\
\newblock
\APACrefYearMonthDay{2016}{}{}.
\newblock
{\BBOQ}\APACrefatitle {{Constraints on energy release in solar flares from
  RHESSI and GOES X-ray observations. I. Physical parameters and scalings}}
  {{Constraints on energy release in solar flares from RHESSI and GOES X-ray
  observations. I. Physical parameters and scalings}}.{\BBCQ}
\newblock
\APACjournalVolNumPages{Astronomy and Astrophysics}{588}{}{A115}.
\newblock
\begin{APACrefDOI} \doi{10.1051/0004-6361/201527474} \end{APACrefDOI}
\PrintBackRefs{\CurrentBib}

\bibitem [\protect \citeauthoryear {%
Winske%
\ \BBA {} Daughton%
}{%
Winske%
\ \BBA {} Daughton%
}{%
{\protect \APACyear {2012}}%
}]{%
Winske&Daughton2012_whistler}
\APACinsertmetastar {%
Winske&Daughton2012_whistler}%
\begin{APACrefauthors}%
Winske, D.%
\BCBT {}\ \BBA {} Daughton, W.%
\end{APACrefauthors}%
\unskip\
\newblock
\APACrefYearMonthDay{2012}{}{}.
\newblock
{\BBOQ}\APACrefatitle {Generation of lower hybrid and whistler waves by an ion
  velocity ring distribution} {Generation of lower hybrid and whistler waves by
  an ion velocity ring distribution}.{\BBCQ}
\newblock
\APACjournalVolNumPages{Physics of Plasmas}{19}{7}{072109}.
\newblock
\begin{APACrefDOI} \doi{10.1063/1.4736983} \end{APACrefDOI}
\PrintBackRefs{\CurrentBib}

\bibitem [\protect \citeauthoryear {%
C\BPBI S.~Wu%
, Yoon%
\BCBL {}\ \BBA {} Freund%
}{%
C\BPBI S.~Wu%
\ \protect \BOthers {.}}{%
{\protect \APACyear {1989}}%
}]{%
wu_1989_a}
\APACinsertmetastar {%
wu_1989_a}%
\begin{APACrefauthors}%
Wu, C\BPBI S.%
, Yoon, P\BPBI H.%
\BCBL {}\ \BBA {} Freund, H\BPBI P.%
\end{APACrefauthors}%
\unskip\
\newblock
\APACrefYearMonthDay{1989}{}{}.
\newblock
{\BBOQ}\APACrefatitle {A theory of electron cyclotron waves generated along
  auroral field lines observed by ground facilities} {A theory of electron
  cyclotron waves generated along auroral field lines observed by ground
  facilities}.{\BBCQ}
\newblock
\APACjournalVolNumPages{Geophysical Research Letters}{16}{}{1461-1464}.
\newblock
\begin{APACrefDOI} \doi{10.1029/gl016i012p01461} \end{APACrefDOI}
\PrintBackRefs{\CurrentBib}

\bibitem [\protect \citeauthoryear {%
M.~Wu%
\ \protect \BOthers {.}}{%
M.~Wu%
\ \protect \BOthers {.}}{%
{\protect \APACyear {2013}}%
}]{%
Wu13}
\APACinsertmetastar {%
Wu13}%
\begin{APACrefauthors}%
Wu, M.%
, Lu, Q.%
, Volwerk, M.%
, Vörös, Z.%
, Zhang, T.%
, Shan, L.%
\BCBL {}\ \BBA {} Huang, C.%
\end{APACrefauthors}%
\unskip\
\newblock
\APACrefYearMonthDay{2013}{}{}.
\newblock
{\BBOQ}\APACrefatitle {A statistical study of electron acceleration behind the
  dipolarization fronts in the magnetotail} {A statistical study of electron
  acceleration behind the dipolarization fronts in the magnetotail}.{\BBCQ}
\newblock
\APACjournalVolNumPages{Journal of Geophysical Research: Space
  Physics}{118}{8}{4804-4810}.
\newblock
\begin{APACrefDOI} \doi{https://doi.org/10.1002/jgra.50456} \end{APACrefDOI}
\PrintBackRefs{\CurrentBib}

\bibitem [\protect \citeauthoryear {%
P.~Wu%
, Fritz%
, Larvaud%
\BCBL {}\ \BBA {} Lucek%
}{%
P.~Wu%
\ \protect \BOthers {.}}{%
{\protect \APACyear {2006}}%
}]{%
Wu06}
\APACinsertmetastar {%
Wu06}%
\begin{APACrefauthors}%
Wu, P.%
, Fritz, T\BPBI A.%
, Larvaud, B.%
\BCBL {}\ \BBA {} Lucek, E.%
\end{APACrefauthors}%
\unskip\
\newblock
\APACrefYearMonthDay{2006}{}{}.
\newblock
{\BBOQ}\APACrefatitle {Substorm associated magnetotail energetic electrons
  pitch angle evolutions and flow reversals: Cluster observation} {Substorm
  associated magnetotail energetic electrons pitch angle evolutions and flow
  reversals: Cluster observation}.{\BBCQ}
\newblock
\APACjournalVolNumPages{Geophysical Research Letters}{33}{17}{{L17101}}.
\newblock
\begin{APACrefDOI} \doi{https://doi.org/10.1029/2006GL026595} \end{APACrefDOI}
\PrintBackRefs{\CurrentBib}

\bibitem [\protect \citeauthoryear {%
S\BPBI B.~Xu%
\ \protect \BOthers {.}}{%
S\BPBI B.~Xu%
\ \protect \BOthers {.}}{%
{\protect \APACyear {2021}}%
}]{%
Xu21}
\APACinsertmetastar {%
Xu21}%
\begin{APACrefauthors}%
Xu, S\BPBI B.%
, Huang, S\BPBI Y.%
, Yuan, Z\BPBI G.%
, Deng, X\BPBI H.%
, Jiang, K.%
, Wei, Y\BPBI Y.%
\BDBL {}Waite, J\BPBI H.%
\end{APACrefauthors}%
\unskip\
\newblock
\APACrefYearMonthDay{2021}{}{}.
\newblock
{\BBOQ}\APACrefatitle {Global Spatial Distribution of Dipolarization Fronts in
  the Saturn's Magnetosphere: Cassini Observations} {Global spatial
  distribution of dipolarization fronts in the saturn's magnetosphere: Cassini
  observations}.{\BBCQ}
\newblock
\APACjournalVolNumPages{Geophysical Research Letters}{48}{17}{e2021GL092701}.
\newblock
\begin{APACrefDOI} \doi{https://doi.org/10.1029/2021GL092701} \end{APACrefDOI}
\PrintBackRefs{\CurrentBib}

\bibitem [\protect \citeauthoryear {%
Y.~Xu%
, Fu%
, Liu%
\BCBL {}\ \BBA {} Wang%
}{%
Y.~Xu%
\ \protect \BOthers {.}}{%
{\protect \APACyear {2018}}%
}]{%
Xu18}
\APACinsertmetastar {%
Xu18}%
\begin{APACrefauthors}%
Xu, Y.%
, Fu, H\BPBI S.%
, Liu, C\BPBI M.%
\BCBL {}\ \BBA {} Wang, T\BPBI Y.%
\end{APACrefauthors}%
\unskip\
\newblock
\APACrefYearMonthDay{2018}{}{}.
\newblock
{\BBOQ}\APACrefatitle {Electron Acceleration by Dipolarization Fronts and
  Magnetic Reconnection: A Quantitative Comparison} {Electron acceleration by
  dipolarization fronts and magnetic reconnection: A quantitative
  comparison}.{\BBCQ}
\newblock
\APACjournalVolNumPages{Ap.~J.~}{853}{}{11}.
\PrintBackRefs{\CurrentBib}

\bibitem [\protect \citeauthoryear {%
Yoo%
\ \protect \BOthers {.}}{%
Yoo%
\ \protect \BOthers {.}}{%
{\protect \APACyear {2019}}%
}]{%
Yoo19}
\APACinsertmetastar {%
Yoo19}%
\begin{APACrefauthors}%
Yoo, J.%
, Wang, S.%
, Yerger, E.%
, Jara-Almonte, J.%
, Ji, H.%
, Yamada, M.%
\BDBL {}Alt, A.%
\end{APACrefauthors}%
\unskip\
\newblock
\APACrefYearMonthDay{2019}{}{}.
\newblock
{\BBOQ}\APACrefatitle {Whistler wave generation by electron temperature
  anisotropy during magnetic reconnection at the magnetopause} {Whistler wave
  generation by electron temperature anisotropy during magnetic reconnection at
  the magnetopause}.{\BBCQ}
\newblock
\APACjournalVolNumPages{Physics of Plasmas}{26}{5}{052902}.
\newblock
\begin{APACrefDOI} \doi{10.1063/1.5094636} \end{APACrefDOI}
\PrintBackRefs{\CurrentBib}

\bibitem [\protect \citeauthoryear {%
Zeiler%
\ \protect \BOthers {.}}{%
Zeiler%
\ \protect \BOthers {.}}{%
{\protect \APACyear {2002}}%
}]{%
zeiler2002}
\APACinsertmetastar {%
zeiler2002}%
\begin{APACrefauthors}%
Zeiler, A.%
, Biskamp, D.%
, Drake, J.%
, Rogers, B.%
, Shay, M.%
\BCBL {}\ \BBA {} Scholer, M.%
\end{APACrefauthors}%
\unskip\
\newblock
\APACrefYearMonthDay{2002}{}{}.
\newblock
{\BBOQ}\APACrefatitle {Three-dimensional particle simulations of collisionless
  magnetic reconnection} {Three-dimensional particle simulations of
  collisionless magnetic reconnection}.{\BBCQ}
\newblock
\APACjournalVolNumPages{Journal of Geophysical Research}{107}{}{1230}.
\newblock
\begin{APACrefDOI} \doi{10.1029/2001ja000287} \end{APACrefDOI}
\PrintBackRefs{\CurrentBib}

\bibitem [\protect \citeauthoryear {%
Zhao%
\ \protect \BOthers {.}}{%
Zhao%
\ \protect \BOthers {.}}{%
{\protect \APACyear {2019}}%
}]{%
Zhao19}
\APACinsertmetastar {%
Zhao19}%
\begin{APACrefauthors}%
Zhao, M\BPBI J.%
, Fu, H\BPBI S.%
, Liu, C\BPBI M.%
, Chen, Z\BPBI Z.%
, Xu, Y.%
, Giles, B\BPBI L.%
\BCBL {}\ \BBA {} Burch, J\BPBI L.%
\end{APACrefauthors}%
\unskip\
\newblock
\APACrefYearMonthDay{2019}{}{}.
\newblock
{\BBOQ}\APACrefatitle {Energy Range of Electron Rolling Pin Distribution Behind
  Dipolarization Front} {Energy range of electron rolling pin distribution
  behind dipolarization front}.{\BBCQ}
\newblock
\APACjournalVolNumPages{Geophysical Research Letters}{46}{5}{2390-2398}.
\newblock
\begin{APACrefDOI} \doi{https://doi.org/10.1029/2019GL082100} \end{APACrefDOI}
\PrintBackRefs{\CurrentBib}

\end{thebibliography}

%
%
%
%
%

\end{document}